\documentclass[fleqn,usenatbib]{mnras}

\usepackage{newtxtext,newtxmath}
\usepackage{graphicx} 
\usepackage{subfigure}

\usepackage[T1]{fontenc}

\DeclareRobustCommand{\VAN}[3]{#2}
\let\VANthebibliography\thebibliography
\def\thebibliography{\DeclareRobustCommand{\VAN}[3]{##3}\VANthebibliography}

\usepackage{amsmath}	
\usepackage{multirow}
\usepackage{longtable}
\usepackage{hyperref}

\usepackage{multicol}

\usepackage{bm} 
\usepackage{ulem} 

\title[C$^{17}$O 1-0 and C$^{18}$O 1-0]{$^{18}$O$/^{17}$O abundance ratio toward a sample of massive star forming regions with parallax distances}

\author[C. Ou et al.]{
Chao Ou,$^{1}$
Junzhi Wang,$^{1}$\thanks{E-mail: junzhiwang@gxu.edu.cn}
Siqi Zheng,$^{2}$
Juan Li,$^{2}$
Donatella Romano,$^{3}$
Zhi-Yu Zhang$^{4,5}$
\\
$^{1}$Guangxi Key Laboratory for Relativistic Astrophysics, Department of Physics, Guangxi University, Nanning 530004, PR China\\
$^{2}$Shanghai Astronomical Observatory, Chinese Academy of Sciences, Shanghai, 200030, PR China\\
$^{3}$INAF, Osservatorio di Astrofisica e Scienza dello Spazio, Via Gobetti 93/3, 40129, Bologna, Italy\\
$^{4}$School  of Astronomy and Space Science, Nanjing University, Nanjing,  210023, PR China\\
$^{5}$Key Laboratory of Modern Astronomy and Astrophysics (Nanjing University), Ministry of Education, Nanjing 210023, PR of China
}

\date{Accepted XXX. Received YYY; in original form ZZZ}

\pubyear{20XX}

\begin{document}
\label{firstpage}
\pagerange{\pageref{firstpage}--\pageref{lastpage}}
\maketitle

\begin{abstract}
The $^{18}$O$/^{17}$O abundance ratio is, in principle, a powerful tool to estimate the relative contributions of massive stars and low- to intermediate-mass stars to the chemical enrichment of galaxies. We present $^{18}$O$/^{17}$O ratios derived from  simultaneous observations of C$^{18}$O and C$^{17}$O 1-0 toward fifty-one massive star forming regions with  the Institut de Radioastronomie Millimétrique (IRAM) 30 meter telescope.   Simultaneous  observations of HC$^{18}$O$^{+}$ 1-0 and HC$^{17}$O$^{+}$ 1-0  with the Yebes 40m telescope toward five sources from this sample were also done to test the consistency of $^{18}$O$/^{17}$O ratios derived from different isotopic pairs.  From our improved measurements, resulting in smaller errors than previous work in the literature, we obtain a clear trend of increasing  $^{18}$O$/^{17}$O ratio with increasing galactocentric distance (D$_{GC}$), which provides a significant constraint on Galactic chemical evolution (GCE) models. Current GCE models have to be improved in order to explain the observed C$^{18}$O/C$^{17}$O 1-0 gradient.  
\end{abstract}




\section{Introduction} \label{sec:intro}
Galactic chemical evolution (GCE) models are useful tools to understand the evolution of stars and  galaxies \citep{1980FCPh....5..287T,1997nceg.book.....P,2021A&ARv..29....5M}. These can be exploited to explore the distribution of chemical elements in galaxies and their formation in stars \citep{2021A&ARv..29....5M}. The interstellar medium (ISM) is often used to study how chemical enrichment proceeds in galaxies \citep{1994ARA&A..32..191W}. The primary chemical components of the ISM, carbon, nitrogen, and oxygen (CNO), are produced in stars \citep{1957RvMP...29..547B,1997RvMP...69..995W}. The triple-$\alpha$ capture process produce $^{12}$C and  then $^{18}$O is synthesised from $^{14}$N  seeds during He burning in stars \citep{1952ApJ...115..326S,2007hic..book.....C}, while $^{13}$C, $^{15}$N and $^{17}$O are produced in H-burning zones via the cold  or hot CNO cycles (see \citealt{2010ARNPS..60..381W}; see also \citealt{2022A&ARv..30....7R}, for a review of the nucleosynthesis of CNO in stars). Stars release their nucleosynthetic products in the surrounding medium at different phases of the evolution, including the asymptotic giant branch (AGB), supernova phase, etc. \citep[see, e.g., ][]{2013ARA&A..51..457N,2016MNRAS.462..395D,2019ApJ...878...49W}.

Measuring isotope abundance ratios is an efficient method for tracking different stellar nucleosynthesis pathways and putting strong constraints to GCE models \citep{2017MNRAS.470..401R,2019MNRAS.490.2838R,2019A&ARv..27....3M,2021A&ARv..29....5M}. Stars of different initial mass and chemical composition, produce isotopes of different species in different amounts on different time scales \citep{1979ApJ...229.1046T}.
The $^{12}$C/$^{13}$C ratio, for example, is the result of contributions from stars of different mass that contribute to the existing radial gradient in the Milky Way on different time scales \citep{2005ApJ...634.1126M,2019ApJ...877..154Y}. The $^{14}$N/$^{15}$N ratio can also be used to study stellar nucleosynthesis and enrichment processes in the ISM \citep{2015ApJ...804L...3R,2022A&A...667A.151C}.

The $^{18}$O/$^{17}$O ratio is, in principle, a useful tracer of stellar nucleosynthesis and metal enrichment processes \citep{1994LNP...439...72H,1994ARA&A..32..191W}. Oxygen isotopes arrive at the surface of a star through convective motions and are eventually ejected in the ISM, unless further nuclear burnings do not destroy them. $^{18}$O is mainly released by massive stars with initial mass in the range 13--25 M$_\odot$ \cite[][and references therein]{2022A&ARv..30....7R}, while  $^{17}$O originates from both low- and intermediate-mass stars \citep{1993A&A...274..730H,1999ApJ...510..232B}. The latter isotope can also be produced in non-negligible quantities during nova outbursts \cite[][and references therein]{2022A&ARv..30....7R}. As a result,  $^{17}$O is injected into the ISM on longer time scales slower than $^{18}$O. If the Milky Way is forming inside-out \citep[e.g.,][and refernces therein]{2012A&A...540A..56P}, the $^{18}$O/$^{17}$O ratio can decrease over time and display a positive gradient with the galactocentric distance (D$_{GC}$). 
 
The abundance ratios of $^{18}$O to $^{17}$O can be derived from the intensity ratio of C$^{18}$O and C$^{17}$O lines \citep{2008A&A...487..237W,2015ApJS..219...28Z,2016RAA....16...47L,2020ApJS..249....6Z}. The intensity ratio of C$^{18}$O/C$^{17}$O lines can be viewed as its molecule abundance ratio, given that they have similar excitation properties and are normally optically thin \citep{2008A&A...487..237W}. The abundance ratio of $^{18}$O/$^{17}$O is then obtained from the abundance ratio of C$^{18}$O/C$^{17}$O as these exhibit similar chemical properties.  
In the end, the chemical fractionation of oxygen should be negligible because of the high first ionization potential that results in a low abundance of ionized oxygen and barely proceeds the charge exchange reactions \citep{1984ApJ...277..581L,2019MNRAS.485.5777L}. The isotope-selective photodissociation effect, which is caused by ultraviolet (UV) radiation, does not considerably alter the abundance ratio of C$^{18}$O/C$^{17}$O because of self-shielding \citep{1982ApJ...255..143B,2009A&A...503..323V}.

The $^{18}$O/$^{17}$O ratio takes a value of 2.8 ± 0.2 in the local ISM, which is smaller than the Solar system value of 5.5 \citep{1981ApJ...243L..47W}. \cite{1981ApJ...249..518P} measured the ratios of $^{18}$O/$^{17}$O in giant molecular clouds ranging from the Galactic center to D$_{GC}$ = 12 kpc and concluded that the ratio has a uniform value of $\sim 3.5$, with variations well within 5\%. \cite{2008A&A...487..237W} instead noted a radial gradient by measuring 2.88 ± 0.11 for the central region, 4.16 ± 0.09 in the galactocentric distance range 4--11 kpc and 5.03 ± 0.46 for the outer Galaxy (16--17 kpc) resting on a sample of 18 sources. \cite{2016RAA....16...47L}  measured the ratio for 13 sources covering a galactocentric distance range of 3 kpc to 16 kpc and arrived at a similar conclusion. Finally, \cite{2020ApJS..249....6Z}  observed 286 molecular clouds in the galactic disk and determined a gradient of  $^{18}$O/$^{17}$O=(0.10 ± 0.03) D$_{GC}$+(2.95 ± 0.30) with an unweighted least-square linear fit to the data. Furthermore, they compared the results with predictions of a GCE model \citep{2019MNRAS.490.2838R} thus confirming the existence of a radial gradient. Though \cite{2020ApJS..249....6Z} used a large  sample to  derive the $^{18}$O/$^{17}$O ratio, the large uncertainties of  their observations for individual sources  prevented a definite conclusion and limited their use as GCE model constraints.

In this publication, we present  the ratios of $^{18}$O/$^{17}$O derived  from the intensity ratios of C$^{18}$O/C$^{17}$O 1-0   observed with the IRAM 30 m telescope toward fifty-one molecular clouds in the galactocentric distance range 0.2--16 kpc.
For five of these sources, the  $^{18}$O/$^{17}$O ratios were derived from HC$^{18}$O$^{+}$/HC$^{17}$O$^{+}$ 1-0 data taken with the Yebes 40 m telescope. The observations with IRAM 30 m and  Yebes 40 m are described in Section \ref{sec:obs}. The main results are  described in Section \ref{sec:results}. Sections \ref{sec:dis} and \ref{sec:sum} follow with the discussion and our conclusions.

\section{Observations and data reduction}
\label{sec:obs}
\subsection{IRAM 30 m observation}

We selected fifty-one  late stage massive star forming regions with  parallactic distances from \cite{2014ApJ...783..130R,2019ApJ...885..131R}. The spectral lines, including C$^{17}$O J=1-0 at 112358.988 MHz, C$^{18}$O J=1-0 at 109782.176 MHz and HC$_{3}$N J=12-11 at 109173.638 MHz, were observed by using  30 meter millimeter telescope of the Institut de Radioastronomie Millimétrique (IRAM) on Pico Veleta Spain on June and October 2016, and August 2017, which were the same as that presented in  \cite{2022MNRAS.512.4934L} for NH$_2$D molecules in these sources,  while G211.59+01.05 was observed in 2020. The 3 mm (E0) band of The Eight Mixer Receiver(EMIR) and the Fourier Transform Spectrometers (FTS) backend  were adopted to cover an 8 GHz bandwidth, with a spectral resolution of 195 kHz and dual polarization. The frequency coverage ranges from 105.8 to 113.6 GHz for the 3mm band (E0). The standard position switching mode with an azimuth offset of -600 arcsec was used. 

 The IRAM 30-m telescope has a beam size of 22.4 arcsec at 110 GHz. The typical system temperature is about 150 K. Every two hours, strong nearby quasi-stellar objects were used to correct the pointing offset. Focus was calibrated before each observation or at sunrise and sunset. The main beam bright temperature ($T_{\rm mb}$) is obtained with  $T_{\rm mb}$ = $T_{\rm A}^{\ast}\cdot$ $F_{\rm eff}/$$B_{\rm eff}$, where $T_{\rm A}^{\ast}$ is the antenna temperature, forward efficiency ($F_{\rm eff}$) is 0.93, and the main beam efficiency ($B_{\rm eff}$) is 0.73. Each scan takes 2 minutes observing time, and the total on-source time is 12--234 minutes for each target.
 
 \subsection{Yebes 40 m observation}
For five sources out the total sample, we observed spectral lines of HC$^{18}$O$^{+}$ J=1-0, HC$^{17}$O$^{+}$ J=1-0, and HC$_{3}$N 9-8 at rest frequencies of 85162.223, 87057.530, 81881.467 MHz, respectively.
The observations were carried out in August 2022 with Yebes 40m telescope, with standard  position switching mode. The W-band receiver covers the frequency range of 72-90 GHz with both horizontal and vertical polarizations.  The spectral backends are Fast Fourier Spectrometers (FFTs) with 2.5 GHz bandwidths and 38.15 kHz resolution. The spectra are smoothed to 305.2 kHz, corresponding to a velocity resolution of 1.1 km s$^{-1}$ at 85 GHz. Pointing corrections were done with SiO v=1 J=2-1 line in VXSGR, V1111OPH, and IRC+00363. The main beam temperature $T_{mb}$ is obtained with  $T_{mb}$ = $T_{\rm A}^{\ast}$/$\eta_{mb}$, where $\eta_{mb}$ is the main beam efficiency, ranging from 0.30 at 72.5 GHz to 0.21 at 88.5 GHz. The integration time ranged from 2.9 hours to 4.3 hours for different sources, with a typical system temperature of 200 K, which gave  an 1$\sigma$ rms noise of 2-6 mK in $T_{\rm A}^{\ast}$ after averaging the data of both polarizations.

\subsection{Data reduction}

\begin{figure*}
 \centering 
\subfigure[]{ \label{fig1:a} 
\includegraphics[width=1\columnwidth]{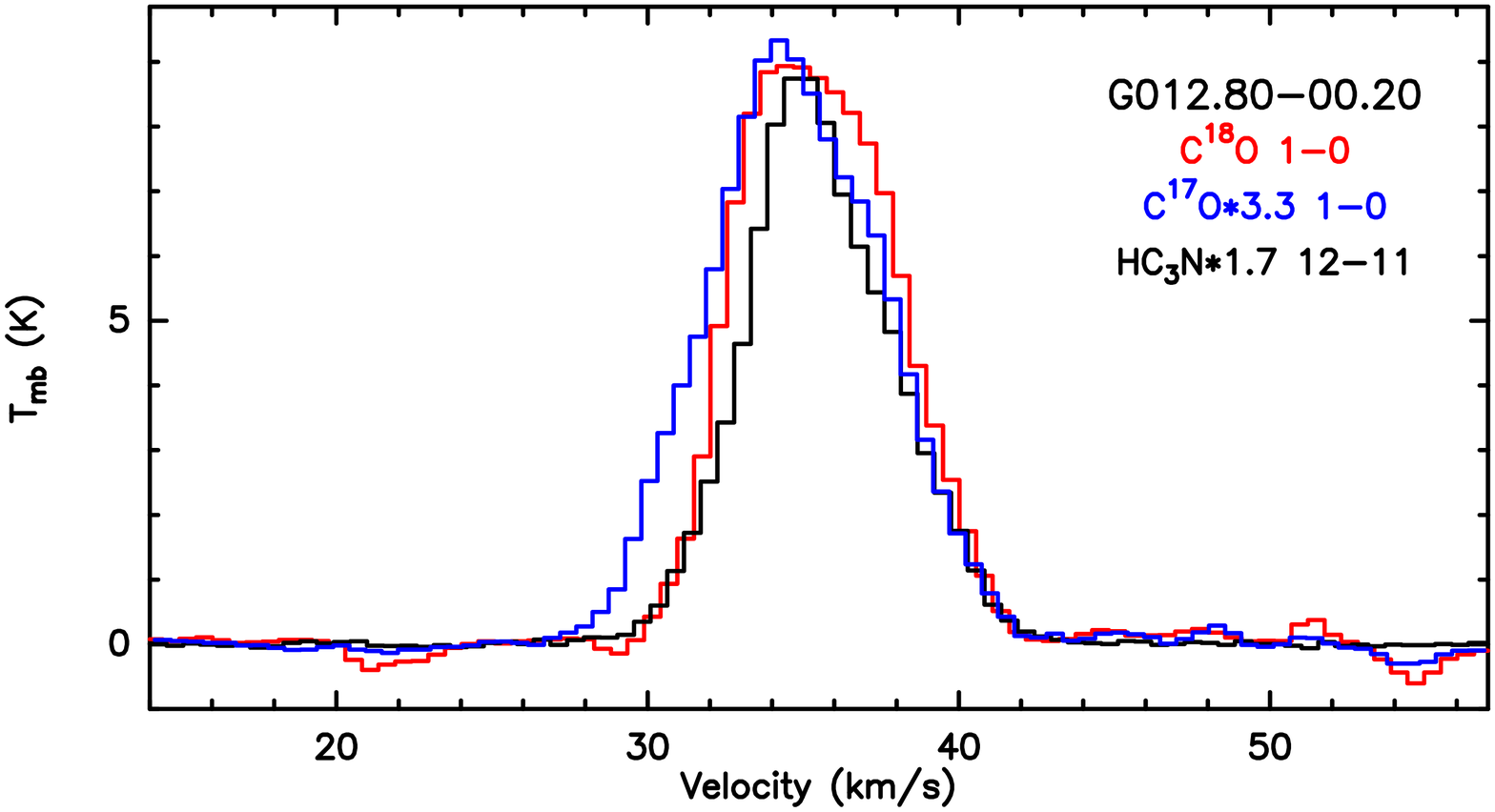} 
} 
\subfigure[]{ \label{fig1:b} 
\includegraphics[width=1\columnwidth]{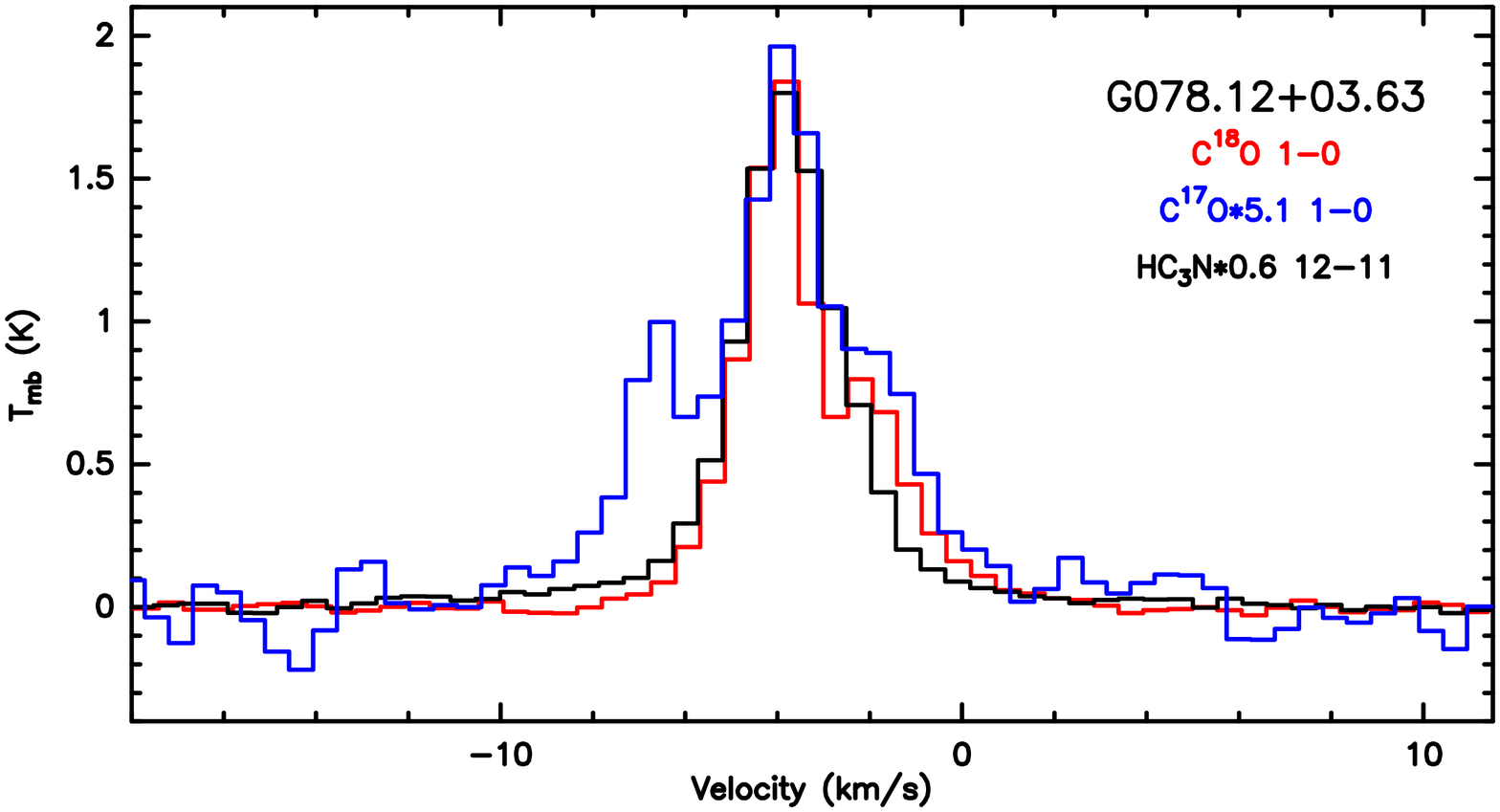} 
} 
\caption{The spectra of  C$^{18}$O 1-0, C$^{17}$O 1-0 and HC$_{3}$N 12-11 for two sources as examples. $\bf left$: C$^{17}$O 1-0 hyperfine structure can not be resolved due to line broadening. $\bf right$: One of the hyperfine can be resolved for source with narrow line width.}
\label{fig:fig1}
\end{figure*}

 Data reduction is based on the CLASS package in the GILDAS\footnote{\url{http://www.iram.fr/IRAMFR/GILDAS}} software. The spectra from all sources were averaged by using the CLASS package and  then the first-order baseline was removed. The velocity integrated line flux densities were obtained  with direct integration of the emission line. The errors  of  velocity integrated line flux densities were calculated by equation $\sigma=rms\sqrt{\delta \rm{v}\cdot\Delta \rm{v}}$, where $\delta \rm{v}$ was channel separation in velocity and $\Delta \rm{v}$ was velocity range for integration, while $rms$ error was obtained for each spectrum with $\delta \rm{v}$ channel spacing. From each spectrum we subtracted a first order baseline and then plotted this on the same velocity scale, as shown in Figure \ref{fig1:a}.  All sources were processed and  shown in Figure \ref{spectrum1}, except for G012.80-00.20 and G078.12+03.63 that are displayed in Figure \ref{fig:fig1}.

\section{Results} \label{sec:results}
C$^{18}$O J=1-0, C$^{17}$O J=1-0 and HC$_{3}$N J=12-11 were detected in all fifty-one sources. Both HC$^{18}$O$^{+}$ 1-0 and HC$^{17}$O$^{+}$ 1-0 were detected in all five sources. The source names, their aliases and equatorial coordinates were listed in Columns 1–3 of Table 1. The heliocentric distance of each source, which was precisely measured by \cite{2014ApJ...783..130R,2019ApJ...885..131R} using parallax method, was listed in Columns 4, while  D$_{GC}$ calculated with  heliocentric distance of each source was listed in Columns 5. The molecular species and their velocity  integrated intensities were listed in Columns 6–7, while the velocity ranges  were listed in Columns 8–9. The rms level derived with baseline fitting, the peak temperature, and the central velocity were respectively listed in Columns 10–12, while the latter two were resulted from Gaussian fitting  of the spectra. Column 13 listed the intensity ratio of C$^{18}$O/C$^{17}$O 1-0 or HC$^{18}$O$^{+}$/HC$^{17}$O$^{+}$ 1-0. 

As shown in  Table \ref{table1}, the velocity integrated intensities of C$^{18}$O 1-0 and C$^{17}$O 1-0 range from 2.11± 0.02 and 0.53 ± 0.02 K km\,s$^{-1}$ in G183.72-03.66 to 89.27 ± 0.26 and 30.79 ± 0.36 K km\,s$^{-1}$ in Sgr B2, with mid-value of 15.10 ± 0.04  and 4.62 ± 0.06 K km\,s$^{-1}$. 
 The velocity integrated intensities of HC$^{18}$O$^{+}$ 1-0 and HC$^{17}$O$^{+}$ 1-0 range from 0.80 ± 0.03 to 3.45 ± 0.07 K$\,$km s$^{-1}$ and 0.21 ± 0.04 to 1.03 ± 0.06 K$\,$km s$^{-1}$, respectively. 

The full width at half maximum (FWHM) of C$^{18}$O 1-0 and C$^{17}$O  1-0  lines in our samples were less than 8 km s$^{-1}$, with the exception of Sgr B2 and W49N.  These had FWHMs of approximately 24 and 14 km s$^{-1}$. C$^{17}$O  1-0 with obvious hyperfine structure (hfs) features was found in some sources, as shown with  blue line in Figure \ref{fig1:b}, which  results from the interaction of the electric quadrupole and magnetic dipole moment of the $^{17}$O nucleus \citep{1981JChPh..74.6990F}. The C$^{17}$O  1-0 spectra exhibited two fairly close  components when the FWHM of C$^{18}$O 1-0 was lower than 3 km s$^{-1}$, such as G059.78+00.06. Additional  velocity component can be found in  several sources, such as W49N, due to multiple  clouds with different distances in the line of sight. Emission at  off positions could result in a fake `absorption'  feature, such as in  G049.48-00.36.  As observations of the two line pairs,  C$^{18}$O/C$^{17}$O  1-0 and HC$^{18}$O$^+$/HC$^{17}$O$^+$ 1-0,   were made simultaneously, the uncertainties of  absolute flux calibration did not cause extra error of the line ratio. The ratios of  C$^{18}$O/C$^{17}$O were above 10$\sigma$ level for all sources. 

\begin{figure*}
 \centering 
\subfigure[]{ \label{fig2:a} 
\includegraphics[width=1\columnwidth]{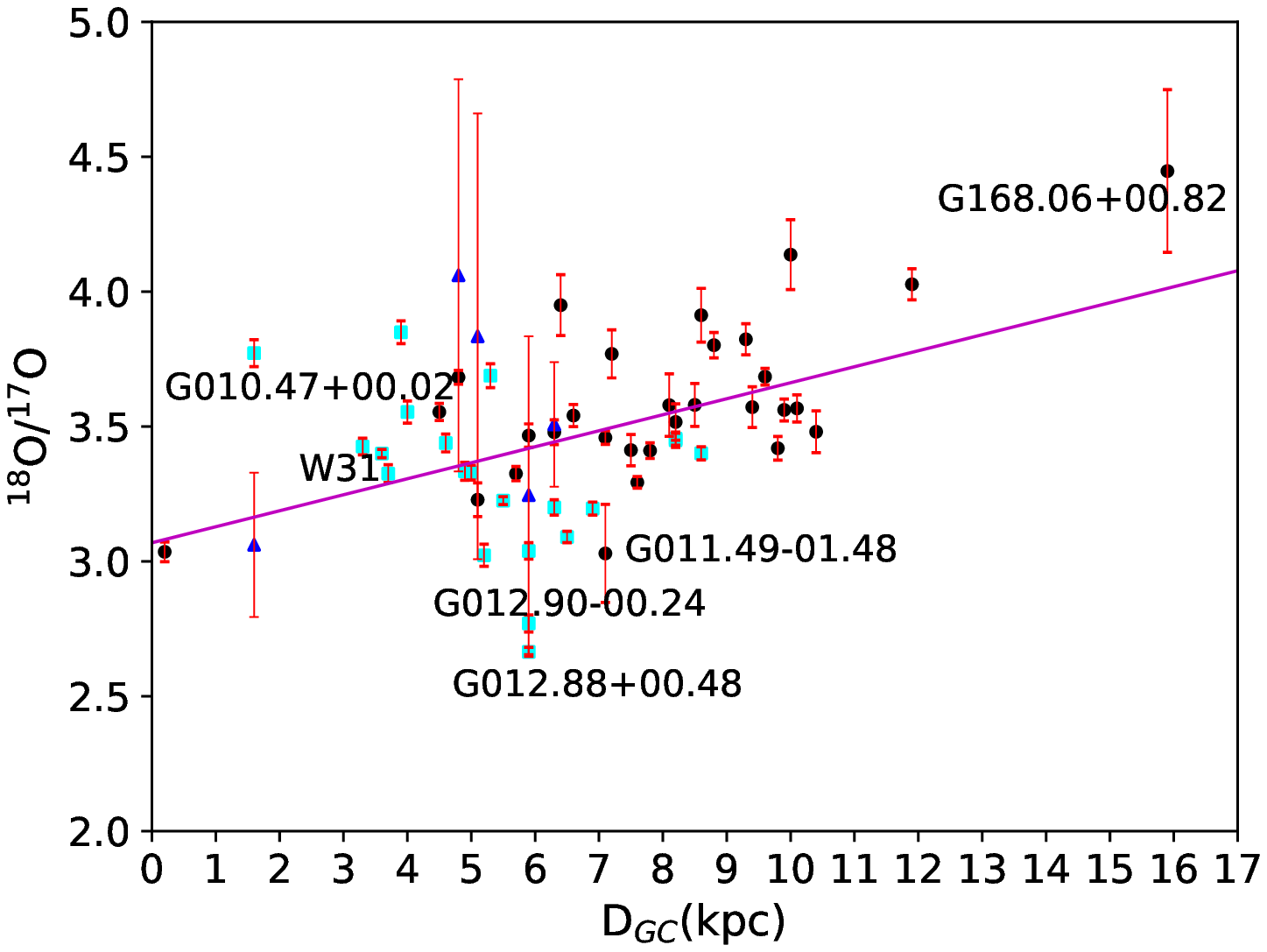} 
} 
\subfigure[]{ \label{fig2:b} 
\includegraphics[width=1\columnwidth]{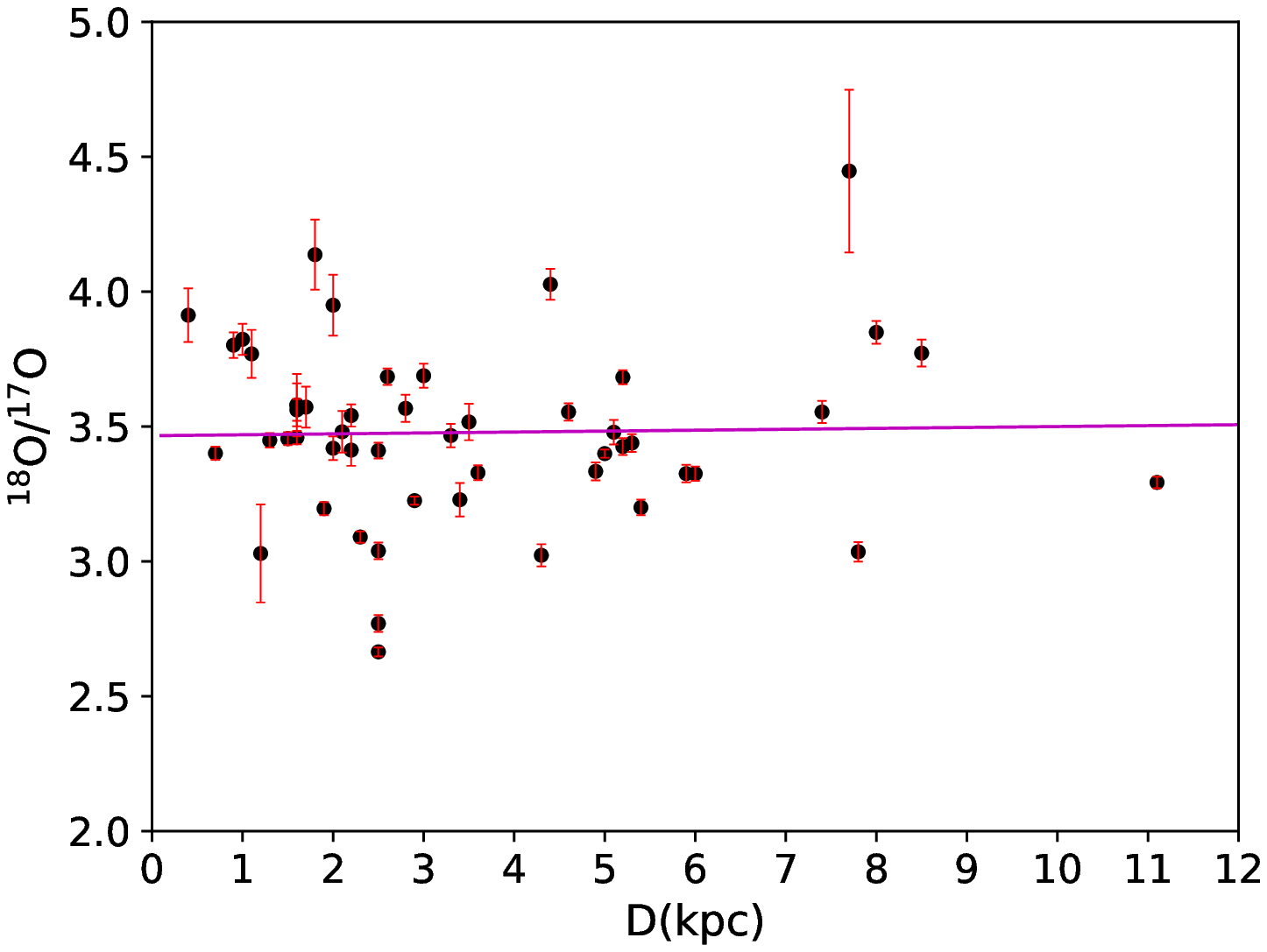} 
} 
\caption{The abundance  ratio of $^{18}$O/$^{17}$O following with  the galactocentric distance (D$_{GC}$)  on the left and  the heliocentric distance on the right. The results from HC$^{18}$O$^{+}$ 1-0 and HC$^{17}$O$^{+}$ 1-0 of 5 sources  are also shown on the left as  blue triangles, while only results from C$^{18}$O/C$^{17}$O  1-0  are on the right. Filled black circles are for sources with C$^{18}$O 1-0 peak $T \rm _{mb}$  lower than 4 K, while filled cyan squares for those higher than 4K. The magenta lines are fitting results for data points derived from C$^{18}$O/C$^{17}$O  1-0.} 
\label{fig:fig2}
\end{figure*}

The abundance  ratios can be derived from the intensity ratios, by multiplying the frequency correction factors of 1.047 derived with $(\nu_{\rm C^{17}O}/\nu_{\rm C^{18}O})^2$ (e.g., \citealt{1977ApJ...214...50L,2020ApJS..249....6Z}). After rectification, $^{18}$O/$^{17}$O ratios with errors were shown in Figure \ref{fig2:a},  taking into account the relationship with D$_{GC}$. The ratio of G000.67-00.03 was 3.04 ± 0.04,  which was consistent  with the values of 3.09 ± 0.10 and 3.06 ± 0.03  reported in  \cite{2015ApJS..219...28Z}  and \cite{2008A&A...487..237W} for Sgr B2. 
 The ratio in W3(OH) was 3.42 ± 0.04, which  was 10\%  lower than  the result of 3.87 ± 0.09 from \cite{2008A&A...487..237W}.  The main reason for this difference may be due to the non-gaussian line profiles of   C$^{18}$O 1-0 and C$^{17}$O  1-0 in W3(OH) (see Figure \ref{spectrum1}), since Gaussian fitting was used in  \cite{2008A&A...487..237W}.
 
The range of corrected $^{18}$O/$^{17}$O ratios derived  from  C$^{18}$O/C$^{17}$O 1-0 was from 2.66$\pm$0.02 in G012.88+00.48 to 4.45$\pm$0.30 in G168.06+00.82, while such ratios could be higher than 5 in many sources reported in \cite{2020ApJS..249....6Z},  which had large error bars.  Extra uncertainties in \cite{2020ApJS..249....6Z} could also be caused by the observations of C$^{18}$O 1-0 and C$^{17}$O 1-0 at different time with ARO 12m.   Even though there were almost no overlapping  sources in our sample and  those in  \cite{2020ApJS..249....6Z}, most of  the  high  $^{18}$O/$^{17}$O ratios in \cite{2020ApJS..249....6Z}  are likely due to the large uncertainties of measurement.  Gaussian fitting in  \cite{2020ApJS..249....6Z} could also cause uncertainties, since C$^{17}$O 1-0 can not be simple gaussian profiles due to  hfs.

The results of HC$^{18}$O$^{+}$ 1-0 and HC$^{17}$O$^{+}$ 1-0 of  five sources (see Table \ref{table1})  were also plotted in Figure \ref{fig2:a} as blue triangles where the ratios increased by the frequency correction factors of 1.045, while the spectra were presented in Figure \ref{spectrum2}. As mentioned in Section \ref{sec:intro}, it was thought that  the effects of chemical and physical processes do not significantly change the ratio of HC$^{18}$O$^{+}$/HC$^{17}$O$^{+}$ from $^{18}$O/$^{17}$O ratio (e.g., \citealt{1982ApJ...263L..89G}). The abundance ratios of $^{18}$O/$^{17}$O derived from HC$^{18}$O$^{+}$/HC$^{17}$O$^{+}$ 1-0 line ratios  are consistent with that derived from  C$^{18}$O/C$^{17}$O 1-0 ratios within the  error  bars, except for G010.47+00.02, which was 3.06$\pm0.27$ derived from  HC$^{18}$O$^{+}$/HC$^{17}$O$^{+}$ 1-0 and  3.77$\pm$0.05 derived from  C$^{18}$O/C$^{17}$O 1-0. 

C$^{18}$O and C$^{17}$O 1-0 lines were assumed to be optically thin for deriving C$^{18}$O/C$^{17}$O abundance ratio. However,   C$^{18}$O 1-0 may be  optically thick in some cases (e.g., \citealt{2001ApJ...562L.185B,2005A&A...430..549W}), which can  cause underestimation of C$^{18}$O/C$^{17}$O ratio.  C$^{18}$O 1-0  lines with higher $T \rm _{mb}$  may have higher optical depths than those with lower  $T \rm _{mb}$. We marked   21 sources with  C$^{18}$O 1-0 peak $T \rm _{mb}$  higher than 4 K as cyan squares in  Figure \ref{fig2:a}, in which $^{18}$O/$^{17}$O ratios may be under estimated due to the optically thick C$^{18}$O 1-0. There were  three sources, G010.62-00.38 (W31), G012.88+00.48 and G012.80-00.20,  with peak $T \rm _{mb}$  of C$^{17}$O 1-0   higher than 2 K, which means that C$^{18}$O 1-0 lines in these three sources  are very likely  optically thick, which  can cause  underestimated $^{18}$O/$^{17}$O ratio.  However,  it is  difficult to correct for optical depth  effects for  C$^{18}$O 1-0 in these  sources. In  Figure \ref{fig2:a}, we have marked those sources which may have C$^{18}$O 1-0 lines that are optically thick.

The relation between derived $^{18}$O/$^{17}$O ratios  and  D$_{GC}$ was plotted in  Figure \ref{fig2:a}.   An unweighted least-squares linear  fitting gave the result of $^{18}$O/$^{17}$O= (0.06±0.02) D$_{GC}$+ (3.07±0.11), with Pearson's correlation coefficient of 0.49. On the other hand, as shown in Figure \ref{fig2:b}, using the heliocentric distance instead of D$_{GC}$, there was  not any correlation. 

\section{discussion}
\label{sec:dis}

As shown in Figure \ref{fig2:a}, there is a  weak trend of increasing $^{18}$O/$^{17}$O ratios with  increasing D$_{GC}$. A similar trend was previously found with a smaller sample and weak correlation  \citep{2008A&A...487..237W}, or with a relatively  large sample but large error bars as well \citep{2020ApJS..249....6Z}.  In this work, we provide a much more solid trend.

 $^{18}$O is synthesised from $^{14}$N seeds during He burning, but can survive successive nuclear  burning only in massive stars  \citep{1952ApJ...115..326S,2007hic..book.....C}. $^{17}$O is produced in H-burning zones via the cold  and hot CNO cycles  \citep{2010ARNPS..60..381W, 2022A&ARv..30....7R}, with low- to intermediate-mass stars, massive stars and novae all contributing to its synthesis \citep[e.g.,][and references therein]{2022A&ARv..30....7R}. Thus, the $^{18}$O/$^{17}$O ratio reflects the relative contributions of massive and low- to  intermediate-mass stars, which in turn are related to the star formation history and initial mass function of the system. 

 \cite{2017MNRAS.470..401R}  predicted the abundance ratios of CNO isotopes at different times and positions within the Galaxy with their GCE model and compared the results with previous observational data in the literature. Later, \cite{2019MNRAS.490.2838R} updated the model to include the production of $^{17}$O and $^{18}$O by AGB stars, novae and massive stars including fast massive rotators.  The GCE models discussed by \cite{2017MNRAS.470..401R,2019MNRAS.490.2838R} belong to the class of the two-infall models, where the thick and thin discs of the Milky Way are formed in a sequence out of two distinct episodes of gas accretion. In particular, the thin disc forms from a mixture of gas of primordial chemical composition and enriched gas left over from the previous thick-disc formation process.
 \cite{2020ApJS..249....6Z} adopted the model results from  \cite{2019MNRAS.490.2838R} to compare with their observational results of C$^{18}$O and C$^{17}$O 1-0. However, due to the large uncertainties in their individual $^{18}$O/$^{17}$O ratio measurements \citep{2020ApJS..249....6Z},  it was hard to provide reliable constraints to the isotope ratios predicted by the GCE model.
 
 \begin{figure*}
 \centering 
\subfigure{ 
\includegraphics[width=2\columnwidth]{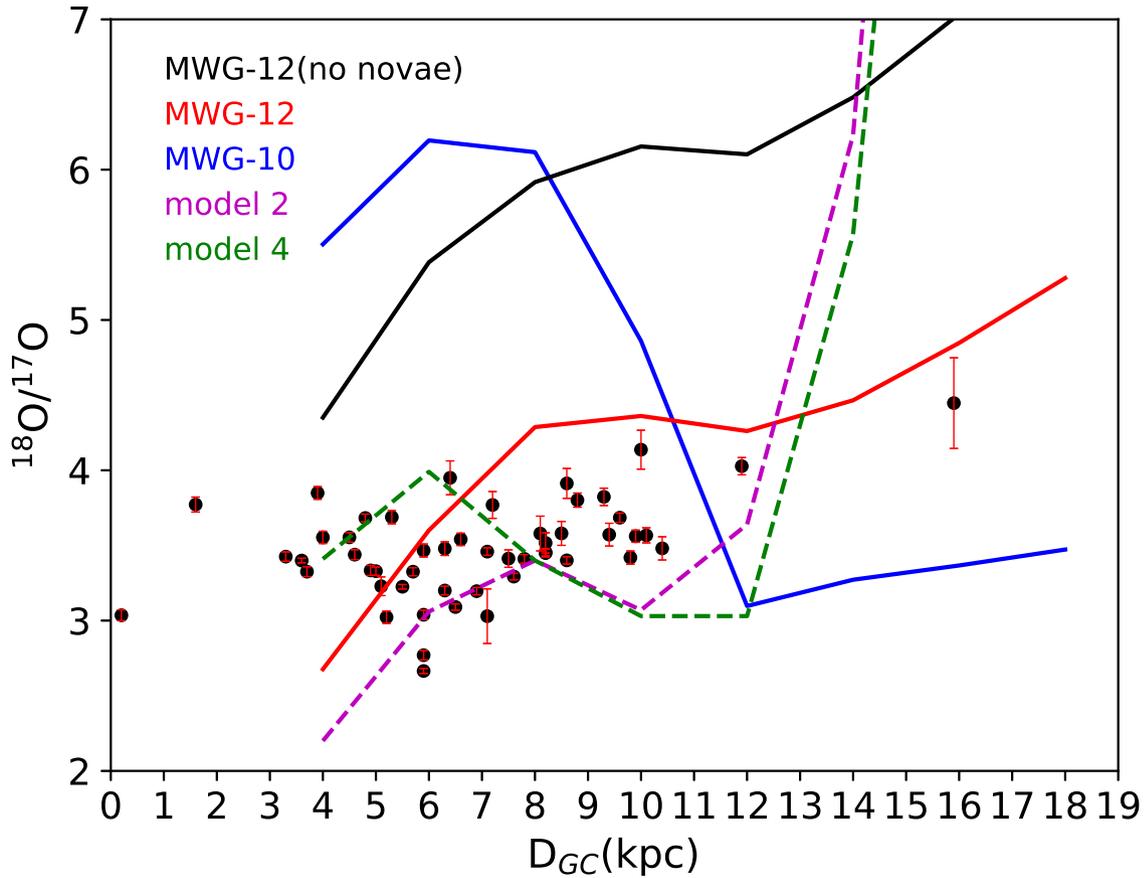} 
}  
\caption{ The  abundance ratio gradient overlaid with GCE model in different condition. Only the ratios obtained from C$^{18}$O 1-0 and C$^{17}$O 1-0 are used.  The model lines are described in Section \ref{sec:dis}. The Milky Way galaxy (MWG) and the number indicates the selected yield set combination (see Table 1 in \citealt{2019MNRAS.490.2838R}).
\label{fig:fig3}}
\end{figure*} 
 
Our  $^{18}$O/$^{17}$O ratios, derived from simultaneous observations of C$^{18}$O and C$^{17}$O 1-0 in fifty-one sources, are plotted in Figure \ref{fig:fig3} as  a function of D$_{GC}$, overlayed to theoretical gradients from five GCE models. The continuous lines refer to the predictions of two-infall models published in \cite{2019MNRAS.490.2838R}: MWG-10, assuming non-rotating stars and no novae (blue line); MWG-12, assuming fast-rotating stars below [Fe/H] = $-1$ and novae (red line); finally the black line shows MWG-12 without novae. The dashed lines are for models 2 and 4 discussed in \cite{2022A&A...667A.151C} that are one-infall models referring to the thin disc of the Galaxy.  In this case, the thick and thin discs of the Milky Way are formed in parallel out of two distinct episodes of infall of gas of primordial chemical composition (see \citealt{2018MNRAS.481.2570G}, for details).  At variance with what happens in the case of the two-infall models, the thin disc hence starts with a null metallicity, which allows an important amount of primary $^{18}$O from low-metallicity, massive fast-rotating stars to be ejected in the ISM. None of the models  can  explain the observational  relation between $^{18}$O/$^{17}$O ratios and  D$_{GC}$.   The MWG-12 model is closer to the observational results for the inner Galaxy than others,  still,  profound revisions are required, especially for  D$_{GC}$ between $\sim$ 6 and  11 kpc.

 In Figure \ref{fig:fig3} the scatter at  a fixed D$_{GC}$ is much larger than  the observational uncertainties in the data points,  which suggests additional factors. The possible underestimation of the $^{18}$O/$^{17}$O ratios due to  optically thick C$^{18}$O 1-0 (points marked as cyan squares in   Figure \ref{fig2:a}) offers a viable explanation. However, the scatter is still present for the sources marked with filled black circles in Figure \ref{fig2:a}, which are thought to be mainly  optically thin for C$^{18}$O 1-0.  So,  the  $^{18}$O/$^{17}$O ratio, which is indicative of the ratio of massive stars  to low- and intermediate-mass stars, is related not only to  D$_{GC}$.  Since $^{17}$O  is synthesised in non-negligible amounts by novae, namely, rare systems characterized by largely variable yields \citep[see][their figure 6]{2017MNRAS.470..401R}, it may well be that part of the scatter is caused by the stochasticity of the enrichment process.

 Further simultaneous observations of C$^{18}$O and C$^{17}$O 1-0 for sources with parallactic distances, such as  those in   \cite{2019ApJ...885..131R}, with high sensitivity so as to secure low uncertainties in the measurements, are necessary to better determine the relation of $^{18}$O/$^{17}$O ratio with D$_{GC}$.   Mapping observations  of sources with high C$^{18}$O 1-0 brightness temperature, which may be due to possible optically thick emission, are also needed. Spatially resolved C$^{18}$O/C$^{17}$O 1-0 ratio can be obtained with such mapping observations. With the assumption of uniform $^{18}$O/$^{17}$O ratio  within one source, optically thick C$^{18}$O 1-0, which should be mainly around  the emission peak,  might be ruled out, while the $^{18}$O/$^{17}$O ratio can be derived with observational data in the outer regions.  C$^{18}$O and C$^{17}$O  2-1 and 3-2 lines can also be used to double check the  $^{18}$O/$^{17}$O ratio obtained by C$^{18}$O and C$^{17}$O 1-0 in individual sources, especially for sources with possible optically thick  C$^{18}$O 1-0 emission.  Sensitive observations  for lines of  other molecules, such as HC$^{18}$O$^+$  and HC$^{17}$O$^+$ 1-0, also provide useful supplementary information to check if chemical fractionation is important or not.

\section{Summary and conclusion remarks} \label{sec:sum}

Using observations done with the IRAM 30 meter telescope, we detected  C$^{18}$O 1-0 and C$^{17}$O 1-0 emission  in  fifty-one massive star forming regions with accurate distances. The abundance ratio of $^{18}$O/$^{17}$O in each source is obtained from the intensity ratio of C$^{18}$O/C$^{17}$O 1-0, both of which are above 10$\sigma$ level in all fifty-one sources.  $^{18}$O/$^{17}$O ratio derived with C$^{18}$O 1-0 and C$^{17}$O 1-0  has a trend in the radial direction of the Galaxy  with $^{18}$O/$^{17}$O= (0.06±0.02) D$_{GC}$+ (3.07±0.11), as obtained from an unweighted least-squares linear fitting with Pearson's correlation coefficient of 0.49.   In contrast, no trend is found  for the $^{18}$O/$^{17}$O ratio with the distance from the sun. 
 The relation between  $^{18}$O/$^{17}$O ratio and   D$_{GC}$ in these sources is compared to GCE model predictions.  Improved GCE models are needed to explain the derived $^{18}$O/$^{17}$O ratio versus  D$_{GC}$  relationship. HC$^{18}$O$^{+}$ 1-0 and HC$^{17}$O$^{+}$ 1-0 in five of fifty-one sources were also observed with Yebes 40 meter telescope.  These provided consistent results of $^{18}$O/$^{17}$O ratios in four sources.  However for G010.47+00.02 the $^{18}$O/$^{17}$O ratio obtained with HC$^{18}$O$^{+}$ 1-0 and HC$^{17}$O$^{+}$ 1-0  observation is lower than that  C$^{18}$O 1-0 and C$^{17}$O 1-0  measurement.

\section*{Acknowledgements}
 We thank the referee, T.L.Wilson, who provided suggestions for the manuscript. This work is supported by  the National Natural Science Foundation of China grant 12173067, 12173016, 12041305, the Program for Innovative Talents, Entrepreneur in Jiangsu and the science research grants from the China Manned Space Project with NOs.CMS-CSST-2021-A08 and CMS-CSST-2021-A07. This work also benefited from the International Space Science Institute (ISSI/ISSI-BJ) in Bern and Beijing, thanks to the funding of the team “Chemical abundances in the ISM: the litmus test of stellar IMF variations in galaxies across cosmic time” (Principal Investigator Donatella Romano and Zhi-Yu Zhang).  This study is based on observations
carried out under project number 012-16, 023-17 and 005-20  with the IRAM 30m telescope and on observations carried out with the Yebes 40m telescope. IRAM is
supported by INSU/CNRS (France), MPG (Germany) and IGN (Spain). The 40m radio telescope at Yebes Observatory is operated by the Spanish Geographic Institute (IGN; Ministerio de Transportes, Movilidad y Agenda Urbana).

\onecolumn
\renewcommand\arraystretch{0.6}
\setlength{\tabcolsep}{3pt}

\begin{longtable}[c]{ccccccrcccccc}

	\caption{Data parameters}
	\label{table1}\\
	\hline 
	\hline
              
	Source          & R.A.        & Decl.        & D     & D$_{GC}$ & Line             & $\int T_{\rm mb}\rm  d\rm{v}$ & \multicolumn{2}{c}{Velocity range} & Rms          & $T_{\rm peak}$   & $V_{\rm LSR}$                      & Ratio$_{corr}$      \\
	Alias           & (hh:mm:ss)  & (dd:mm:ss)   & (kpc) & (kpc)    & (1-0)            & (K·km\,s$^{-1}$)      & \multicolumn{2}{c}{(km\,s$^{-1}$)}    & (10$^{-2}$K) & (K)                        & (km\,s$^{-1}$) 
	                 &           \\

\hline

\endfirsthead
\caption[]{(continued)}\\
	         
	\hline 
	\hline

	Source          & R.A.        & Decl.        & D     & D$_{GC}$ & Line             & $\int T_{\rm mb}\rm  d\rm{v}$ & \multicolumn{2}{c}{Velocity range} & Rms          & $T_{\rm peak}$   & $V_{\rm LSR}$                      & Ratio$_{corr}$      \\
	Alias           & (hh:mm:ss)  & (dd:mm:ss)   & (kpc) & (kpc)    & (1-0)            & (K·km\,s$^{-1}$)      & \multicolumn{2}{c}{(km\,s$^{-1}$)}    & (10$^{-2}$K) & (K)                        & (km\,s$^{-1}$)               &        \\

	         \hline
  
    \endhead
\hline
\endfoot

G121.29+00.65   & 00:36:47.35 & +63:29:02.20 & 0.9  & 8.8  & C$^{18}$O        & 7.06(0.01)  & -23  & -12   & 0.6 & 2.51 & -17.5(0.1) & 3.80(0.05)  \\
L 1287          &             &              &      &      & C$^{17}$O        & 1.95(0.02)  & -24  & -13   & 1.0 & 0.40  & -18.1(0.1) &            \\
G123.06-06.30   & 00:52:24.70 & +56:33:50.50 & 2.8  & 10.1 & C$^{18}$O        & 5.87(0.02)  & -38  & -22   & 0.7 & 1.62 & -30.0(0.1) & 3.57(0.05) \\
NGC 281         &             &              &      &      & C$^{17}$O        & 1.72(0.02)  & -38  & -26   & 0.9 & 0.33 & -30.7(0.1) &            \\
G133.94+01.06   & 02:27:03.82 & +61:52:25.20 & 2.0  & 9.8  & C$^{18}$O        & 15.10(0.04) & -54  & -39   & 1.5 & 3.28 & -47.3(0.1) & 3.42(0.04) \\
W 3OH           &             &              &      &      & C$^{17}$O        & 4.62(0.06)  & -58  & -41   & 1.9 & 0.79 & -48.0(0.1) &            \\
G168.06+00.82   & 05:17:13.74 & +39:22:19.90 & 7.7  & 15.9 & C$^{18}$O        & 2.69(0.04)  & -29  & -21   & 1.8 & 1.15 & -25.3(0.1) & 4.45(0.30) \\
IRAS 05137+3919 &             &              &      &      & C$^{17}$O        & 0.63(0.04)  & -31  & -22.4 & 2.0 & 0.16 & -25.8(0.1) &            \\
G209.00-19.38   & 05:35:15.80 & -05:23:14.10 & 0.4  & 8.6  & C$^{18}$O        & 3.27(0.02)  & 3    & 15    & 0.6 & 1.07 & 8.9(0.1)   & 3.91(0.10) \\
Orion Nebula    &             &              &      &      & C$^{17}$O        & 0.88(0.02)  & 1    & 18    & 0.7 & 0.18 & 8.4(0.1)   &            \\
G176.51+00.20   & 05:37:52.14 & +32:00:03.90 & 1.0  & 9.3  & C$^{18}$O        & 4.54(0.02)  & -25  & -12   & 0.6 & 1.83 & -17.7(0.1) & 3.82(0.06) \\
                &             &              &      &      & C$^{17}$O        & 1.24(0.02)  & -26  & -11   & 0.7 & 0.27 & -18.2(0.1) &            \\
G183.72-03.66   & 05:40:24.23 & +23:50:54.70 & 1.8  & 10.0 & C$^{18}$O        & 2.11(0.02)  & -5   & 10    & 0.6 & 0.57 & 2.3(0.1)   & 4.14(0.13) \\
                &             &              &      &      & C$^{17}$O        & 0.53(0.02)  & -4   & 8     & 0.6 & 0.11 & 1.8(0.1)   &            \\
G188.94+00.88   & 06:08:53.35 & +21:38:28.70 & 2.1  & 10.4 & C$^{18}$O        & 4.71(0.02)  & -4   & 10    & 0.8 & 1.39 & 3.2(0.1)   & 3.48(0.08) \\
S 252           &             &              &      &      & C$^{17}$O        & 1.42(0.03)  & -5   & 10    & 1.1 & 0.26 & 2.5(0.1)   &            \\
G192.60-00.04   & 06:12:54.02 & +17:59:23.30 & 1.6  & 9.9  & C$^{18}$O        & 7.77(0.02)  & 1    & 14    & 0.8 & 2.30  & 7.2(0.1)   & 3.56(0.04) \\
S 255           &             &              &      &      & C$^{17}$O        & 2.28(0.03)  & -1   & 13.5  & 0.9 & 0.45 & 6.8(0.1)   &            \\
G211.59+01.05   & 06:52:45.32 & +01:40:23.10 & 4.4  & 11.9 & C$^{18}$O        & 3.83(0.01)  & 40   & 50    & 0.4 & 1.00  & 45.4(0.1)  & 4.03(0.06) \\
                &             &              &      &      & C$^{17}$O        & 0.99(0.01)  & 38   & 49    & 0.6 & 0.19 & 44.8(0.1)  &            \\
G232.62+00.99   & 07:32:09.78 & -16:58:12.80 & 1.7  & 9.4  & C$^{18}$O        & 5.12(0.03)  & 6    & 24    & 0.8 & 1.32 & 16.6(0.1)  & 3.57(0.08) \\
                &             &              &      &      & C$^{17}$O        & 1.50(0.03)  & 8    & 23    & 1.1 & 0.26 & 16.0(0.1)  &            \\
G000.67-00.03   & 17:47:20.00 & -28:22:40.00 & 7.8  & 0.2  & C$^{18}$O        & 89.27(0.26) & 38   & 110   & 4.3 & 3.42 & 66.9(0.2)  & 3.04(0.04) \\
Sgr B2          &             &              &      &      & C$^{17}$O        & 30.79(0.36) & 35   & 114   & 5.6 & 1.22 & 65.8(0.1)  &            \\
G005.88-00.39   & 18:00:30.31 & -24:04:04.50 & 3.0  & 5.3  & C$^{18}$O        & 30.56(0.13) & -9   & 23    & 3.1 & 5.96 & 9.0(0.1)   & 3.69(0.04) \\
                &             &              &      &      & C$^{17}$O        & 8.68(0.10)  & -1   & 17    & 3.2 & 1.33 & 8.3(0.1)   &            \\
G009.62+00.19   & 18:06:15.00 & -20:31:31.70 & 5.2  & 3.3  & C$^{18}$O        & 38.77(0.07) & -6   & 16    & 2.1 & 6.92 & 3.7(0.1)   & 3.43(0.03) \\
                &             &              &      &      & C$^{17}$O        & 11.85(0.10) & -6   & 14    & 3.2 & 1.91 & 3.0(0.1)   &            \\
G010.47+00.02   & 18:08:38.00 & -19:51:50.30 & 8.5  & 1.6  & C$^{18}$O        & 31.72(0.11) & 53   & 78    & 3.0 & 4.56 & 66.9(0.1)  & 3.77(0.05) \\
                &             &              &      &      & C$^{17}$O        & 8.8(0.11)   & 57   & 76    & 3.6 & 1.18 & 66.2(0.1)  &            \\
                &             &              &      &      & HC$^{18}$O$^{+}$ & 2.09(0.06)  & 58   & 76    & 1.5 & 0.32 & 66.9(0.1)  & 3.06(0.27) \\
                &             &              &      &      & HC$^{17}$O$^{+}$ & 0.71(0.06)  & 58   & 76    & 1.3 & 0.06 & 65.9(0.1)  &            \\
G010.62-00.38   & 18:10:28.55 & -19:55:48.60 & 5.0  & 3.6  & C$^{18}$O        & 76.11(0.10) & -14  & 10    & 2.7 & 11.7 & -2.8(0.1)  & 3.40(0.02) \\
W 31            &             &              &      &      & C$^{17}$O        & 23.44(0.10) & -13  & 7     & 3.1 & 3.21 & -3.5(0.1)  &            \\
G012.88+00.48   & 18:11:51.42 & -17:31:29.00 & 2.5  & 5.9  & C$^{18}$O        & 28.73(0.05) & 28   & 39    & 2.0 & 8.21 & 33.3(0.1)  & 2.66(0.02) \\
IRAS 18089-1732 &             &              &      &      & C$^{17}$O        & 11.29(0.07) & 26   & 39    & 2.6 & 2.28 & 32.7(0.1)  &            \\
G011.91-00.61   & 18:13:58.12 & -18:54:20.30 & 3.4  & 5.1  & C$^{18}$O        & 10.33(0.06) & 33.4 & 47    & 2.1 & 3.16 & 36.4(0.1)  & 3.23(0.06) \\
                &             &              &      &      & C$^{17}$O        & 3.35(0.06)  & 30   & 43    & 2.4 & 0.72 & 36.1(0.1)  &            \\
                &             &              &      &      & HC$^{18}$O$^{+}$ & 1.05(0.06)  & 30   & 43    & 1.7 & 0.28 & 36.3(0.1)  & 3.83(0.83) \\
                &             &              &      &      & HC$^{17}$O$^{+}$ & 0.29(0.06)  & 30   & 43    & 1.6 & 0.06 & 36.9(0.1)  &            \\
G012.80-00.20   & 18:14:14.23 & -17:55:40.50 & 2.9  & 5.5  & C$^{18}$O        & 59.28(0.05) & 27   & 43    & 1.7 & 9.52 & 35.4(0.1)  & 3.23(0.01) \\
                &             &              &      &      & C$^{17}$O        & 19.24(0.09) & 26   & 43    & 2.9 & 2.74 & 34.7(0.1)  &            \\
G012.90-00.24   & 18:14:34.42 & -17:51:51.90 & 2.5  & 5.9  & C$^{18}$O        & 27.06(0.11) & 28   & 45    & 3.6 & 5.98 & 36.4(0.1)  & 2.77(0.03) \\
                &             &              &      &      & C$^{17}$O        & 10.23(0.11) & 28   & 42    & 4.0 & 1.73 & 35.5(0.1)  &            \\
G012.90-00.26   & 18:14:39.57 & -17:52:00.40 & 2.5  & 5.9  & C$^{18}$O        & 31.34(0.08) & 32   & 48    & 2.7 & 5.82 & 36.0(0.1)  & 3.04(0.03) \\
                &             &              &      &      & C$^{17}$O        & 10.80(0.11) & 28   & 50    & 3.1 & 1.66 & 35.3(0.1)  &            \\
                &             &              &      &      & HC$^{18}$O$^{+}$ & 1.62(0.10)  & 20   & 50    & 1.8 & 0.46 & 36.9(0.1)  & 3.25(0.59) \\
                &             &              &      &      & HC$^{17}$O$^{+}$ & 0.52(0.09)  & 20   & 50    & 1.6 & 0.06 & 37.7(0.1)  &            \\
G011.49-01.48   & 18:16:22.13 & -19:41:27.20 & 1.2  & 7.1  & C$^{18}$O        & 5.05(0.08)  & 3    & 18    & 2.8 & 1.22 & 10.3(0.1)  & 3.03(0.18) \\
                &             &              &      &      & C$^{17}$O        & 1.74(0.10)  & 1    & 15.6  & 3.7 & 0.27 & 9.3(0.4)   &            \\
G014.33-00.64   & 18:18:54.67 & -16:47:50.30 & 1.1  & 7.2  & C$^{18}$O        & 10.70(0.07) & 19.7 & 27    & 3.3 & 3.98 & 22.3(0.1)  & 3.77(0.09) \\
                &             &              &      &      & C$^{17}$O        & 2.97(0.07)  & 17.3 & 25.6  & 3.3 & 0.79 & 22.1(0.1)  &            \\
G015.03-00.67   & 18:20:24.81 & -16:11:35.30 & 2.0  & 6.4  & C$^{18}$O        & 7.01(0.05)  & 14   & 27    & 1.8 & 2.03 & 19.5(0.1)  & 3.95(0.11) \\
M 17            &             &              &      &      & C$^{17}$O        & 1.86(0.05)  & 13   & 25    & 2.1 & 0.34 & 19.0(0.1)  &            \\
G016.58-00.05   & 18:21:09.08 & -14:31:48.80 & 3.6  & 5.0  & C$^{18}$O        & 21.66(0.05) & 53   & 65    & 2.0 & 6.44 & 59.2(0.1)  & 3.33(0.03) \\
                &             &              &      &      & C$^{17}$O        & 6.81(0.05)  & 51   & 65    & 2.0 & 1.39 & 58.6(0.1)  &            \\
G023.44-00.18   & 18:34:39.29 & -08:31:25.40 & 5.9  & 3.7  & C$^{18}$O        & 35.02(0.10) & 90   & 115   & 2.6 & 5.98 & 101.1(0.1) & 3.33(0.03) \\
                &             &              &      &      & C$^{17}$O        & 11.02(0.10) & 90   & 114   & 2.9 & 1.57 & 100.2(0.1) &            \\
G023.00-00.41   & 18:34:40.20 & -09:00:37.00 & 4.6  & 4.5  & C$^{18}$O        & 27.57(0.06) & 64   & 92    & 1.6 & 3.43 & 76.2(0.1)  & 3.55(0.03) \\
                &             &              &      &      & C$^{17}$O        & 8.12(0.07)  & 63.6 & 91    & 1.9 & 0.93 & 75.5(0.1)  &            \\
G027.36-00.16   & 18:41:51.06 & -05:01:43.40 & 8.0  & 3.9  & C$^{18}$O        & 18.88(0.04) & 86   & 98    & 1.5 & 4.61 & 91.2(0.1)  & 3.85(0.04) \\
                &             &              &      &      & C$^{17}$O        & 5.14(0.06)  & 83   & 98    & 2.0 & 0.98 & 90.6(0.1)  &            \\
G028.86+00.06   & 18:43:46.22 & -03:35:29.60 & 7.4  & 4.0  & C$^{18}$O        & 22.21(0.04) & 94   & 110   & 1.5 & 5.18 & 103.2(0.1) & 3.55(0.04) \\
                &             &              &      &      & C$^{17}$O        & 6.54(0.07)  & 91   & 110   & 2.4 & 1.02 & 102.1(0.1) &            \\
G029.95-00.01   & 18:46:03.74 & -02:39:22.30 & 5.3  & 4.6  & C$^{18}$O        & 28.51(0.06) & 90   & 108   & 2.1 & 6.21 & 97.5(0.1)  & 3.44(0.03) \\
W 43S           &             &              &      &      & C$^{17}$O        & 8.68(0.08)  & 88   & 108   & 2.5 & 1.34 & 97.0(0.1)  &            \\
G031.28+00.06   & 18:48:12.39 & -01:26:30.70 & 4.3  & 5.2  & C$^{18}$O        & 21.51(0.06) & 100  & 117   & 2.0 & 5.21 & 109.0(0.1) & 3.02(0.04) \\
                &             &              &      &      & C$^{17}$O        & 7.45(0.10)  & 98   & 115   & 3.3 & 1.38 & 108.3(0.1) &            \\
G031.58+00.07   & 18:48:41.68 & -01:09:59.00 & 4.9  & 4.9  & C$^{18}$O        & 19.21(0.06) & 87   & 104   & 1.8 & 5.13 & 96.3(0.1)  & 3.33(0.03) \\
W 43Main        &             &              &      &      & C$^{17}$O        & 6.03(0.06)  & 88   & 102   & 2.1 & 1.14 & 95.7(0.1)  &            \\
G032.04+00.05   & 18:49:36.58 & -00:45:46.90 & 5.2  & 4.8  & C$^{18}$O        & 17.95(0.04) & 86   & 105   & 1.2 & 3.70  & 95.6(0.1)  & 3.68(0.03) \\
                &             &              &      &      & C$^{17}$O        & 5.10(0.03)  & 85   & 104   & 1.1 & 0.84 & 95.0(0.1)  &            \\
                &             &              &      &      & HC$^{18}$O$^{+}$ & 0.80(0.03)  & 92   & 102   & 0.7 & 0.22 & 95.4(0.1)  & 4.06(0.73) \\
                &             &              &      &      & HC$^{17}$O$^{+}$ & 0.21(0.04)  & 92   & 102   & 0.8 & 0.03 & 96.5(0.1)  &            \\
G034.39+00.22   & 18:53:19.00 & +01:24:08.80 & 1.6  & 7.1  & C$^{18}$O        & 13.16(0.03) & 50   & 64    & 1.2 & 3.06 & 57.1(0.1)  & 3.46(0.02) \\
                &             &              &      &      & C$^{17}$O        & 3.98(0.03)  & 50   & 63    & 1.0 & 0.73 & 56.6(0.1)  &            \\
G035.02+00.34   & 18:54:00.67 & +02:01:19.20 & 2.3  & 6.5  & C$^{18}$O        & 12.95(0.04) & 46   & 55.8  & 1.6 & 4.48 & 52.4(0.1)  & 3.09(0.02) \\
                &             &              &      &      & C$^{17}$O        & 4.39(0.03)  & 44   & 57    & 1.0 & 0.92 & 51.8(0.1)  &            \\
G037.43+01.51   & 18:54:14.35 & +04:41:41.70 & 1.9  & 6.9  & C$^{18}$O        & 11.68(0.02) & 39   & 49    & 0.9 & 4.09 & 44.2(0.1)  & 3.20(0.02) \\
                &             &              &      &      & C$^{17}$O        & 3.83(0.03)  & 38   & 50    & 1.1 & 0.79 & 43.7(0.1)  &            \\
G035.19-00.74   & 18:58:13.05 & +01:40:35.70 & 2.2  & 6.6  & C$^{18}$O        & 15.98(0.07) & 24   & 44    & 2.3 & 3.94 & 34.0(0.1)  & 3.54(0.04) \\
                &             &              &      &      & C$^{17}$O        & 4.73(0.05)  & 26   & 44    & 1.6 & 0.84 & 33.4(0.1)  &            \\
G035.20-01.73   & 19:01:45.54 & +01:13:32.50 & 3.3  & 5.9  & C$^{18}$O        & 8.51(0.03)  & 40.4 & 50    & 1.3 & 2.15 & 44.0(0.2)  & 3.47(0.04) \\
                &             &              &      &      & C$^{17}$O        & 2.57(0.03)  & 38.5 & 49    & 1.3 & 0.53 & 43.6(0.1)  &            \\
G043.16+00.01   & 19:10:13.41 & +09:06:12.80 & 11.1 & 7.6  & C$^{18}$O        & 46.06(0.06) & -16  & 25    & 1.2 & 3.23 & 7.4(0.1)   & 3.29(0.02) \\
W 49N           &             &              &      &      & C$^{17}$O        & 14.64(0.09) & -16  & 30    & 1.9 & 0.99 & 6.6(0.1)   &            \\
G043.79-00.12   & 19:11:53.99 & +09:35:50.30 & 6.0  & 5.7  & C$^{18}$O        & 18.12(0.06) & 32   & 54    & 1.7 & 3.01 & 44.1(0.1)  & 3.33(0.03) \\
OH 43.8-0.1     &             &              &      &      & C$^{17}$O        & 5.71(0.04)  & 31   & 54    & 1.2 & 0.81 & 43.3(0.1)  &            \\
G049.48-00.36   & 19:23:39.82 & +14:31:05.00 & 5.1  & 6.3  & C$^{18}$O        & 29.55(0.11) & 54   & 76    & 3.4 & 3.70  & 61.2(0.1)  & 3.48(0.05) \\
W 51 IRS2       &             &              &      &      & C$^{17}$O        & 8.89(0.11)  & 53   & 71    & 3.6 & 1.02 & 60.5(0.1)  &            \\
G049.48-00.38   & 19:23:43.87 & +14:30:29.50 & 5.4  & 6.3  & C$^{18}$O        & 55.92(0.17) & 48   & 74    & 4.5 & 6.42 & 56.2(0.1)  & 3.20(0.03) \\
W 51M           &             &              &      &      & C$^{17}$O        & 18.30(0.15) & 46   & 81    & 3.6 & 1.89 & 55.6(0.1)  &            \\
                &             &              &      &      & HC$^{18}$O$^{+}$ & 3.45(0.07)  & 45   & 70    & 1.3 & 0.37 & 56.7(0.1)  & 3.51(0.23) \\
                &             &              &      &      & HC$^{17}$O$^{+}$ & 1.03(0.06)  & 45   & 70    & 1.3 & 0.10  & 55.9(0.1)  &            \\
G059.78+00.06   & 19:43:11.25 & +23:44:03.30 & 2.2  & 7.5  & C$^{18}$O        & 6.04(0.03)  & 16   & 27    & 1.1 & 2.16 & 22.7(0.1)  & 3.41(0.06) \\
                &             &              &      &      & C$^{17}$O        & 1.85(0.03)  & 15   & 30    & 1.1 & 0.36 & 22.1(0.1)  &            \\
G069.54-00.97   & 20:10:09.07 & +31:31:36.00 & 2.5  & 7.8  & C$^{18}$O        & 14.57(0.03) & 3    & 19    & 1.0 & 3.59 & 11.4(0.1)  & 3.41(0.03) \\
ON 1            &             &              &      &      & C$^{17}$O        & 4.47(0.04)  & 2    & 20    & 1.2 & 0.81 & 10.7(0.1)  &            \\
G078.12+03.63   & 20:14:26.07 & +41:13:32.70 & 1.6  & 8.1  & C$^{18}$O        & 5.10(0.03)  & -9   & 4     & 1.3 & 1.45 & -3.7(0.1)  & 3.58(0.12) \\
IRAS 20126+4104 &             &              &      &      & C$^{17}$O        & 1.49(0.05)  & -11  & 2     & 1.8 & 0.28 & -3.9(0.1)  &            \\
G075.76+00.33   & 20:21:41.09 & +37:25:29.30 & 3.5  & 8.2  & C$^{18}$O        & 7.39(0.04)  & -10  & 7     & 1.3 & 1.72 & -1.6(0.1)  & 3.52(0.07) \\
                &             &              &      &      & C$^{17}$O        & 2.20(0.04)  & -11  & 4     & 1.5 & 0.38 & -2.3(0.1)  &            \\
G081.87+00.78   & 20:38:36.43 & +42:37:34.80 & 1.3  & 8.2  & C$^{18}$O        & 21.18(0.05) & 0    & 20    & 1.6 & 5.14 & 9.7(0.1)   & 3.45(0.03) \\
W 75N           &             &              &      &      & C$^{17}$O        & 6.43(0.05)  & 0    & 19    & 1.5 & 1.12 & 9.0(0.1)   &            \\
G081.75+00.59   & 20:39:01.99 & +42:24:59.30 & 1.5  & 8.2  & C$^{18}$O        & 13.92(0.03) & -9   & -1    & 1.6 & 5.68 & -4.0(0.1)  & 3.46(0.02) \\
DR 21           &             &              &      &      & C$^{17}$O        & 4.22(0.03)  & -12  & 0     & 1.1 & 0.92 & -4.5(0.1)  &            \\
G092.67+03.07   & 21:09:21.73 & +52:22:37.10 & 1.6  & 8.5  & C$^{18}$O        & 5.38(0.02)  & -11  & -2    & 1.1 & 2.01 & -6.1(0.1)  & 3.58(0.08) \\
                &             &              &      &      & C$^{17}$O        & 1.57(0.03)  & -13  & -1    & 1.4 & 0.35 & -6.4(0.1)  &            \\
G109.87+02.11   & 22:56:18.10 & +62:01:49.50 & 0.7  & 8.6  & C$^{18}$O        & 19.14(0.03) & -18  & -2    & 1.0 & 5.08 & -10.7(0.1) & 3.40(0.02)  \\
Cep A           &             &              &      &      & C$^{17}$O        & 5.89(0.04)  & -21  & -2    & 1.3 & 1.09 & -11.4(0.1) &            \\
G111.54+00.77   & 23:13:45.36 & +61:28:10.60 & 2.6  & 9.6  & C$^{18}$O        & 11.99(0.03) & -68  & -45   & 0.7 & 2.37 & -57.1(0.1) & 3.68(0.03) \\
NGC 7538        &             &              &      &      & C$^{17}$O        & 3.41(0.03)  & -70  & -46   & 0.8 & 0.55 & -57.7(0.1) &                       \\ \hline 
\end{longtable}	






\onecolumn
\centering
\appendix
\section{The spectral lines of all sample}
\begin{figure*}
\centering
\includegraphics[width=0.45\textwidth]{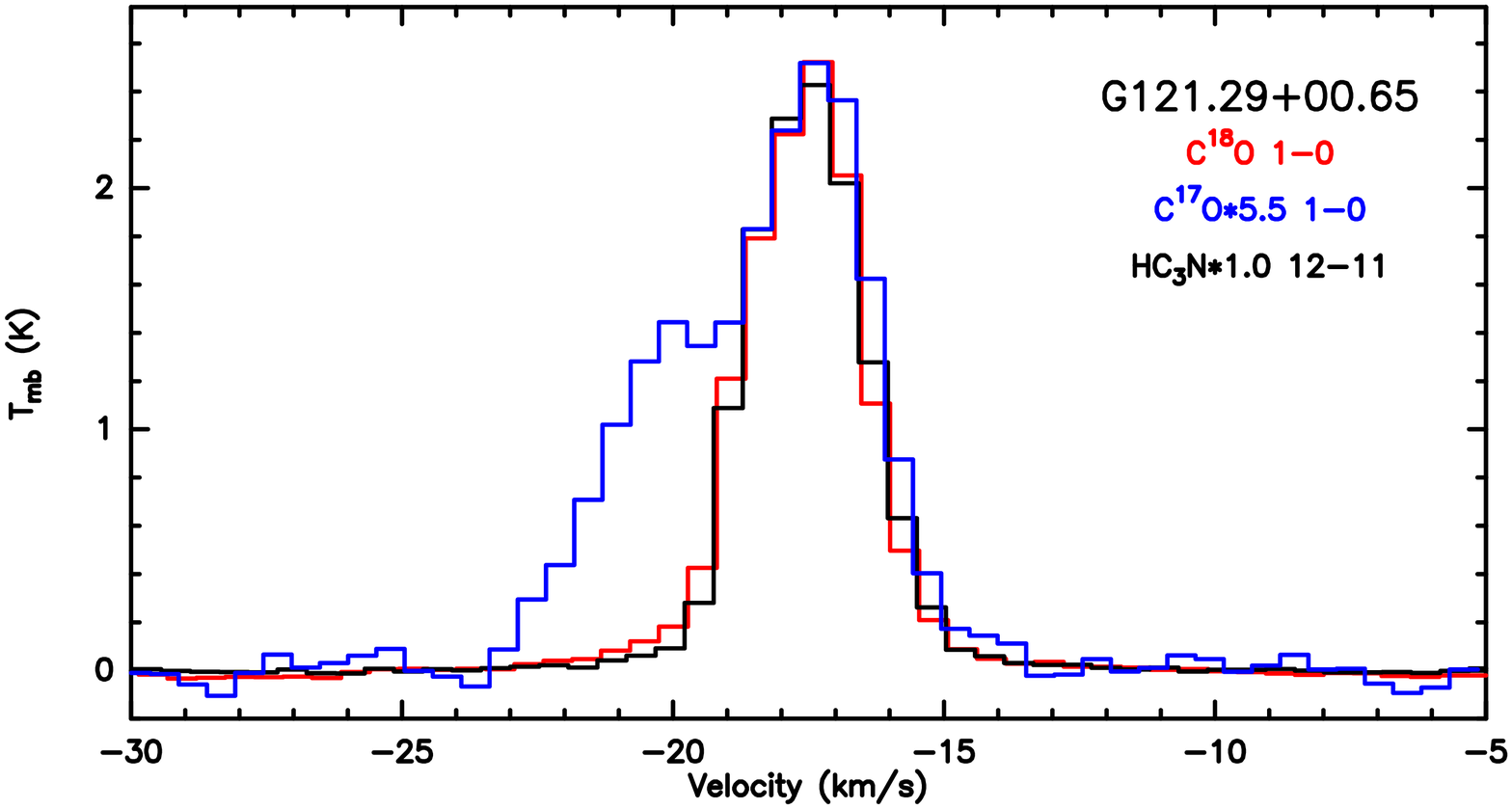}\includegraphics[width=0.45\textwidth]{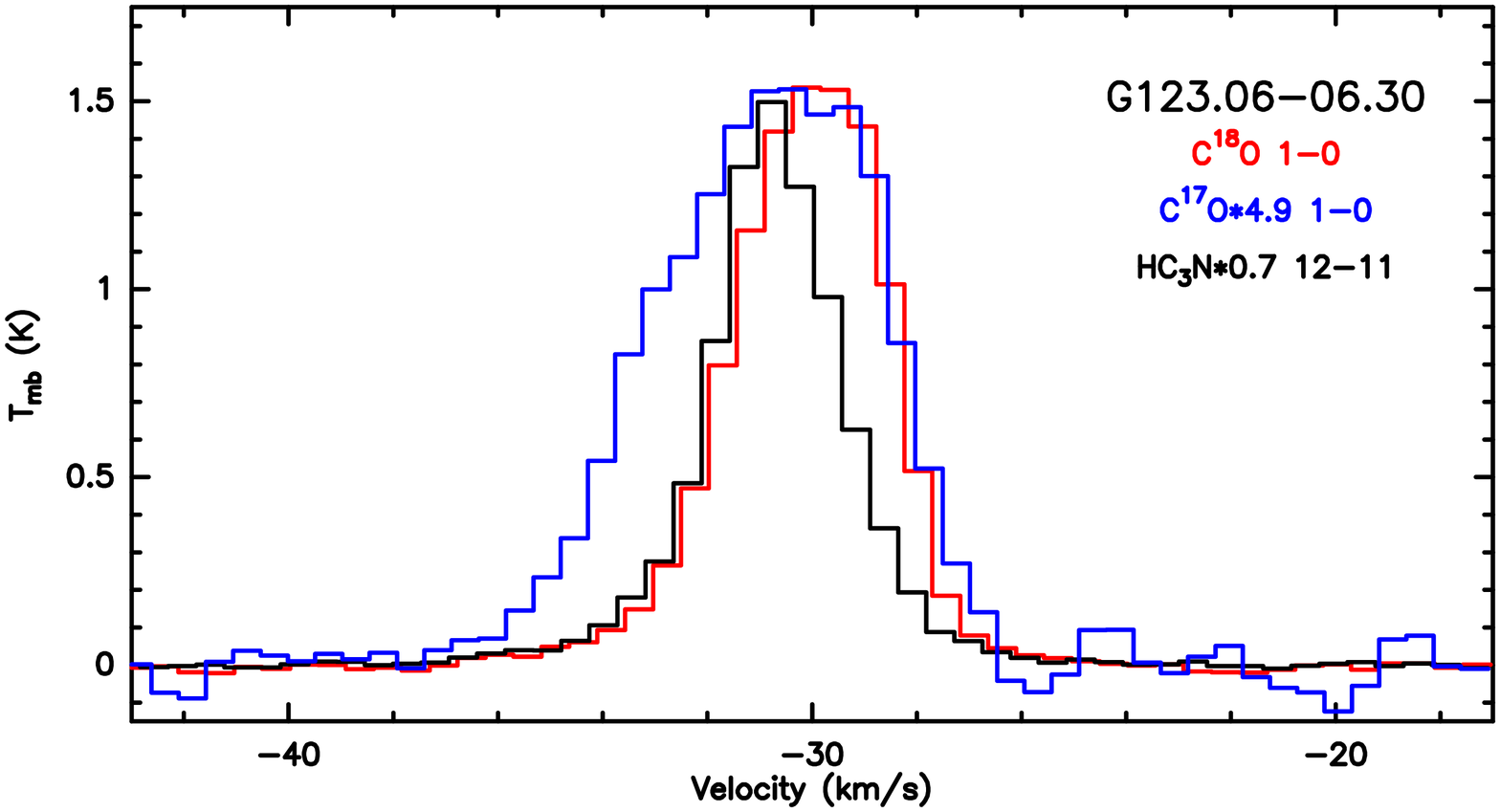}\\
\includegraphics[width=0.45\textwidth]{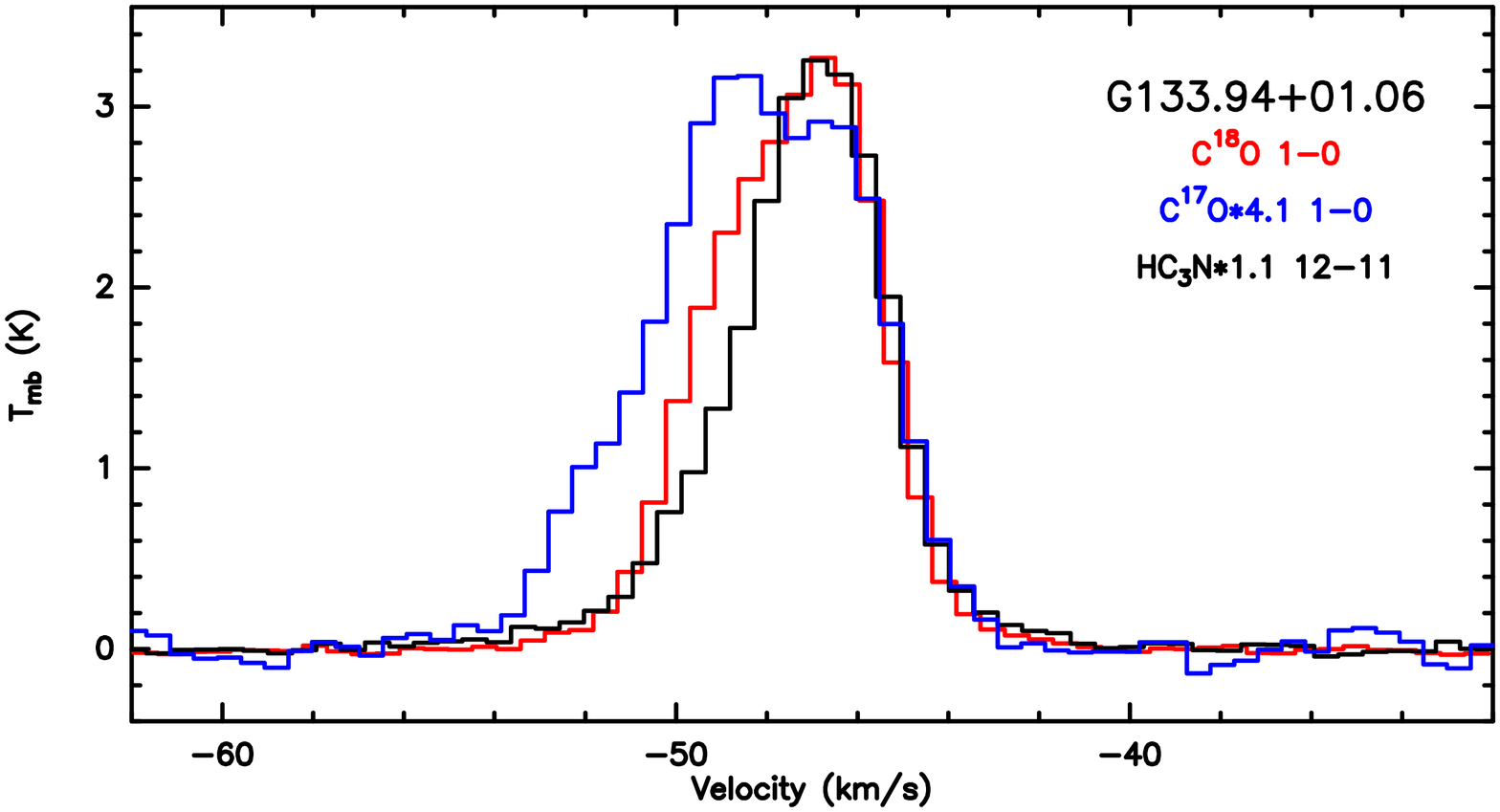}\includegraphics[width=0.45\textwidth]{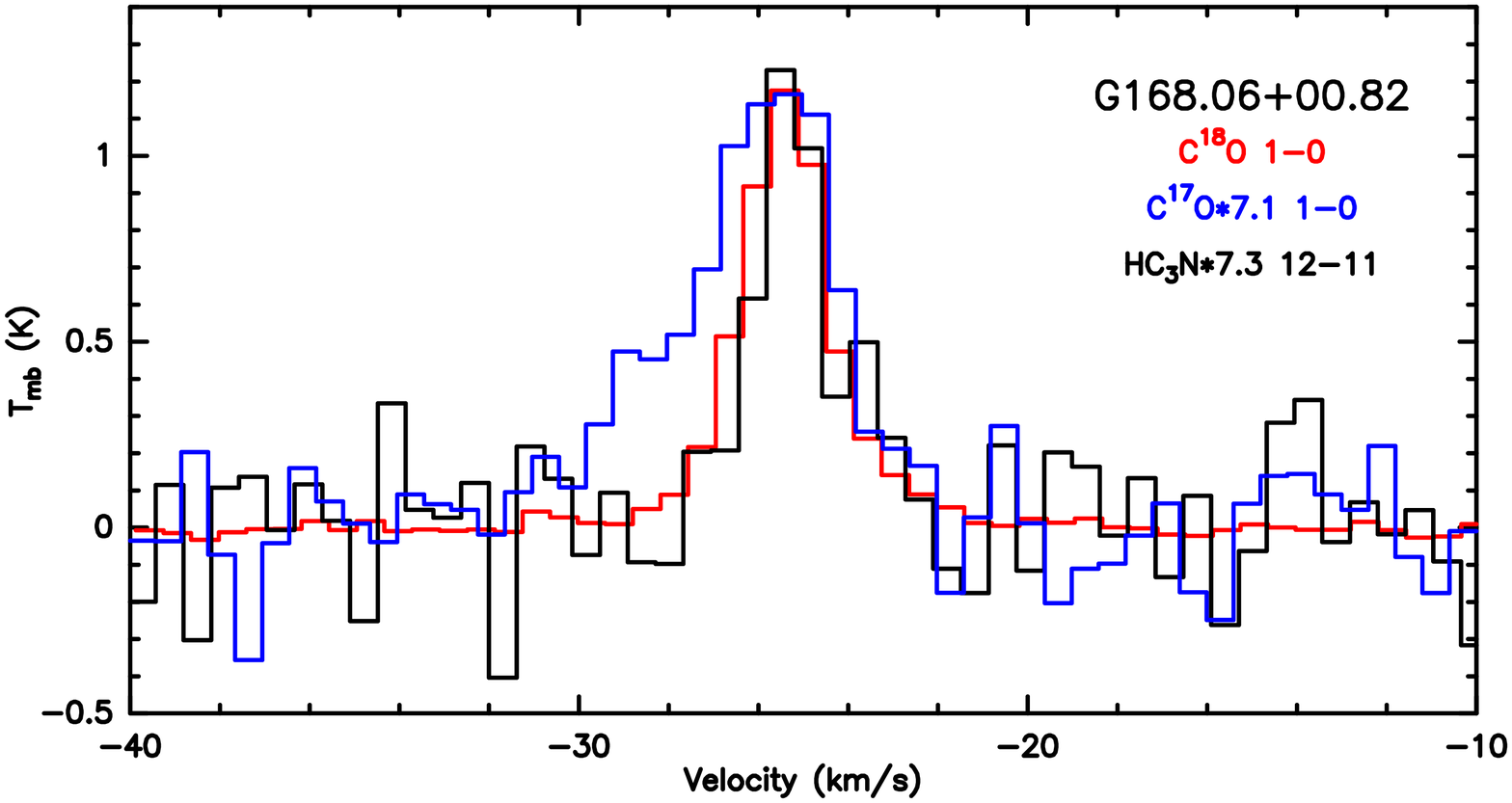}\\
\includegraphics[width=0.45\textwidth]{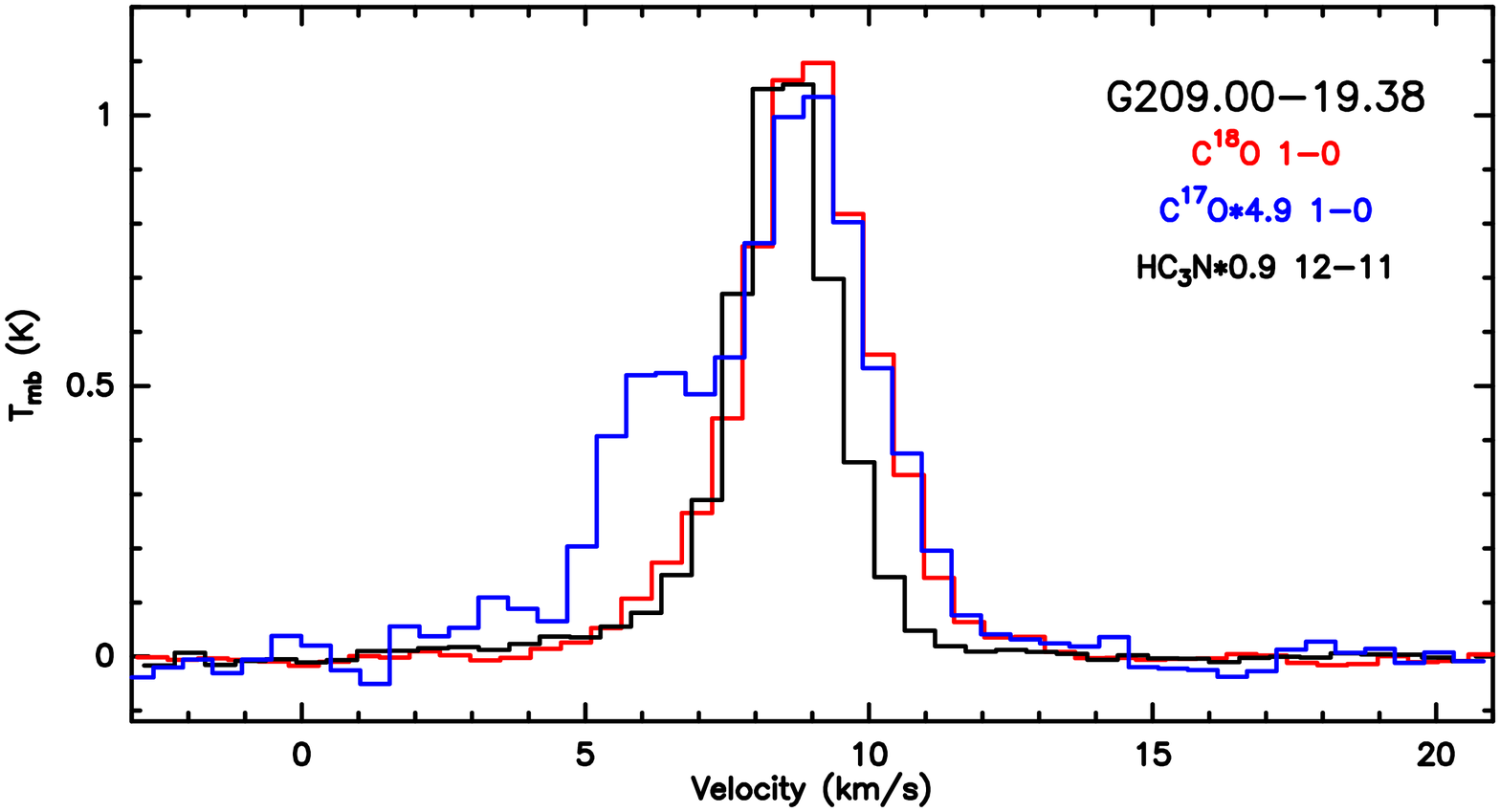}\includegraphics[width=0.45\textwidth]{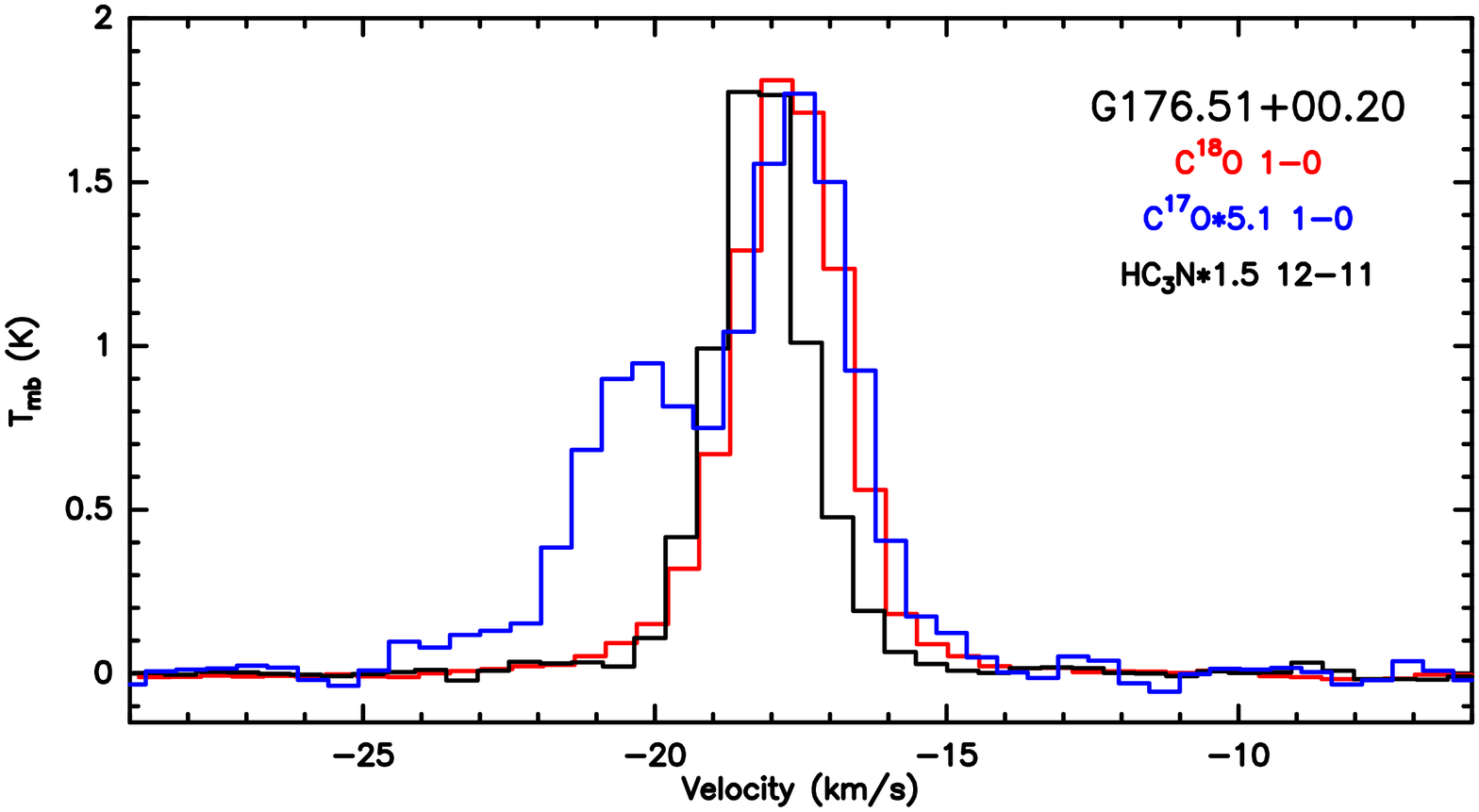}\\
\includegraphics[width=0.45\textwidth]{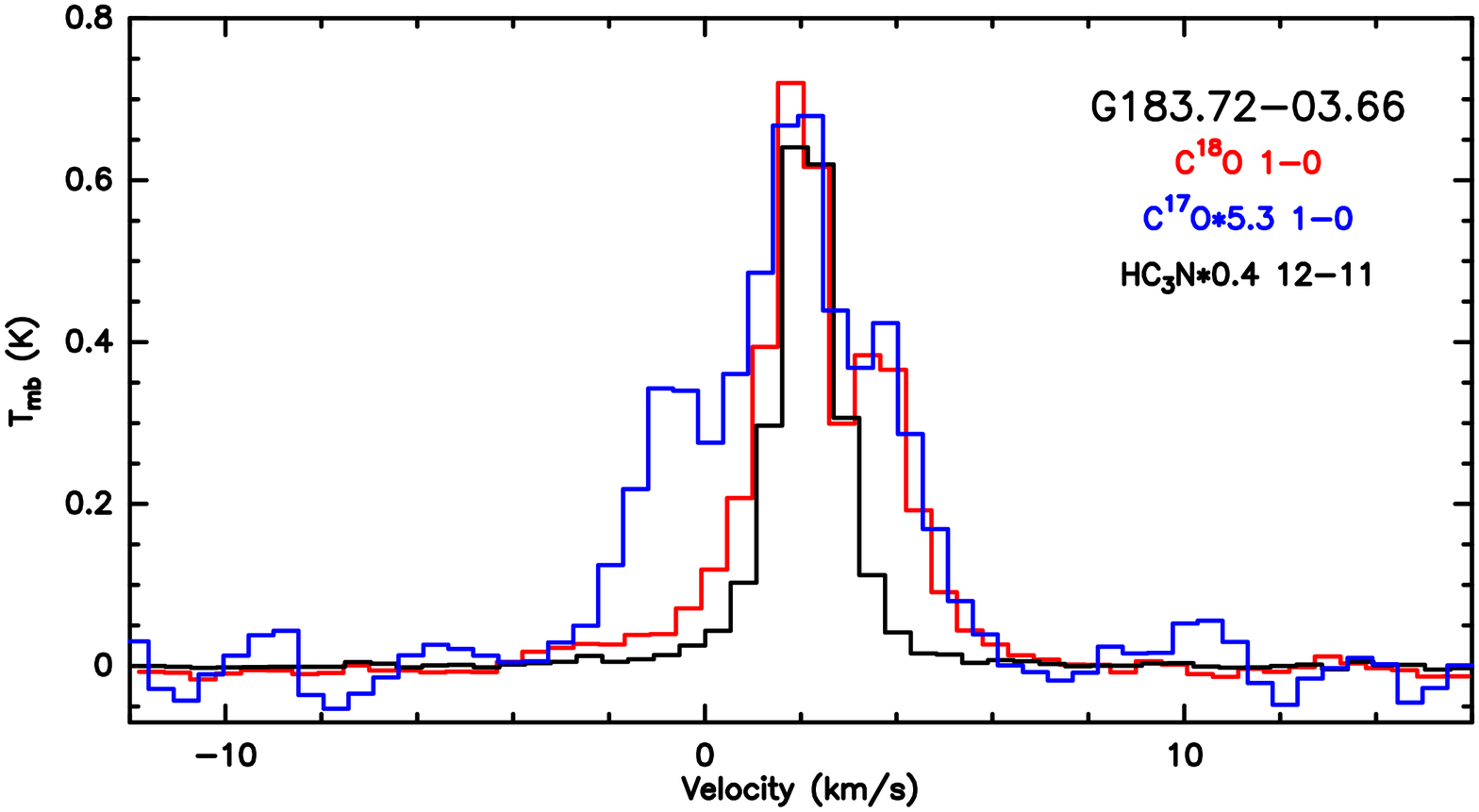}\includegraphics[width=0.45\textwidth]{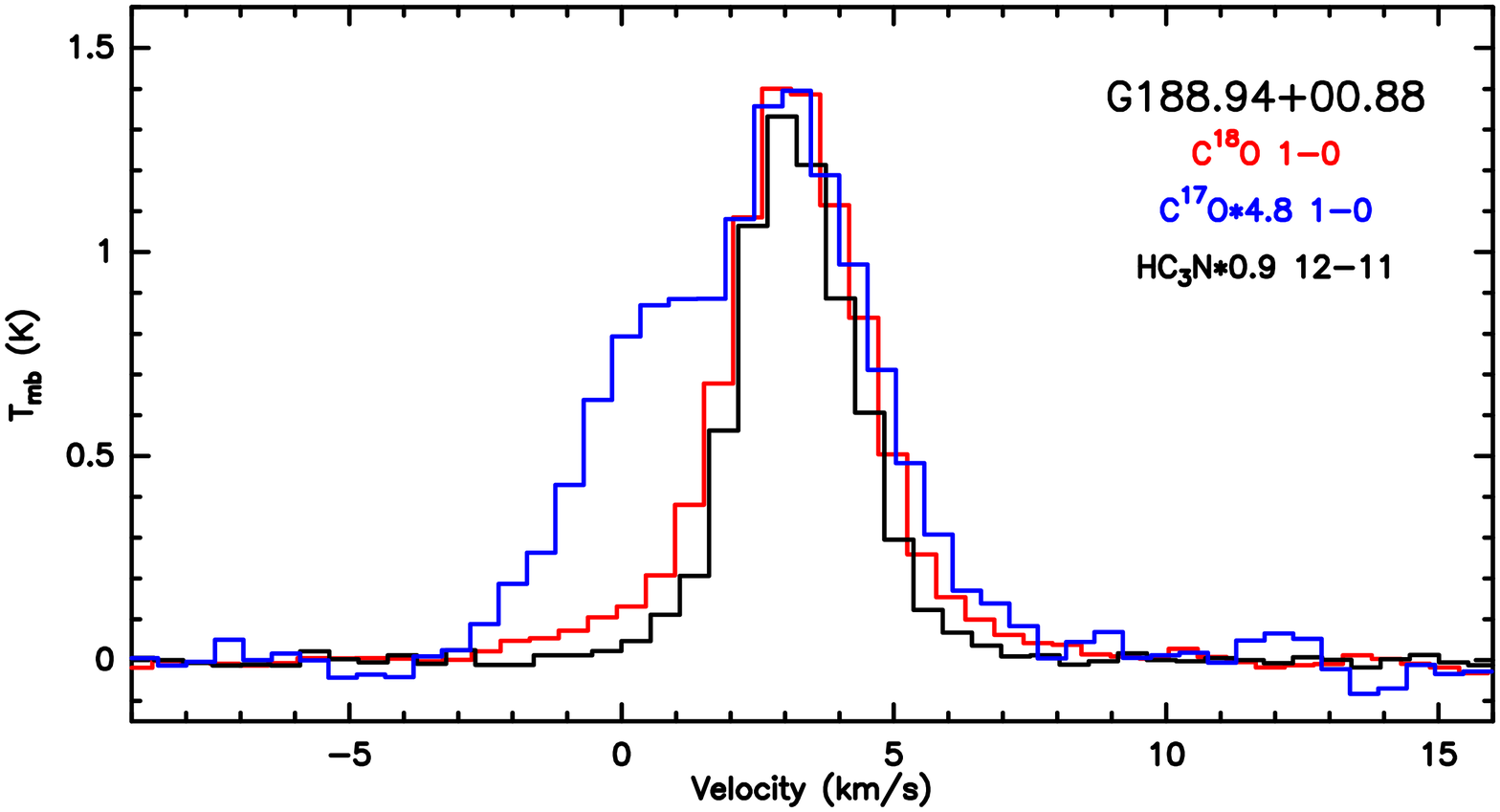}\\
\includegraphics[width=0.45\textwidth]{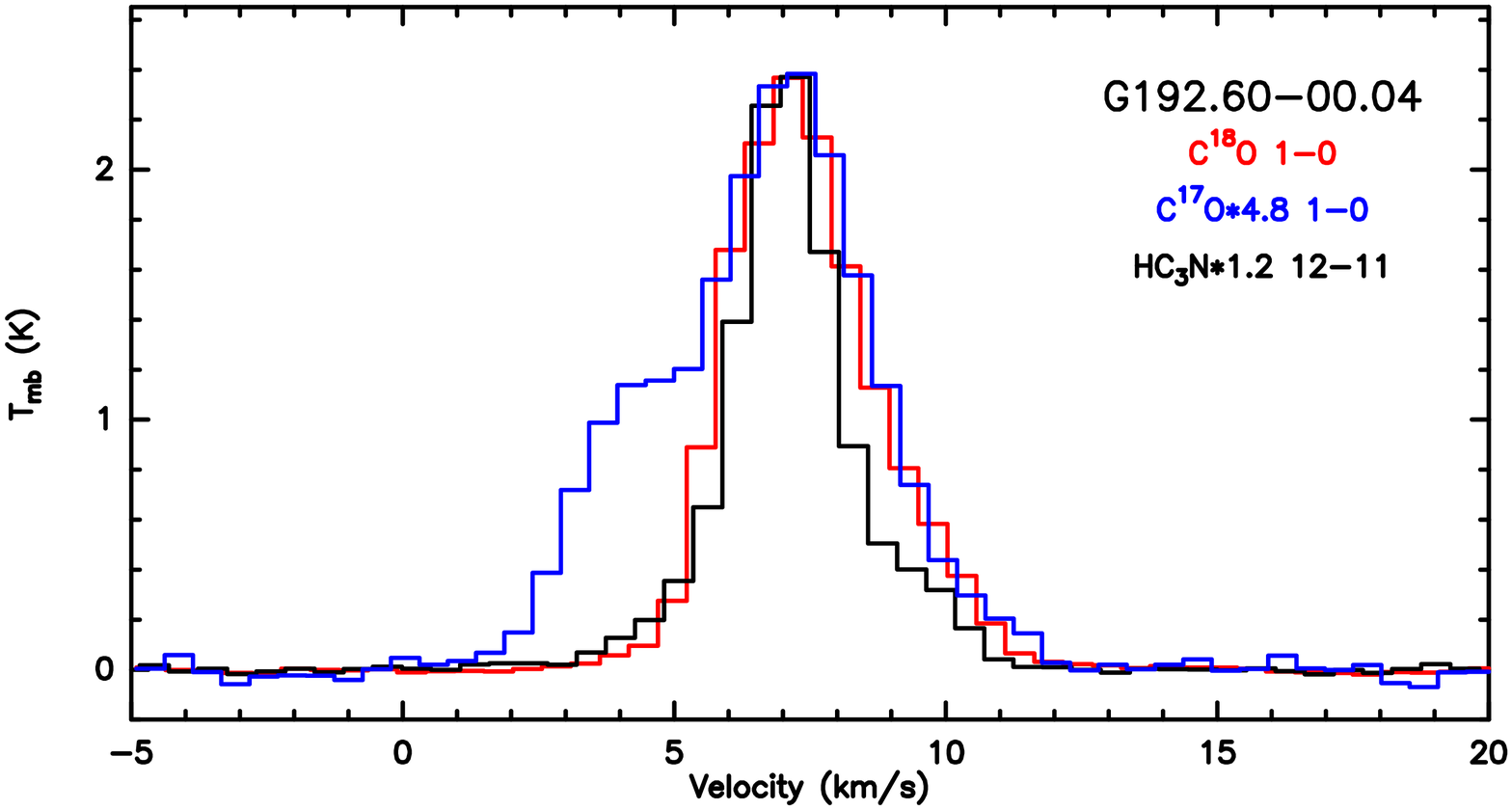}\includegraphics[width=0.45\textwidth]{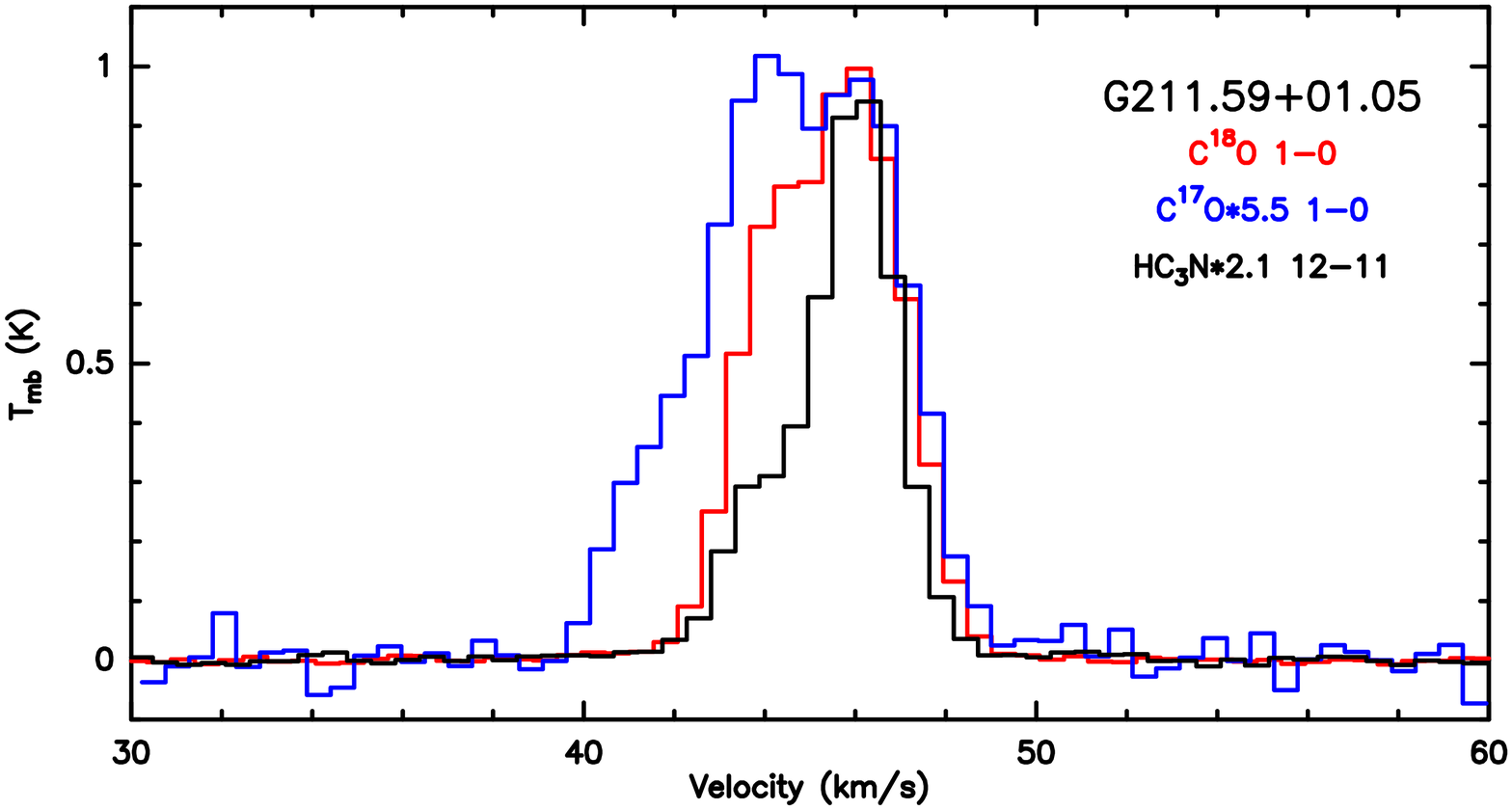}\\
\caption{The observational results of spectra for C$^{18}$O 1-0, C$^{17}$O 1-0 and HC$_{3}$N 12-11.}
\label{spectrum1}
\end{figure*}
\begin{figure*}
\addtocounter{figure}{-1}
\centering
\includegraphics[width=0.45\textwidth]{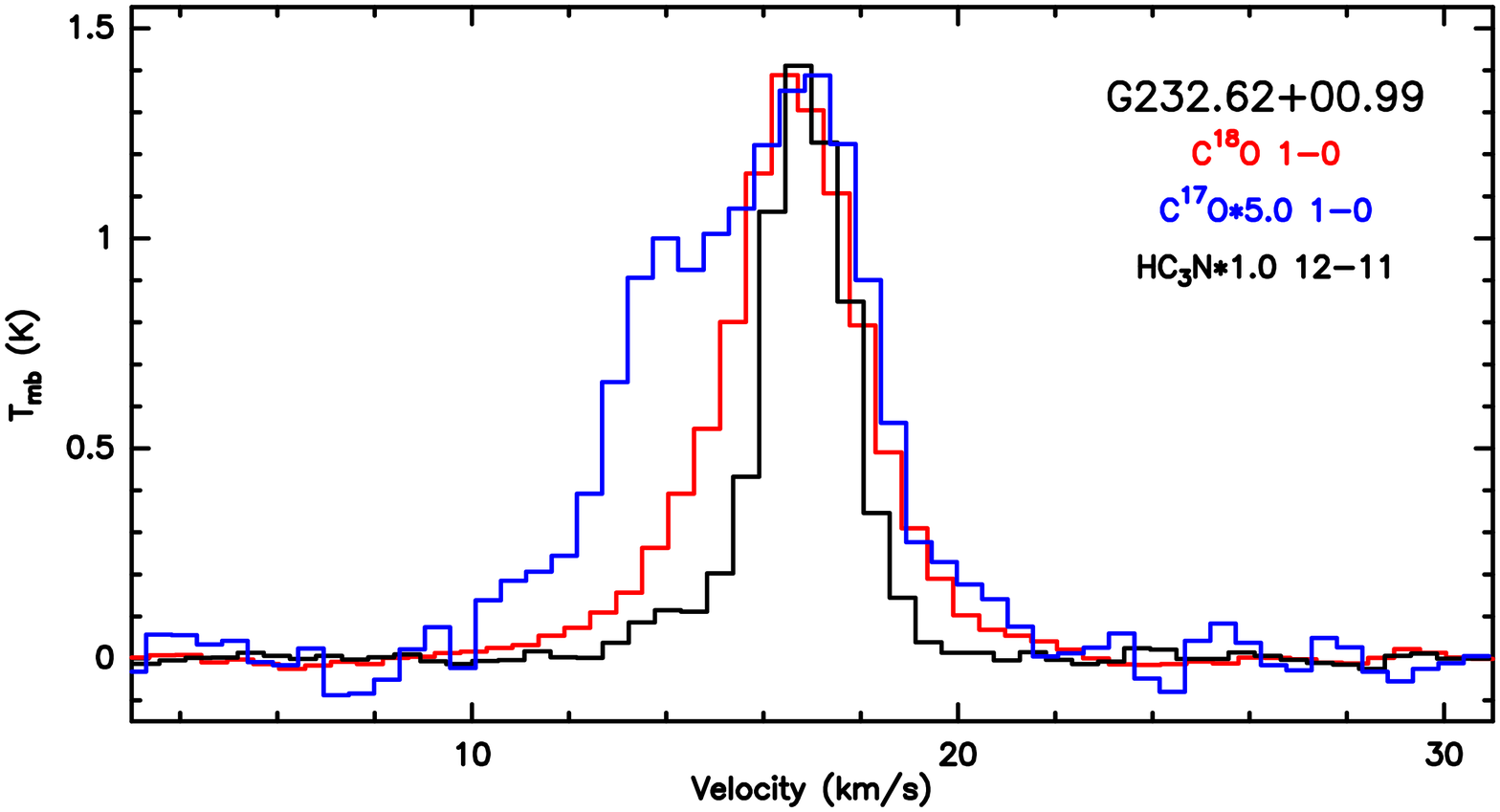}\includegraphics[width=0.45\textwidth]{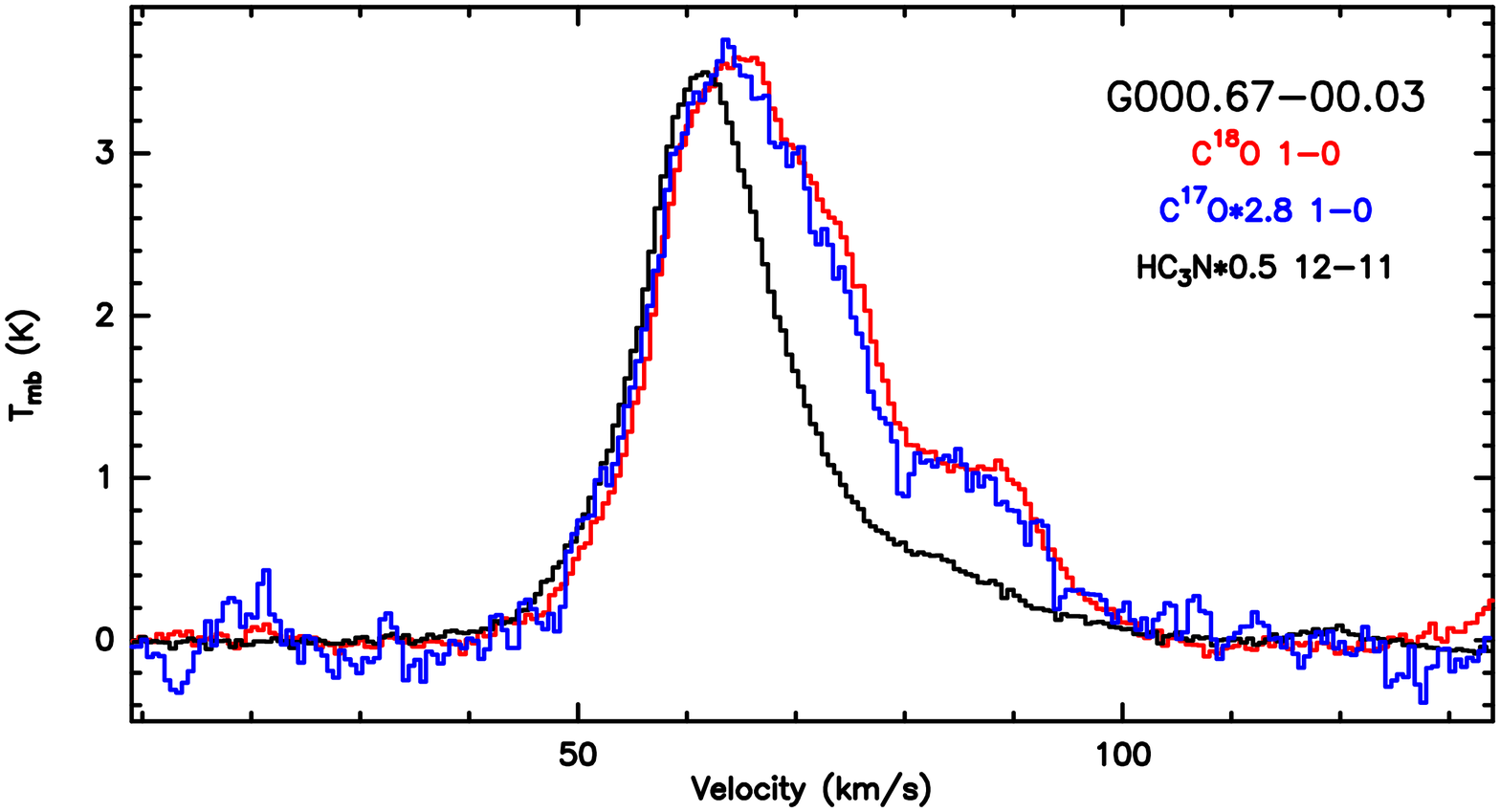}\\
\includegraphics[width=0.45\textwidth]{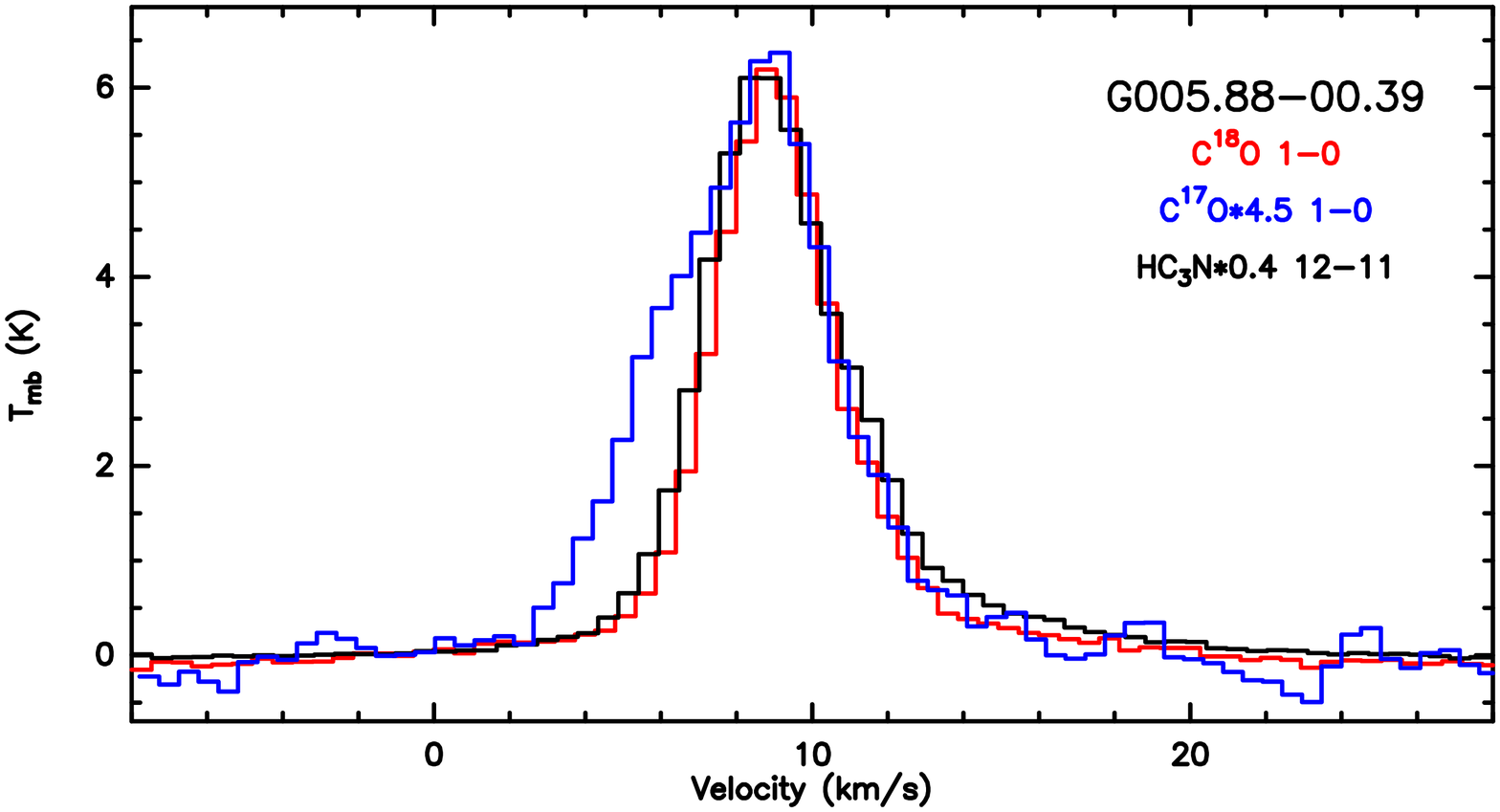}\includegraphics[width=0.45\textwidth]{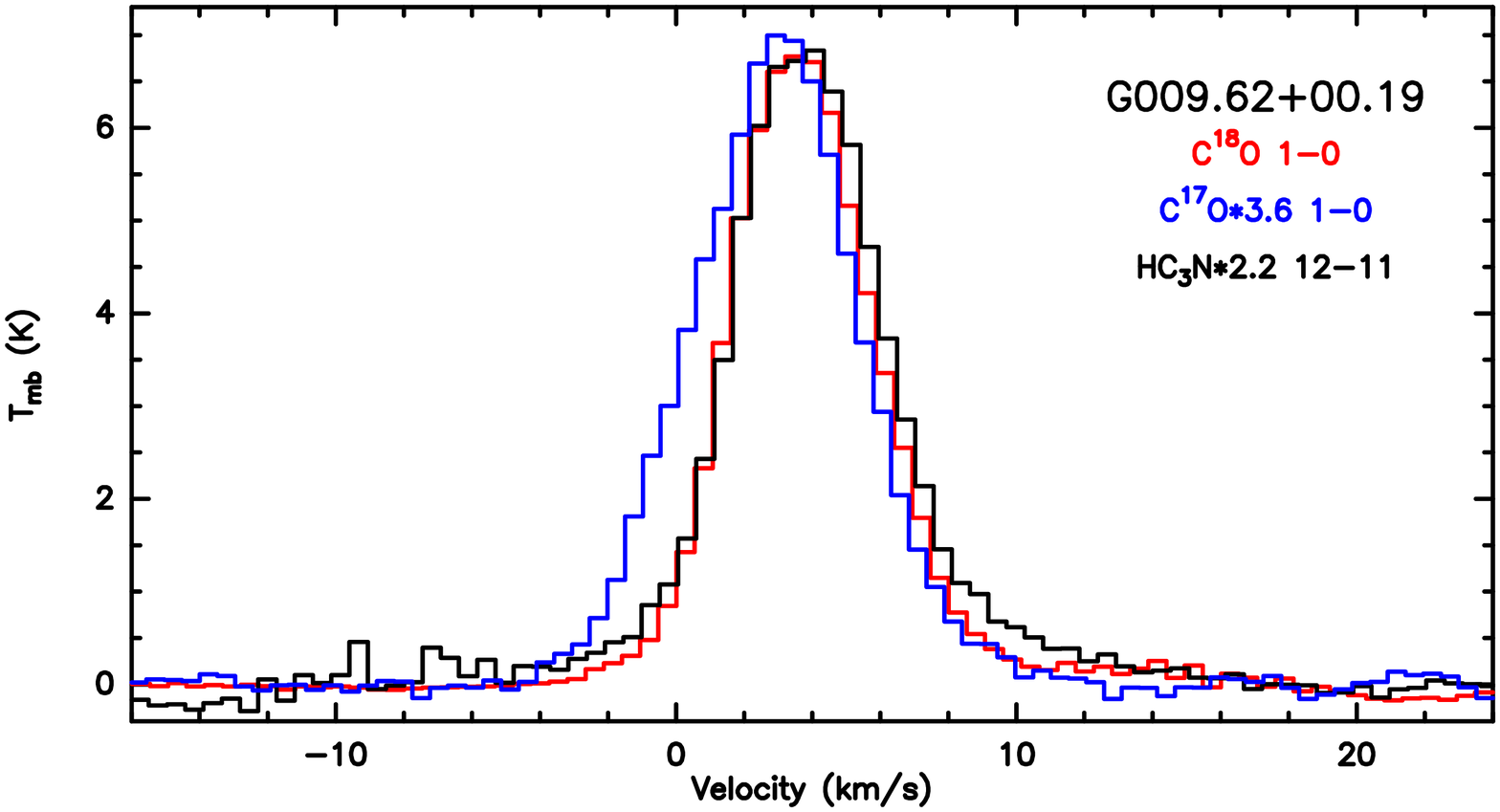}\\
\includegraphics[width=0.45\textwidth]{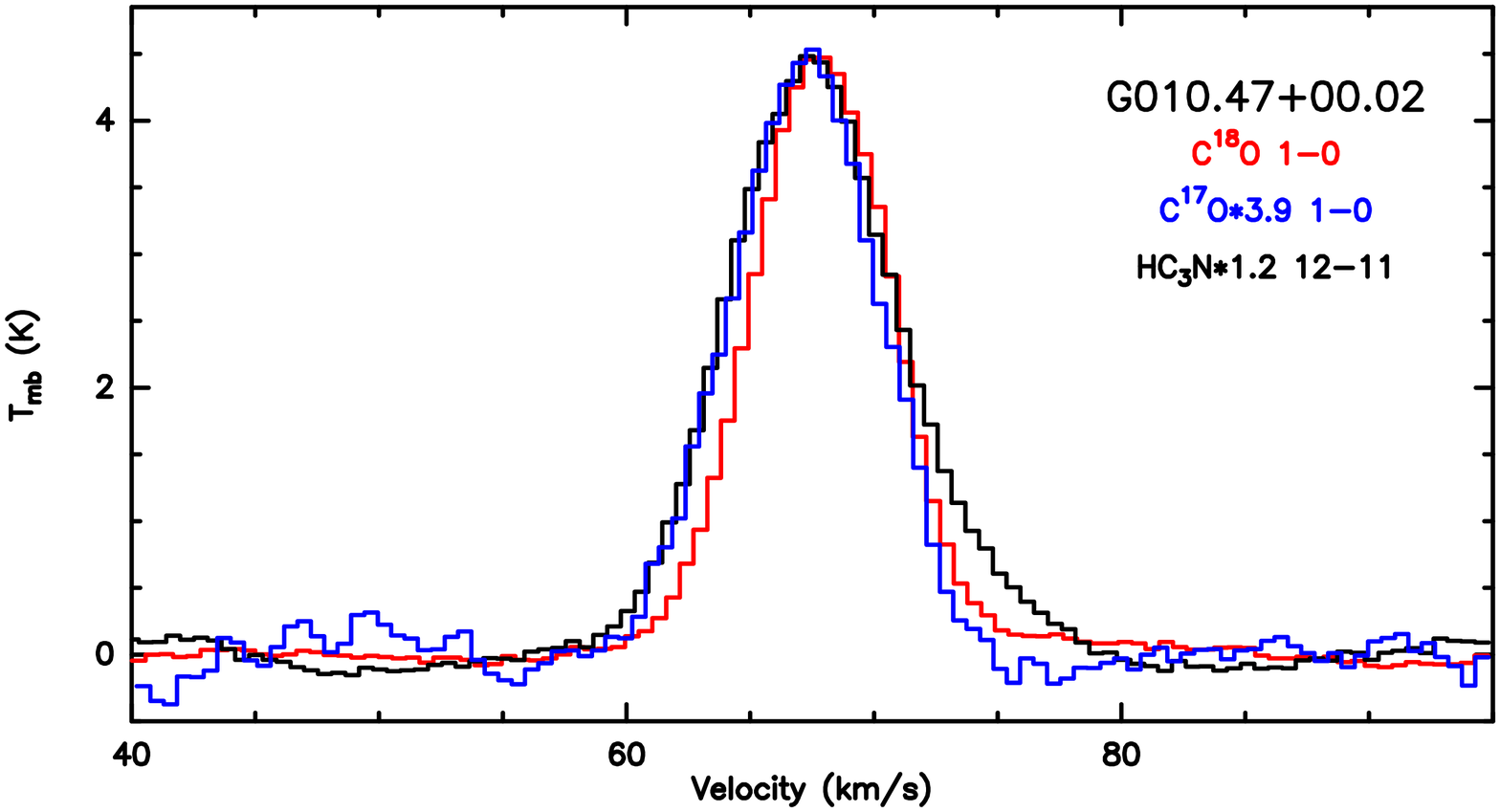}\includegraphics[width=0.45\textwidth]{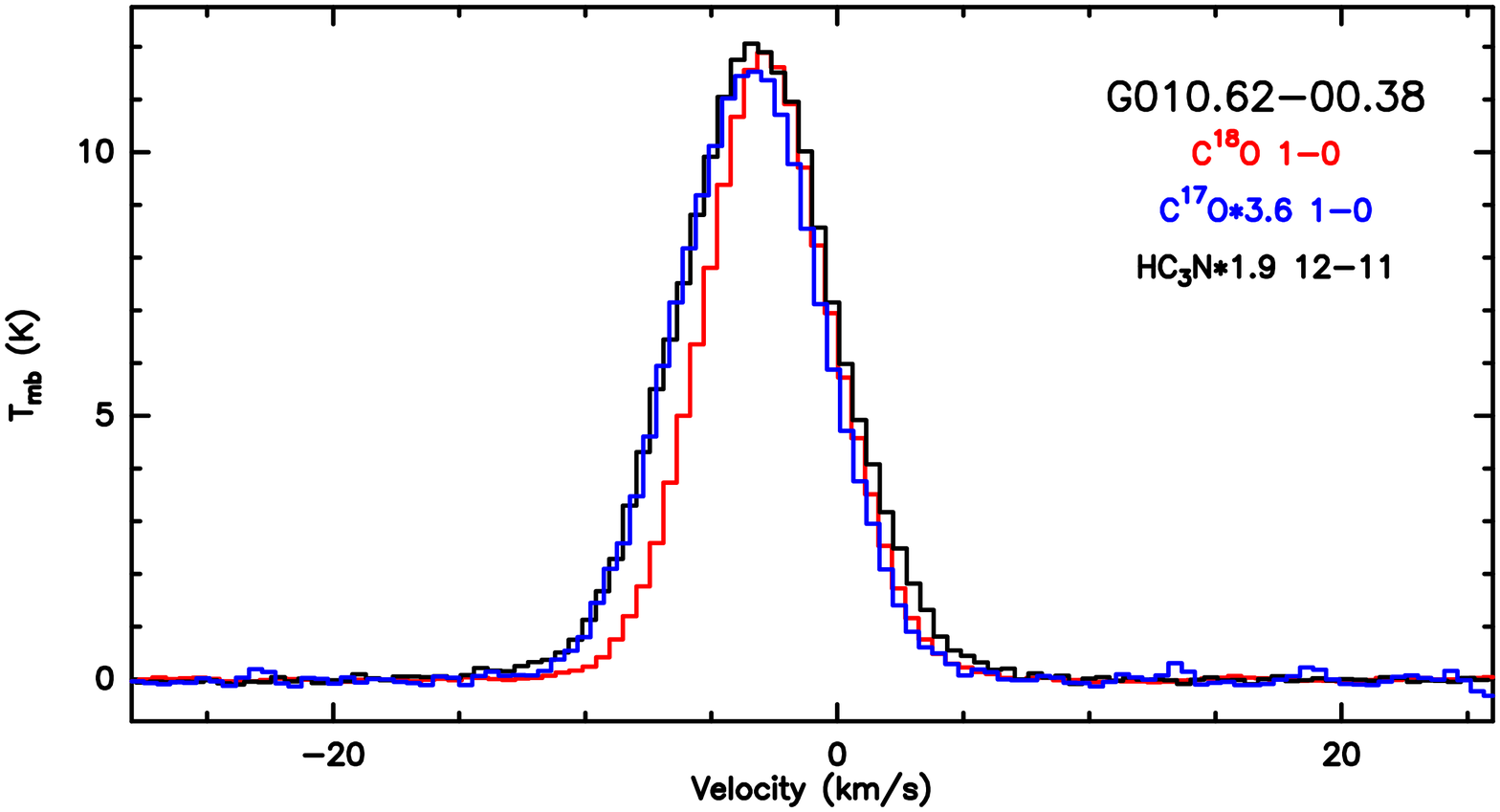}\\
\includegraphics[width=0.45\textwidth]{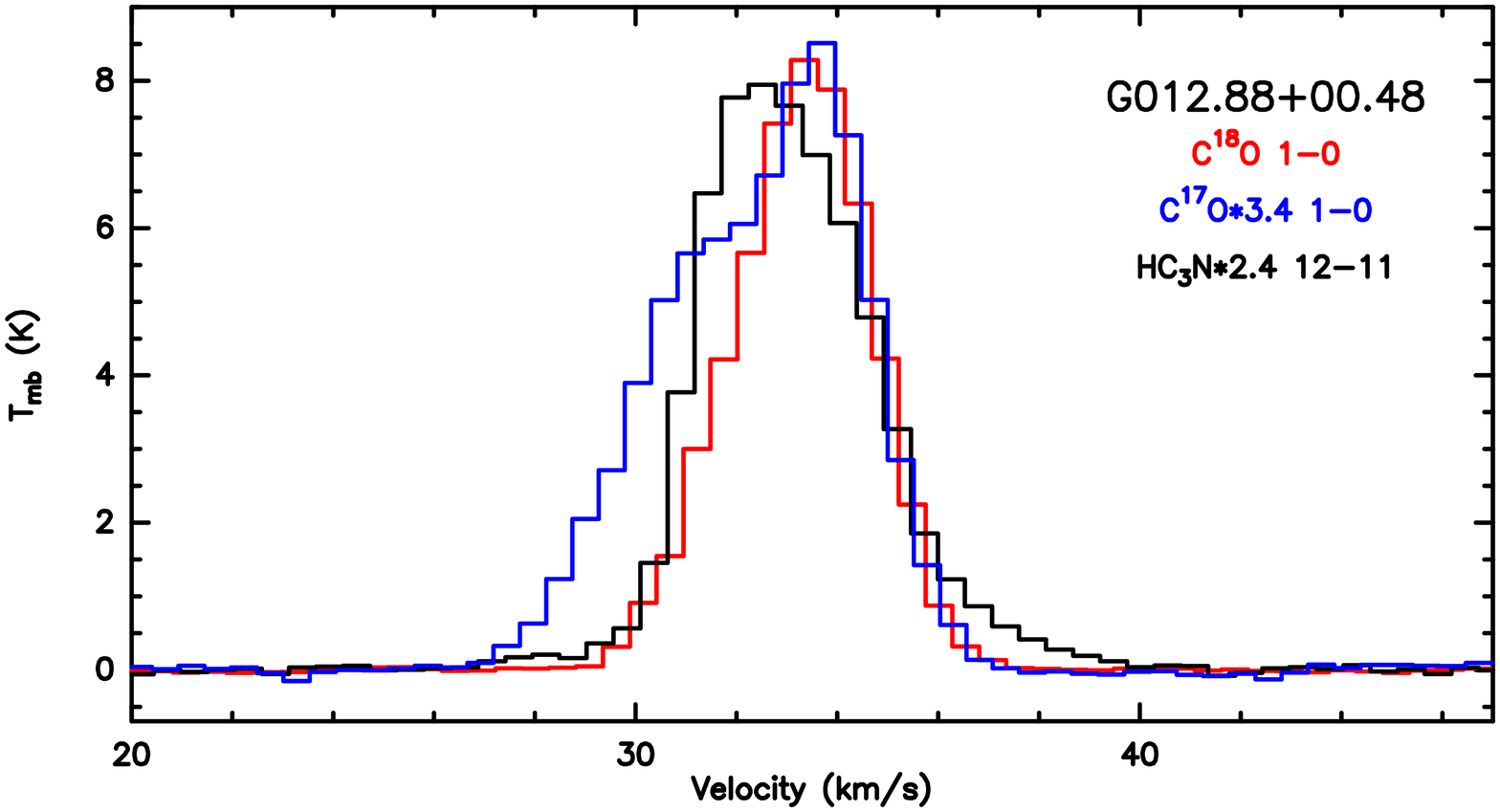}\includegraphics[width=0.45\textwidth]{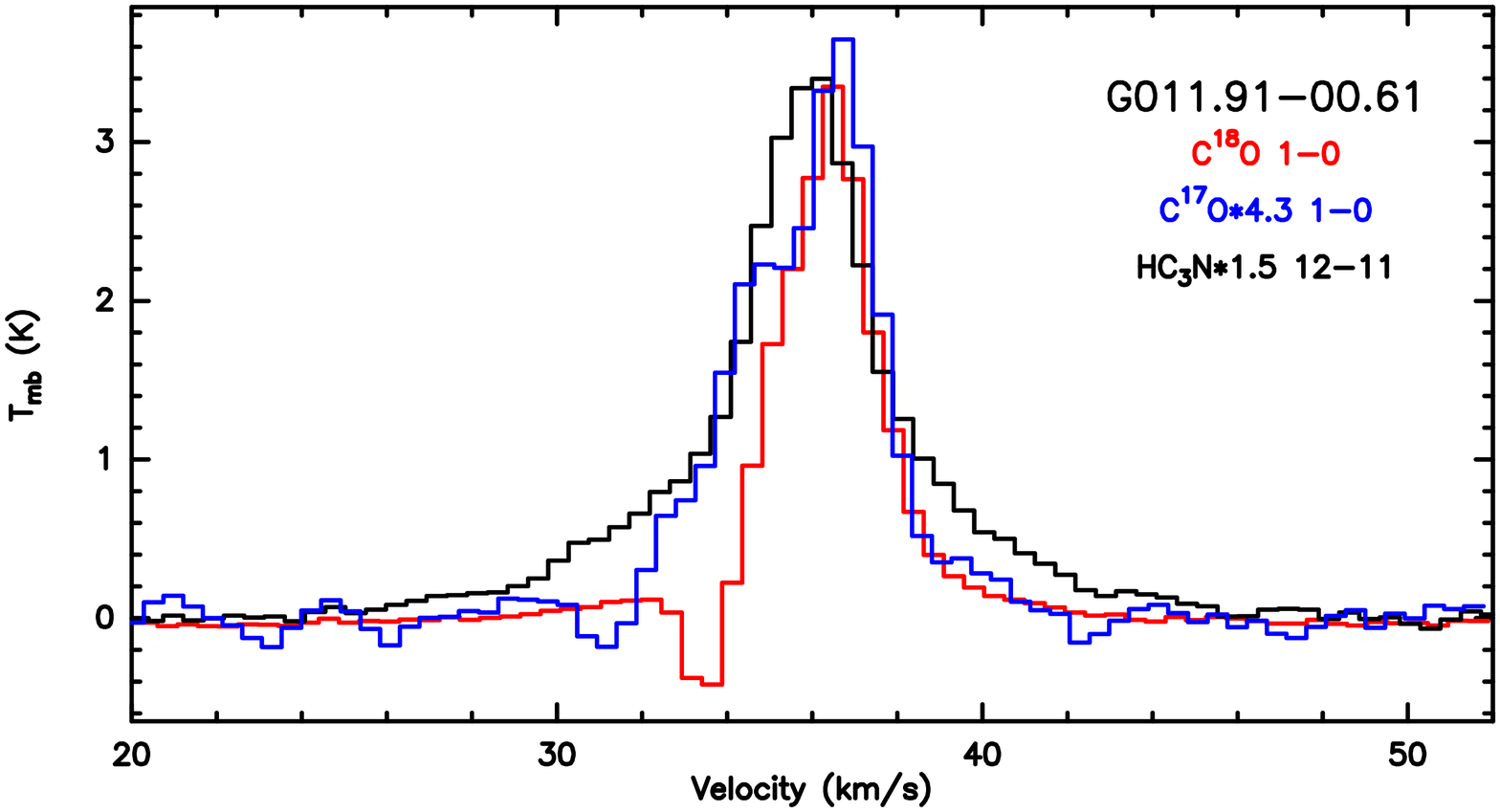}\\
\includegraphics[width=0.45\textwidth]{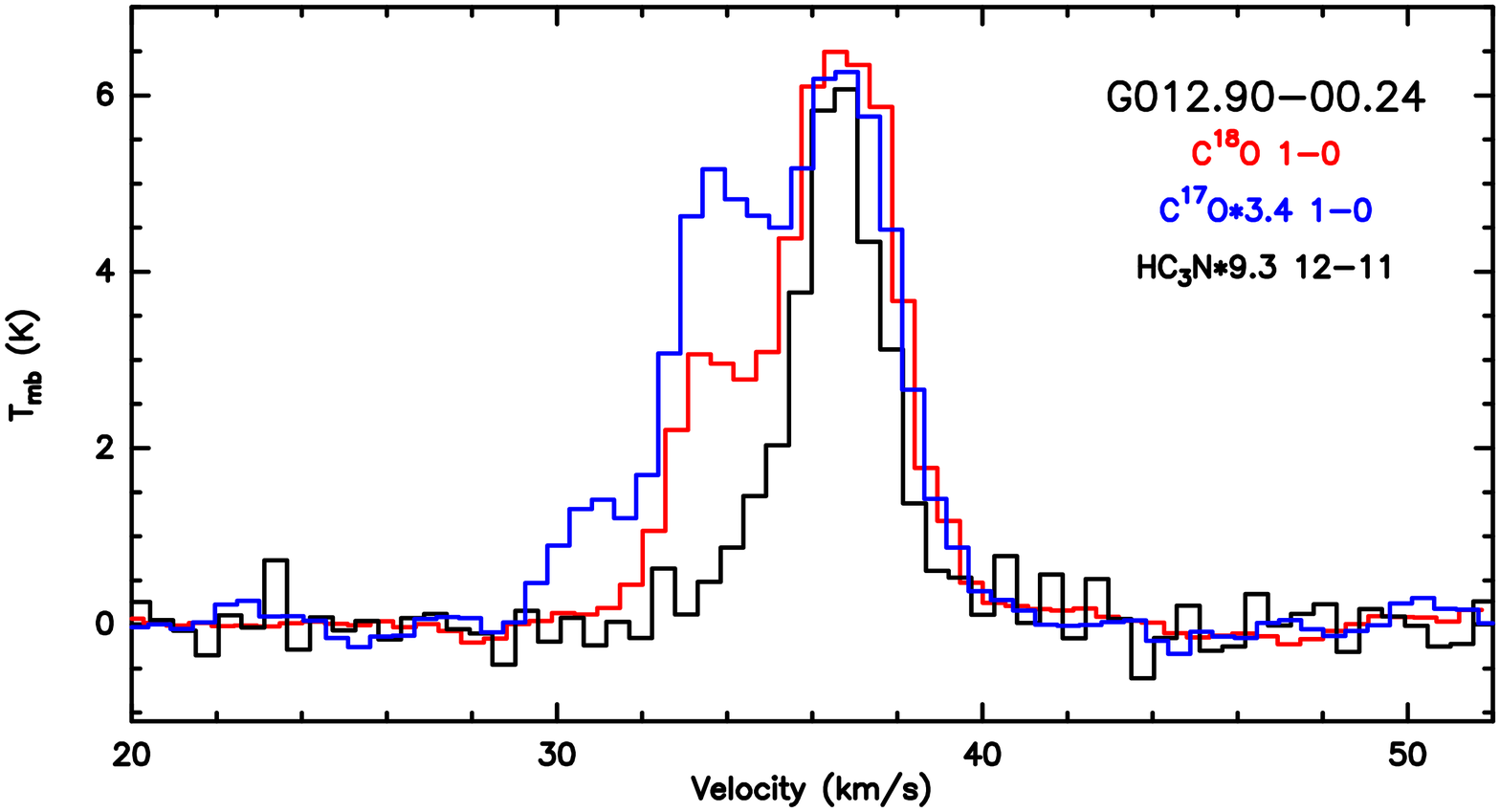}\includegraphics[width=0.45\textwidth]{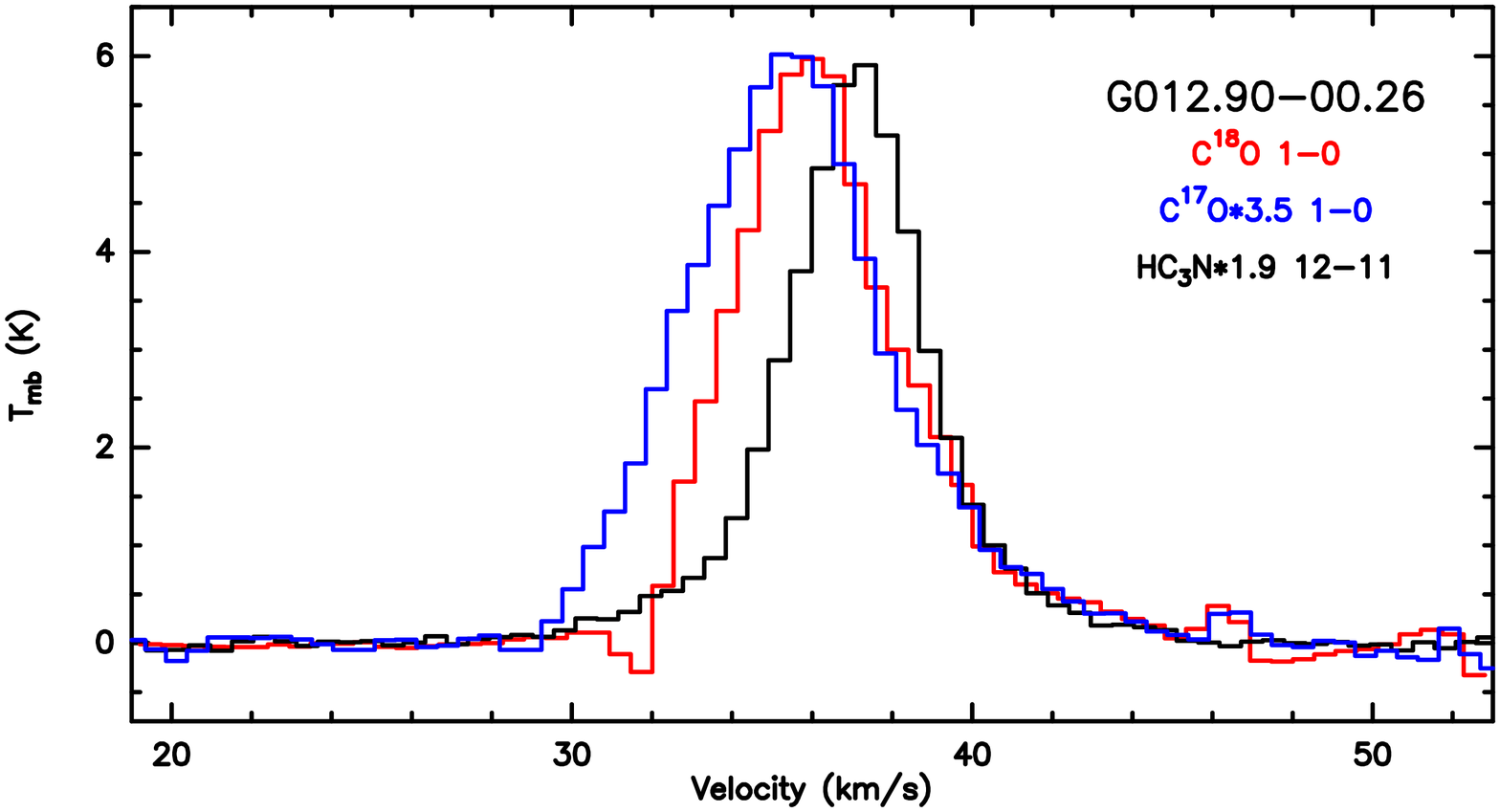}\\
\caption{Continued.}
\label{spectrum1}
\end{figure*}
\begin{figure}
\centering
\addtocounter{figure}{-1}
\includegraphics[width=0.45\textwidth]{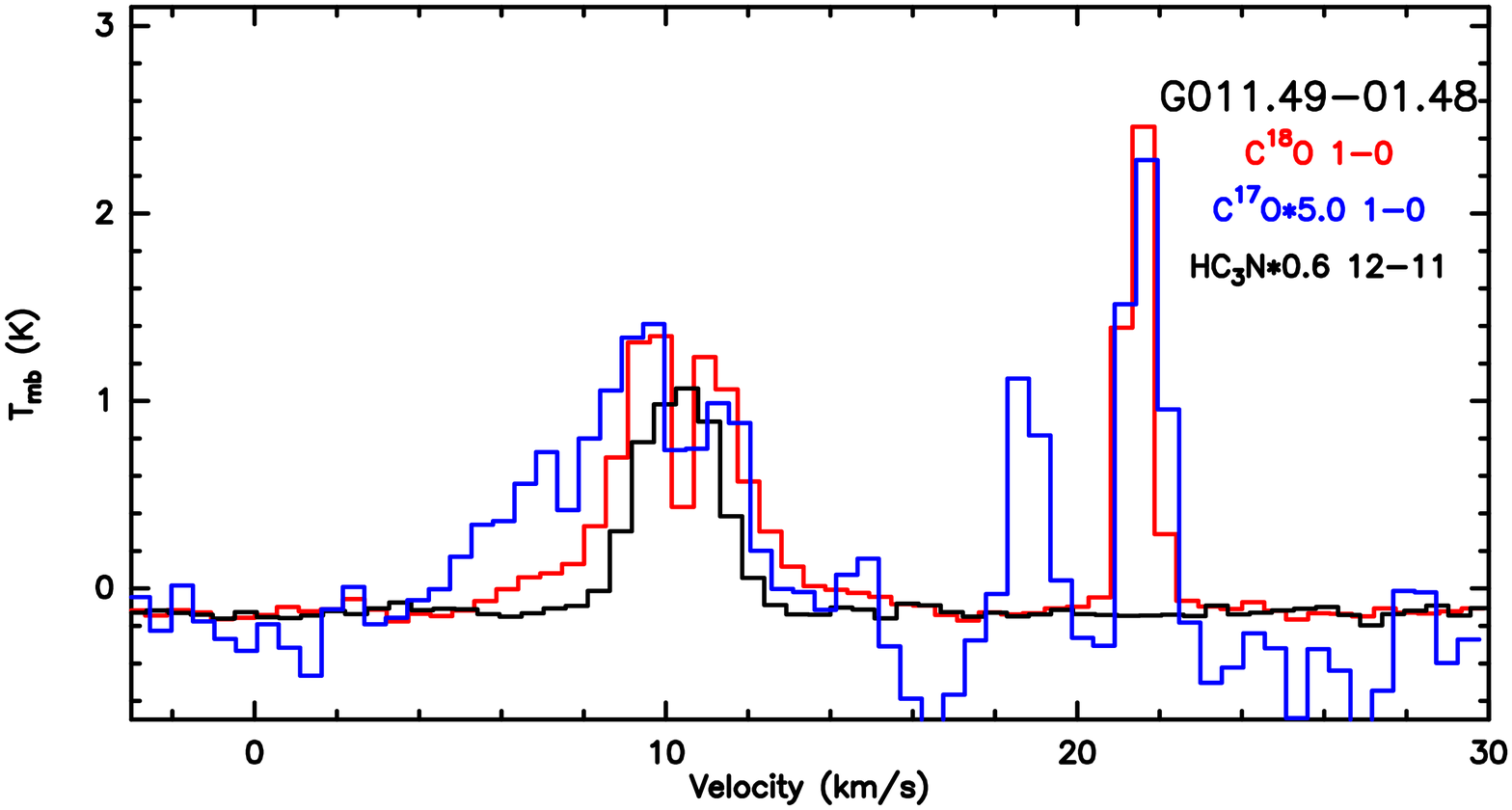}\includegraphics[width=0.45\textwidth]{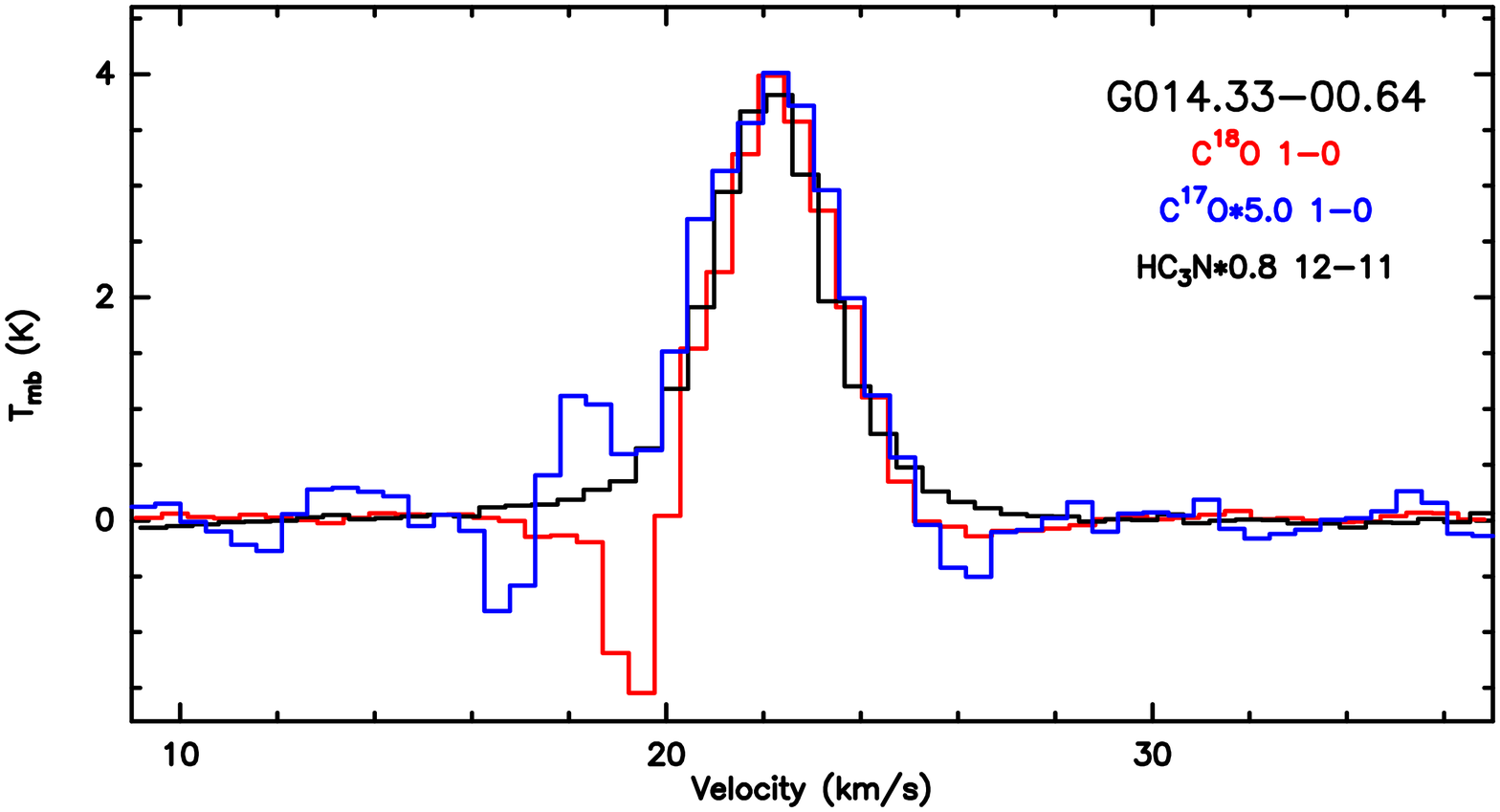}\\
\includegraphics[width=0.45\textwidth]{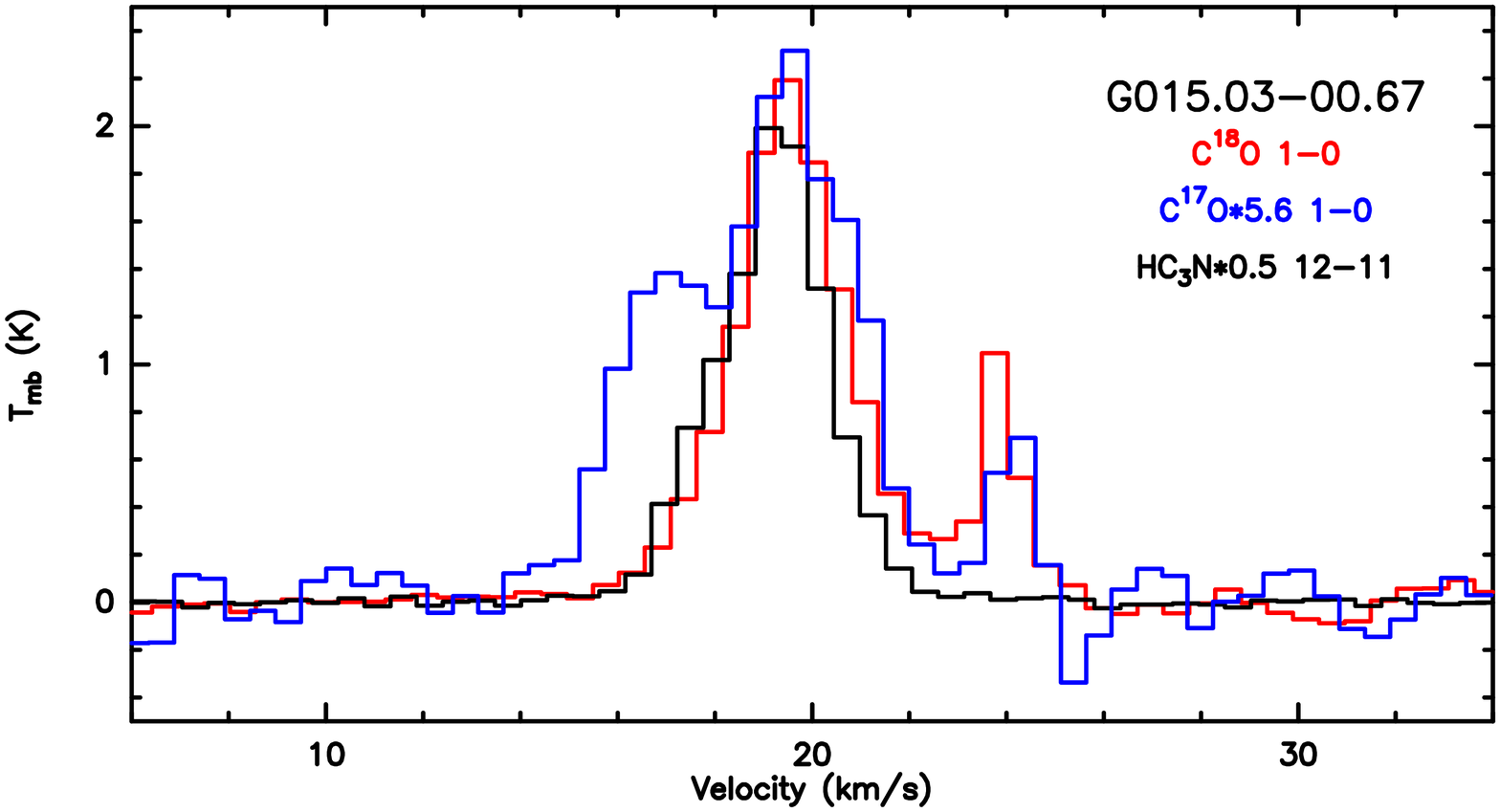}\includegraphics[width=0.45\textwidth]{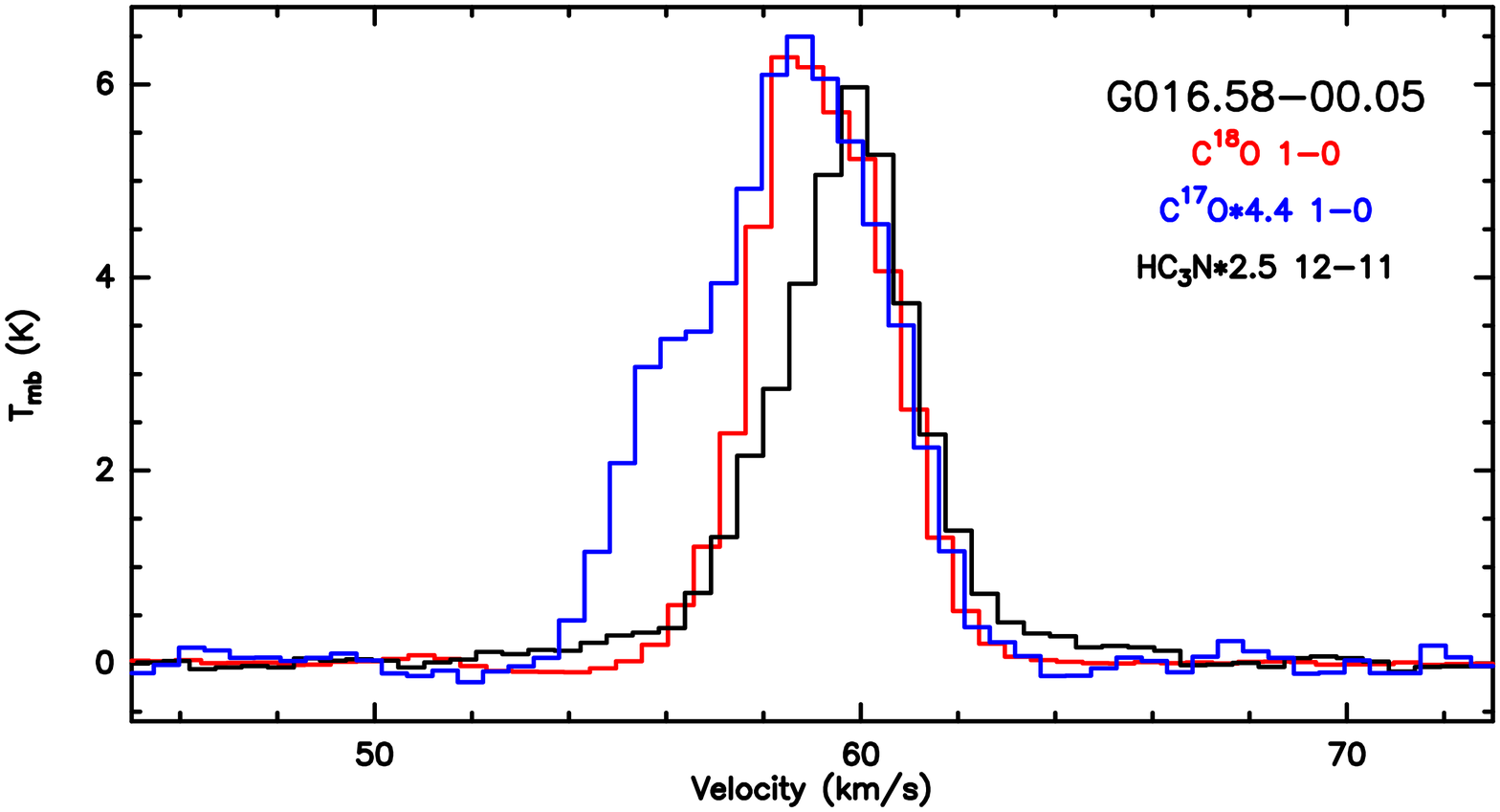}\\
\includegraphics[width=0.45\textwidth]{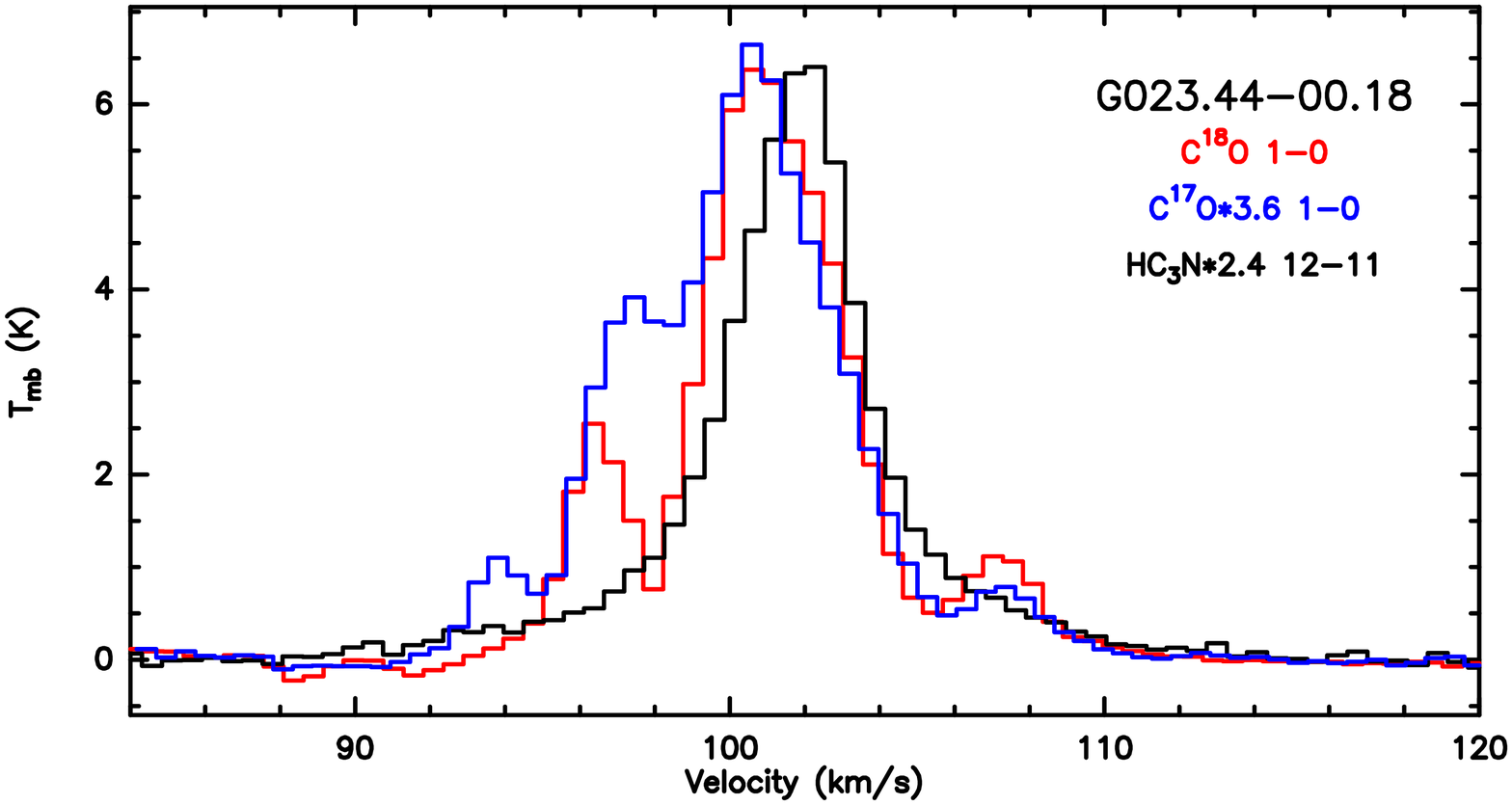}\includegraphics[width=0.45\textwidth]{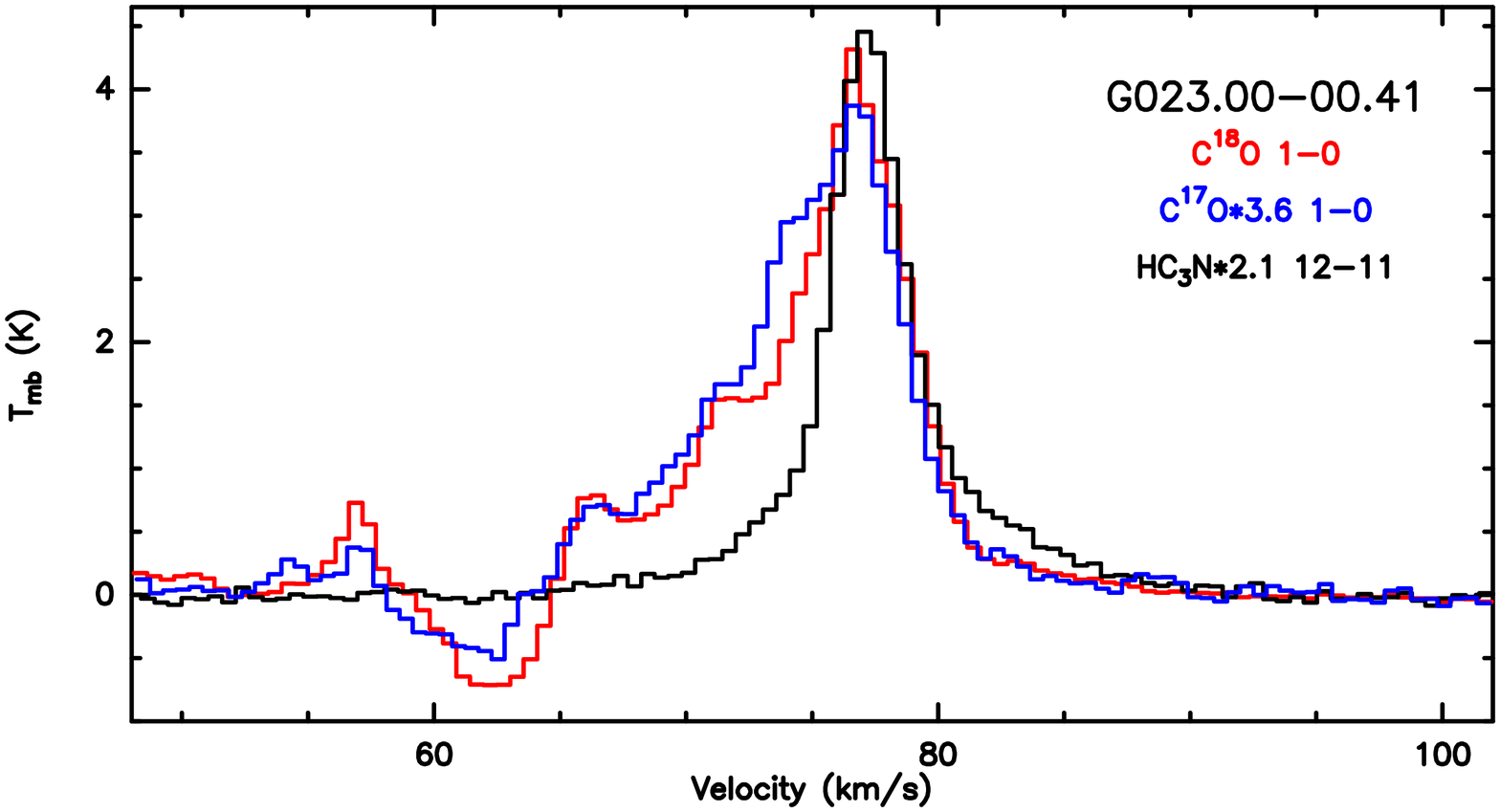}\\
\includegraphics[width=0.45\textwidth]{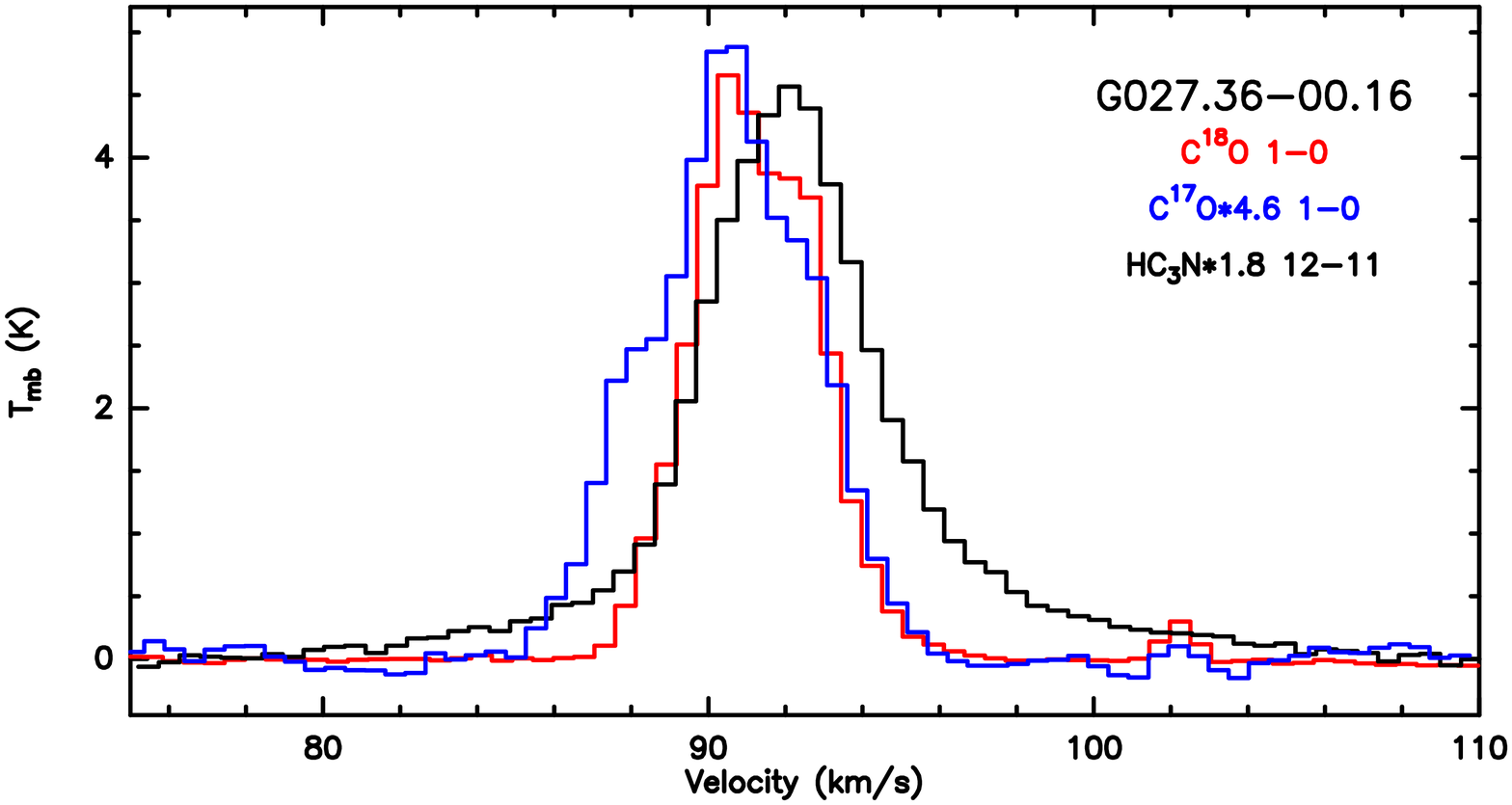}\includegraphics[width=0.45\textwidth]{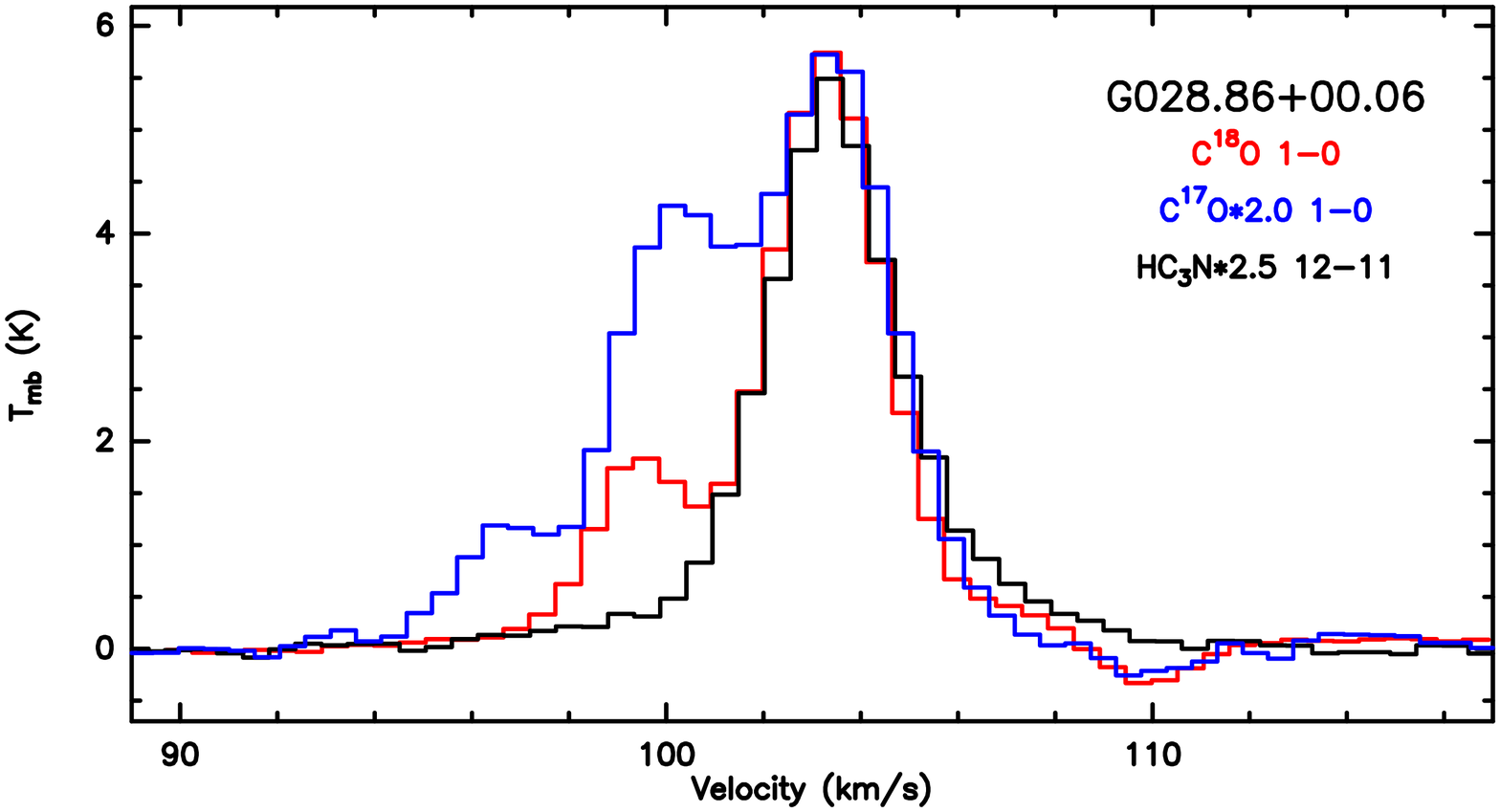}\\
\includegraphics[width=0.45\textwidth]{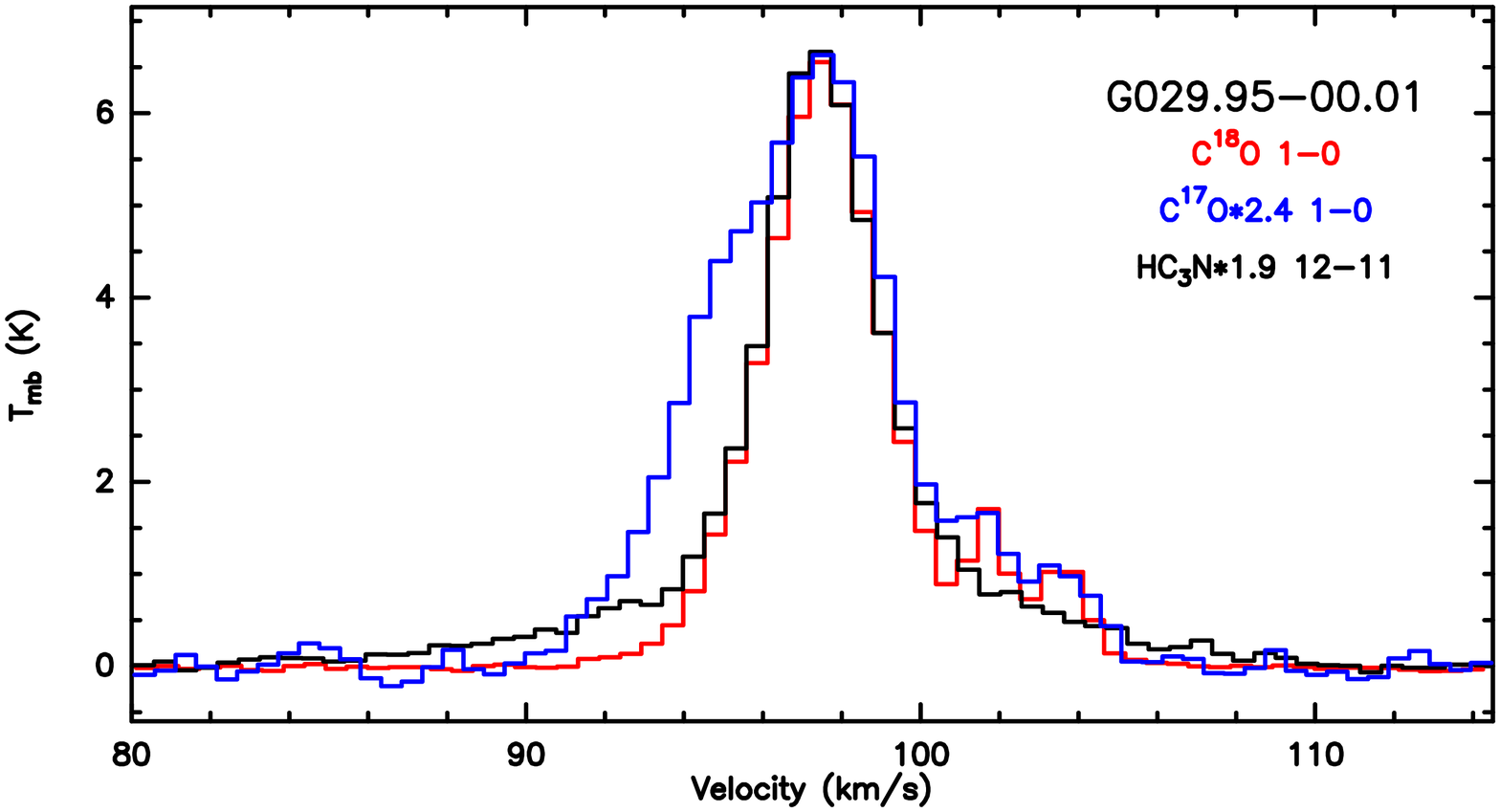}\includegraphics[width=0.45\textwidth]{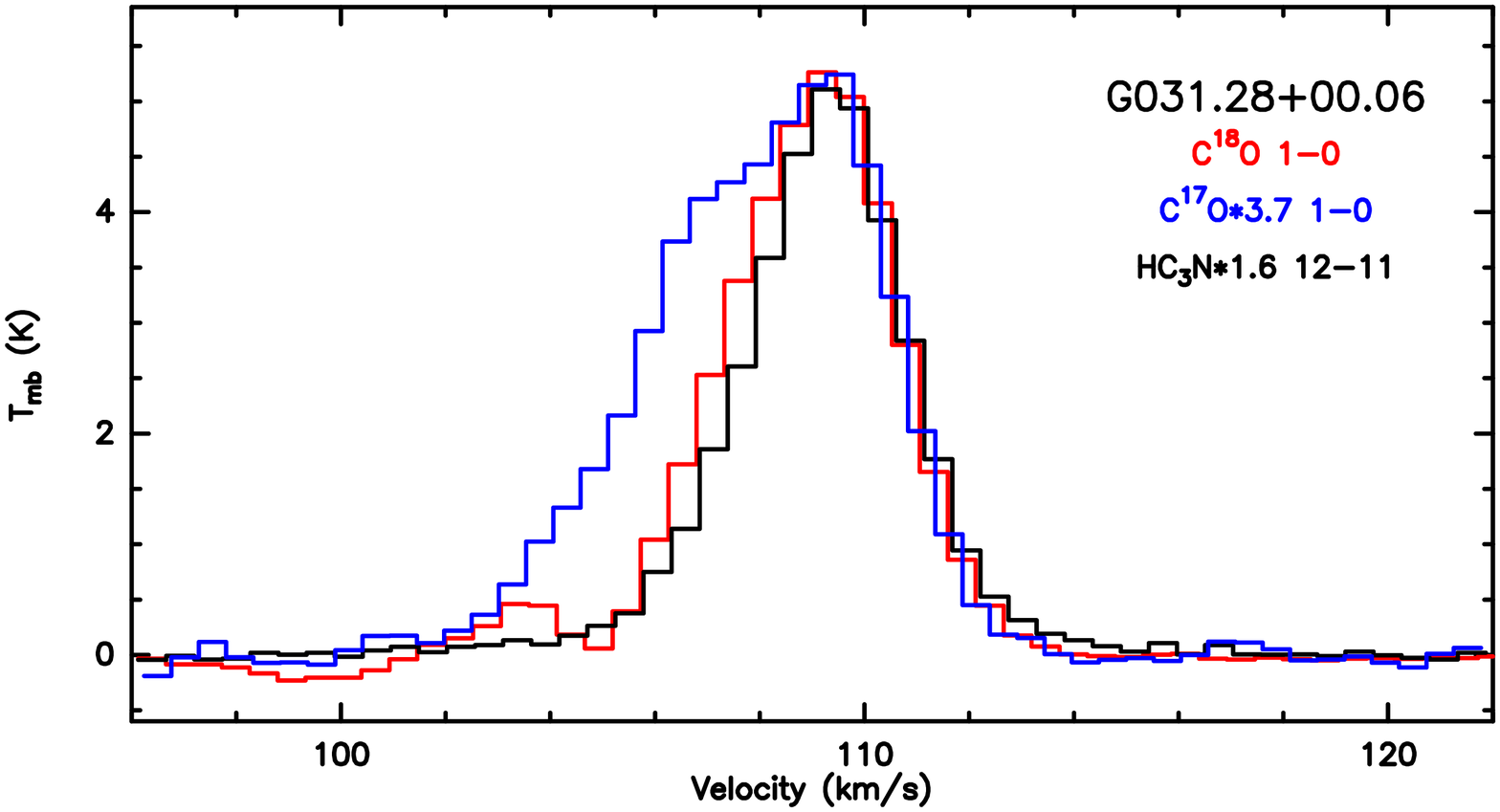}\\
\caption{Continued.}
\label{spectrum1}
\end{figure}
\begin{figure}
\centering
\addtocounter{figure}{-1}
\includegraphics[width=0.45\textwidth]{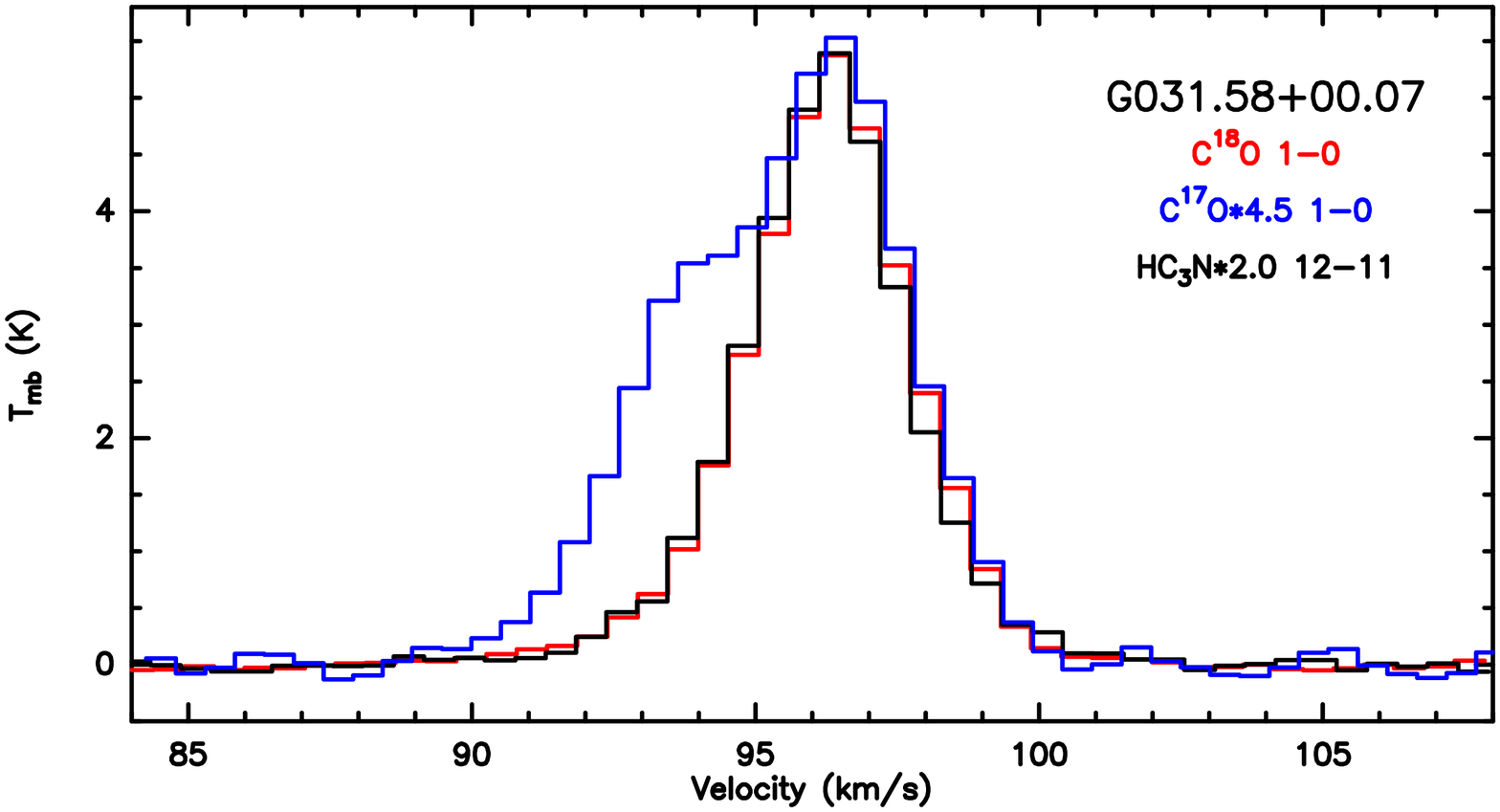}\includegraphics[width=0.45\textwidth]{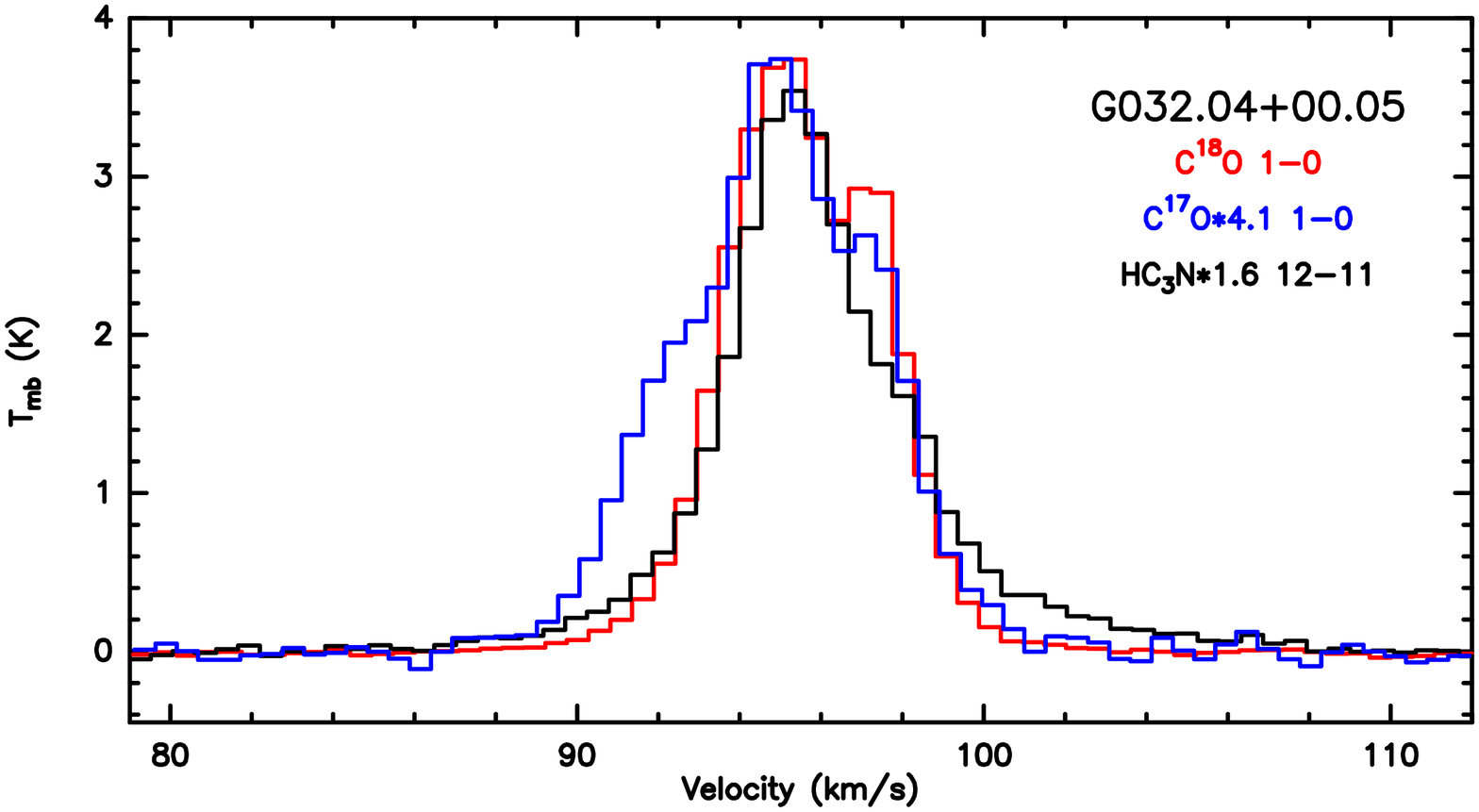}\\
\includegraphics[width=0.45\textwidth]{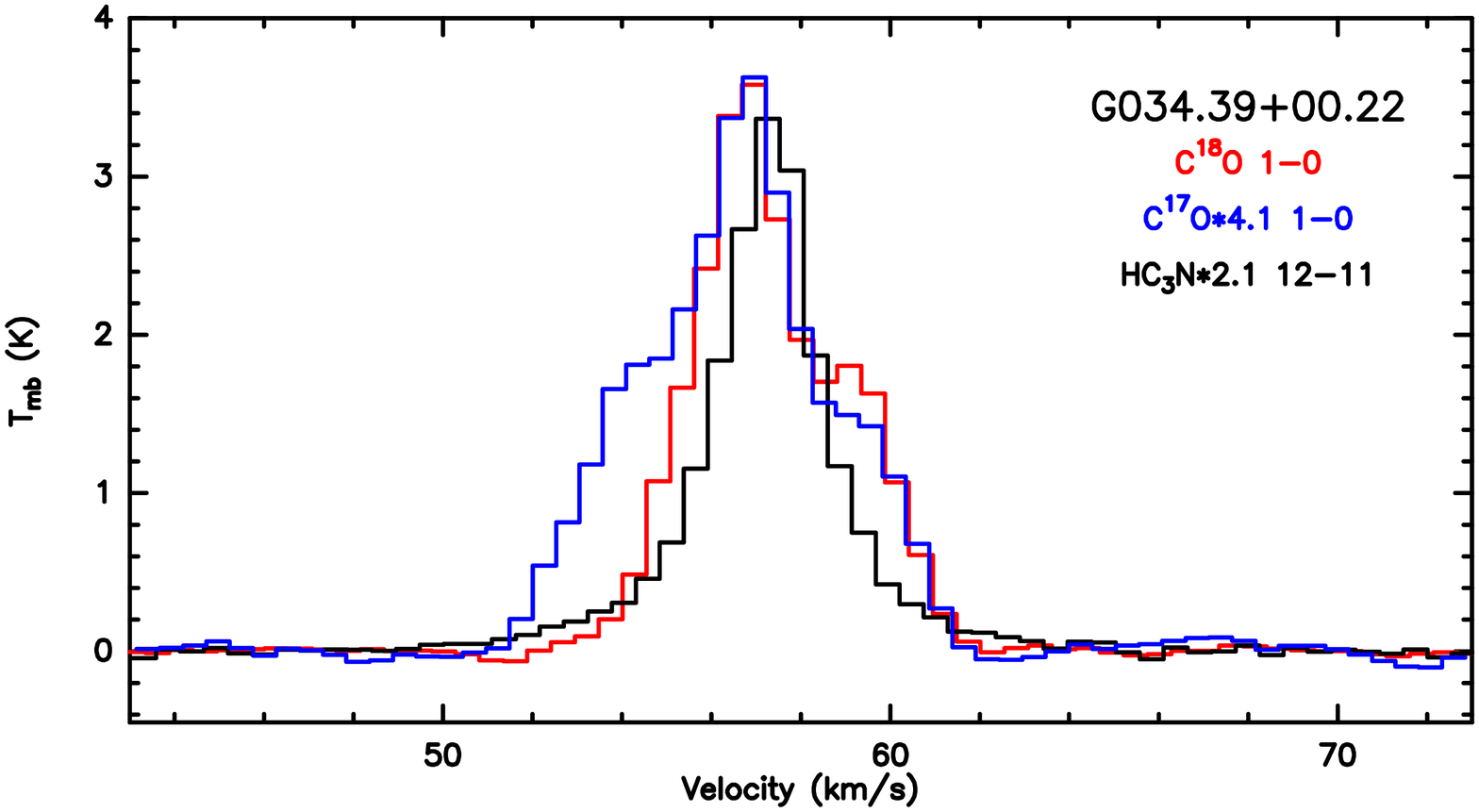}\includegraphics[width=0.45\textwidth]{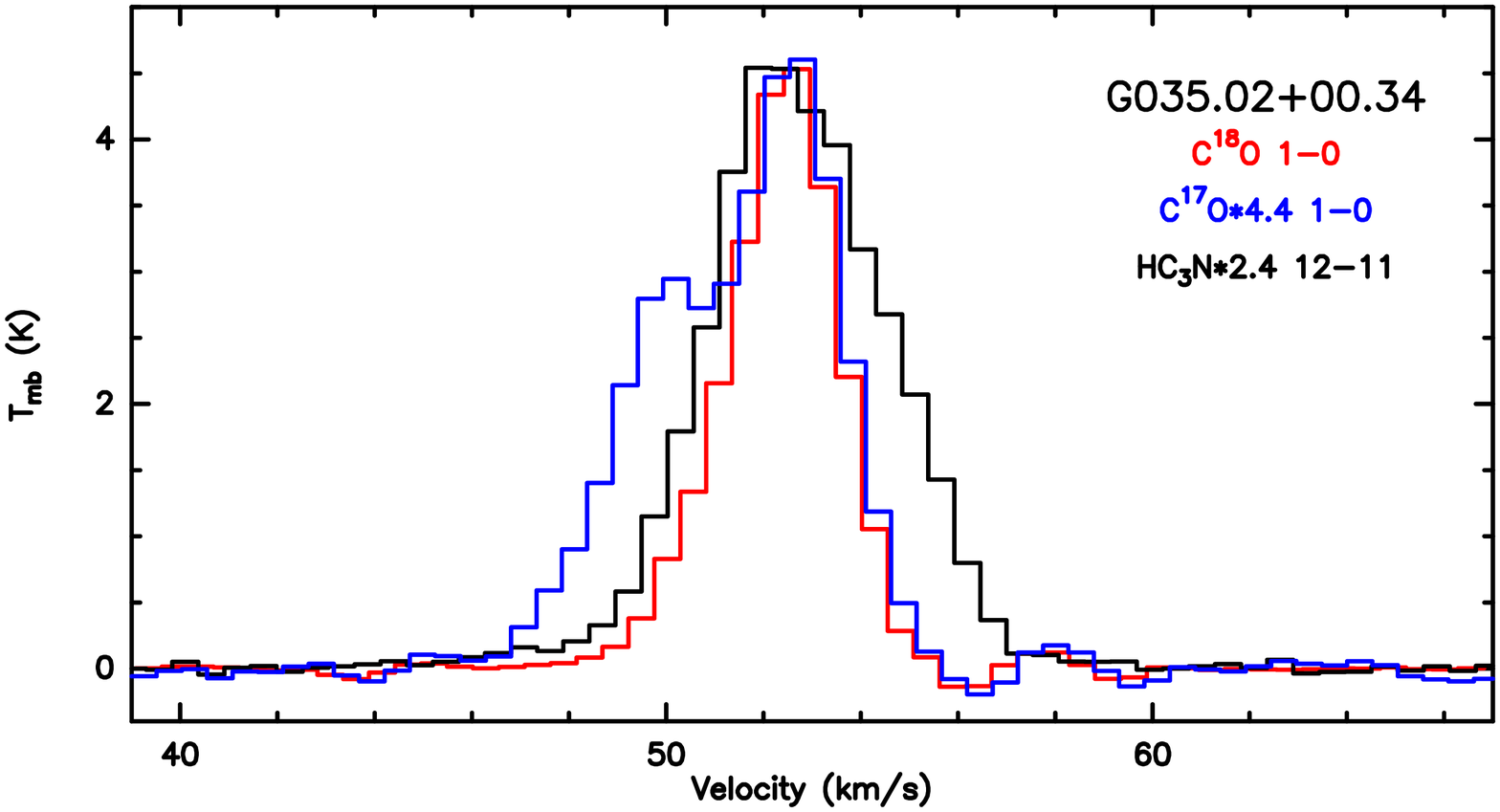}\\
\includegraphics[width=0.45\textwidth]{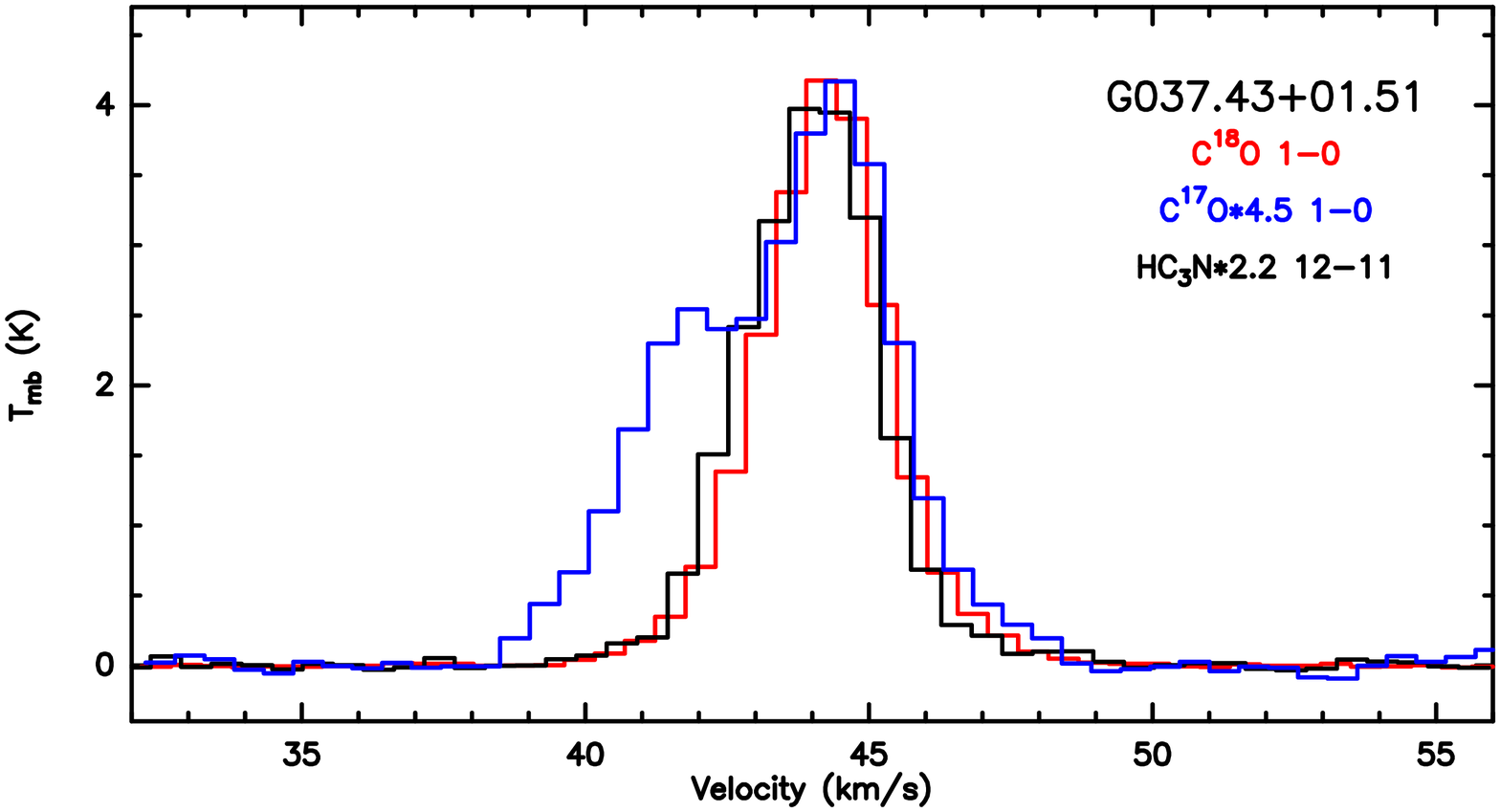}\includegraphics[width=0.45\textwidth]{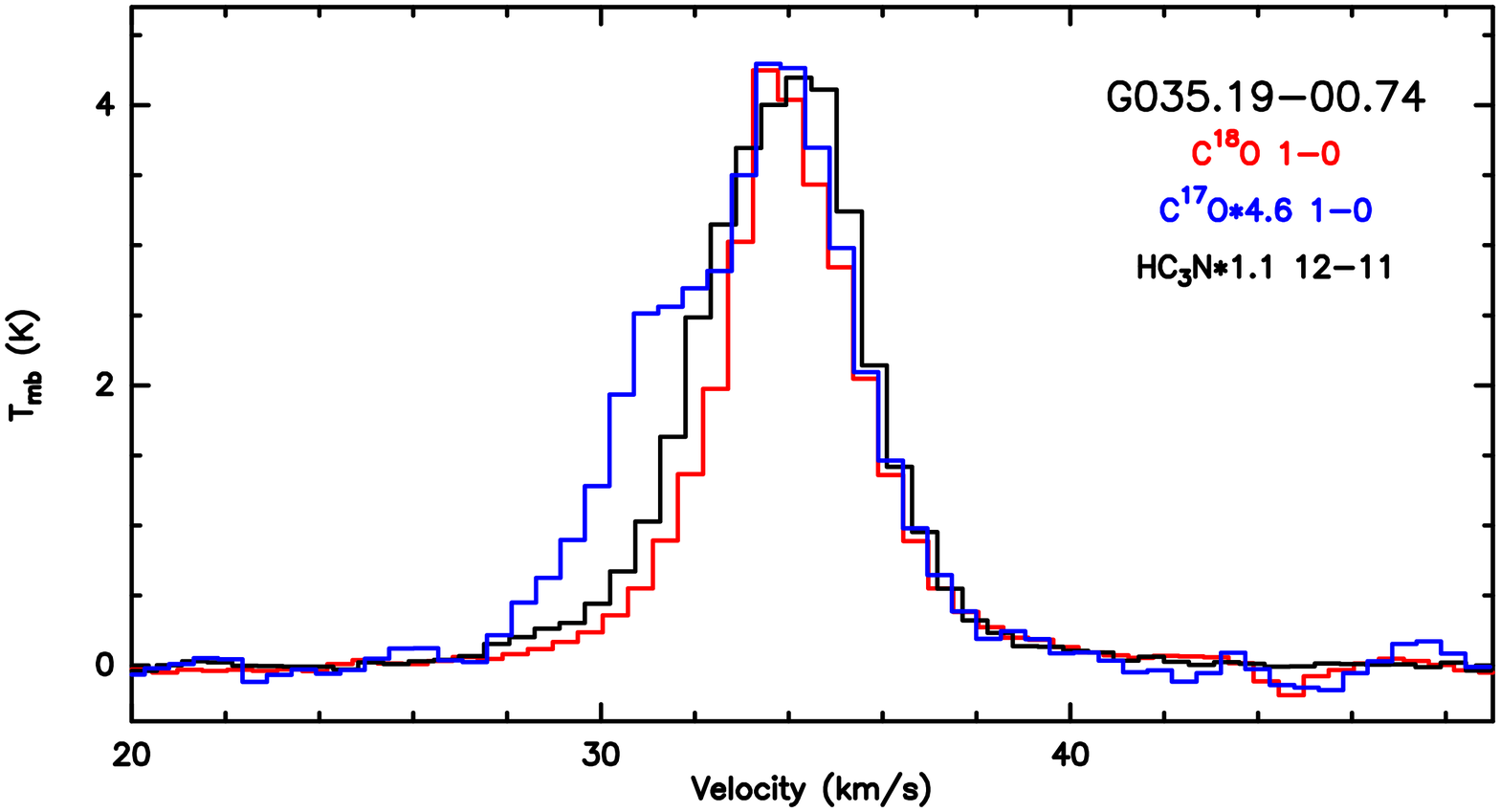}\\
\includegraphics[width=0.45\textwidth]{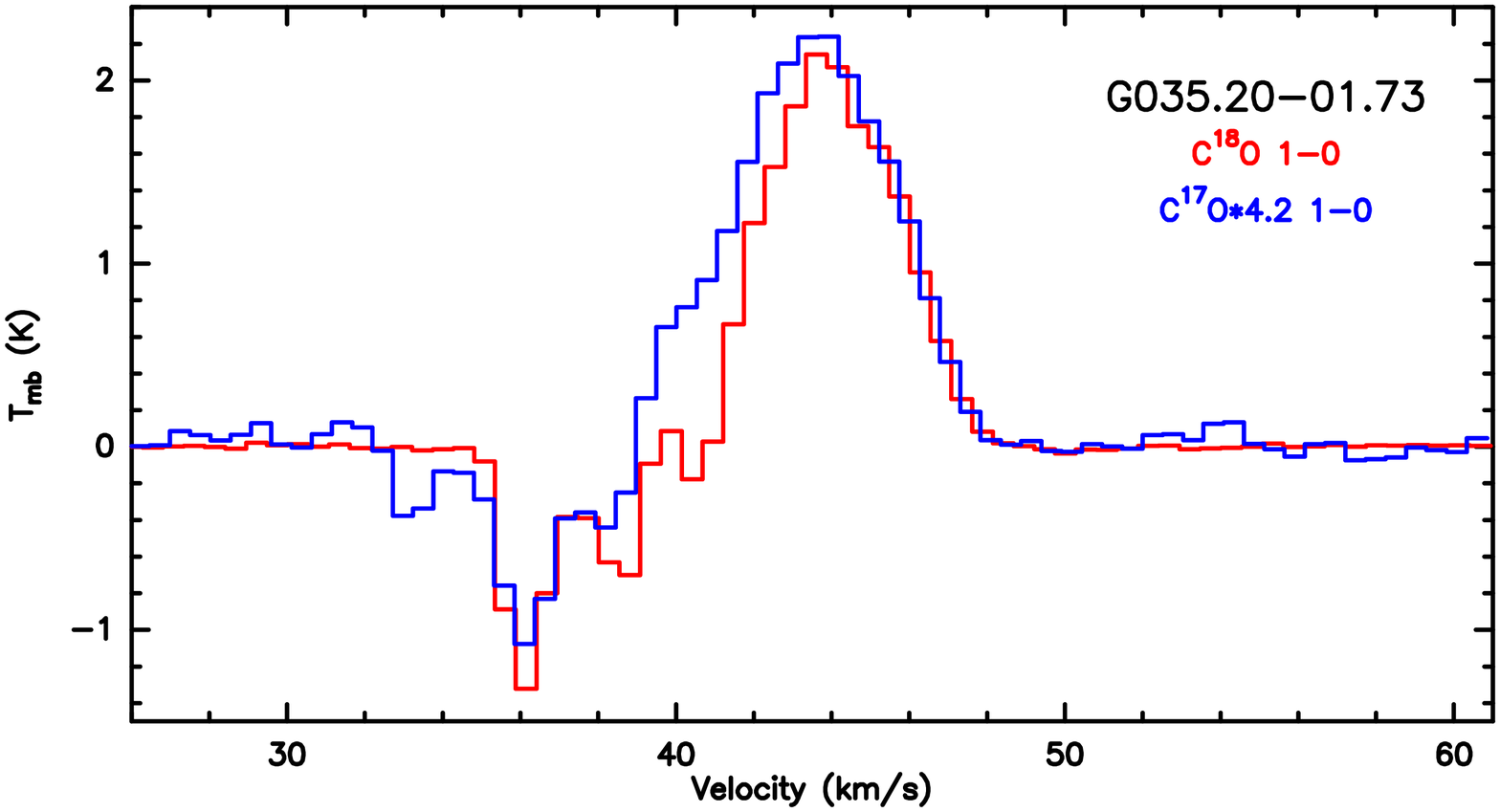}\includegraphics[width=0.45\textwidth]{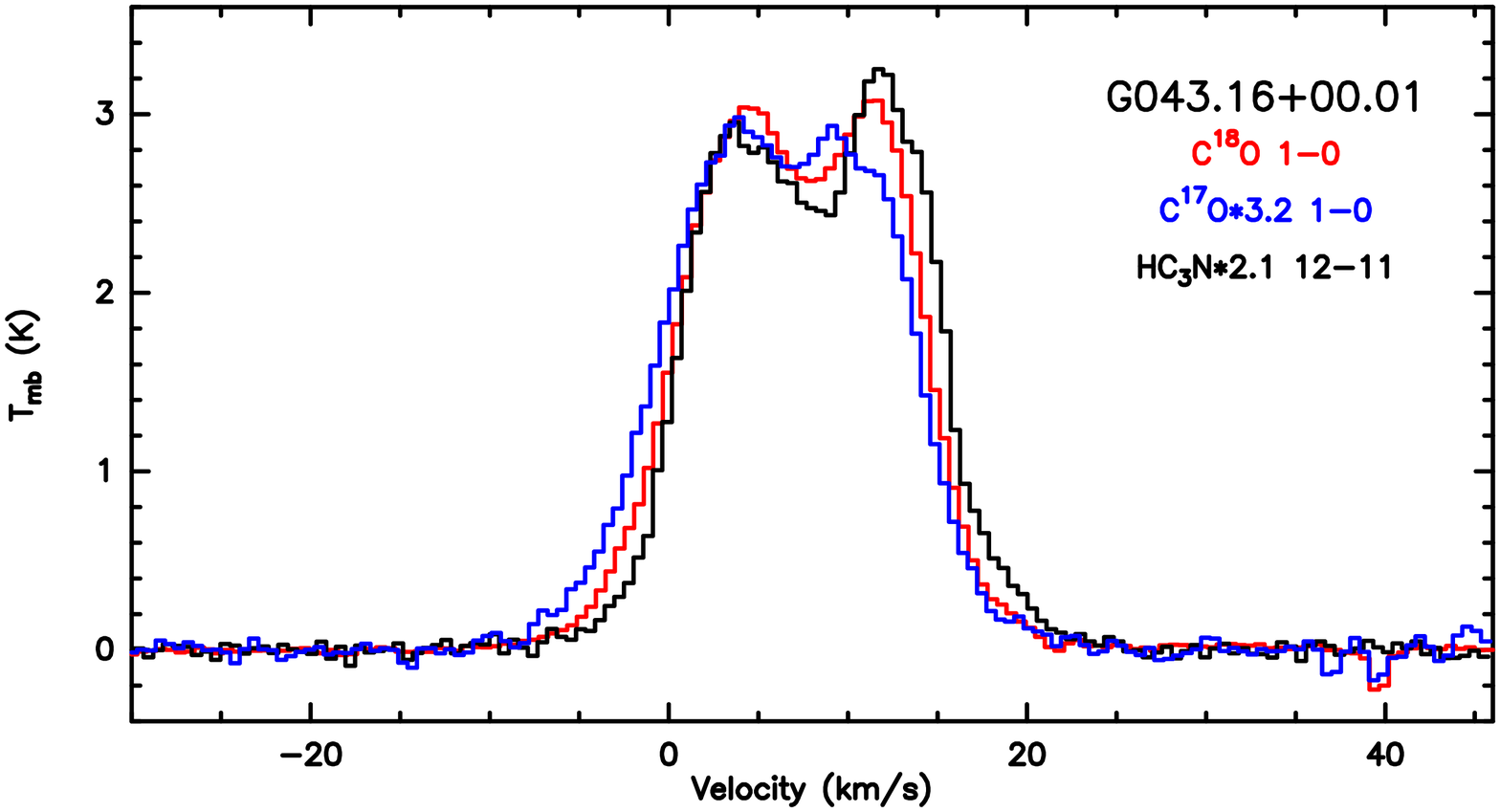}\\
\includegraphics[width=0.45\textwidth]{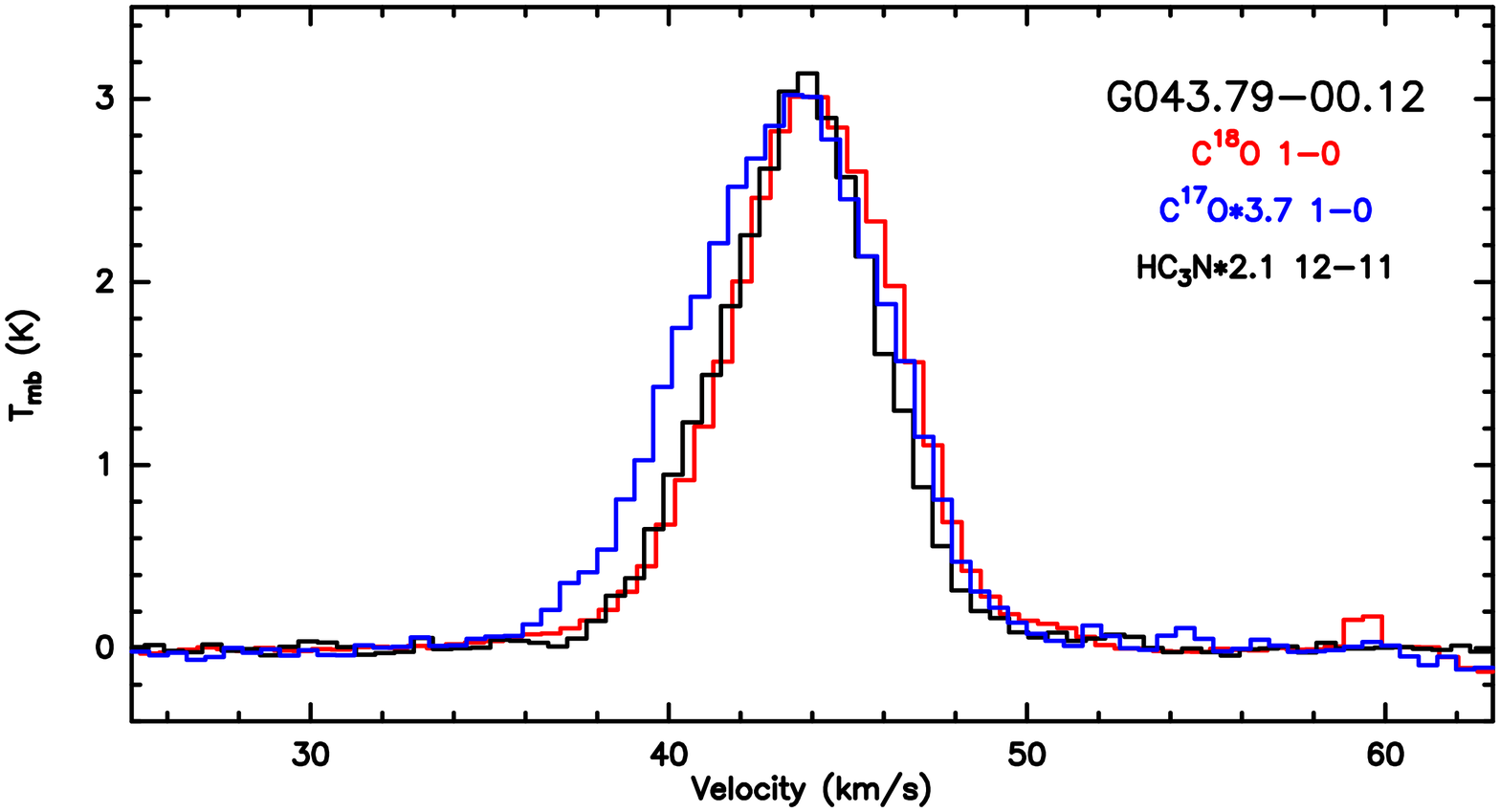}\includegraphics[width=0.45\textwidth]{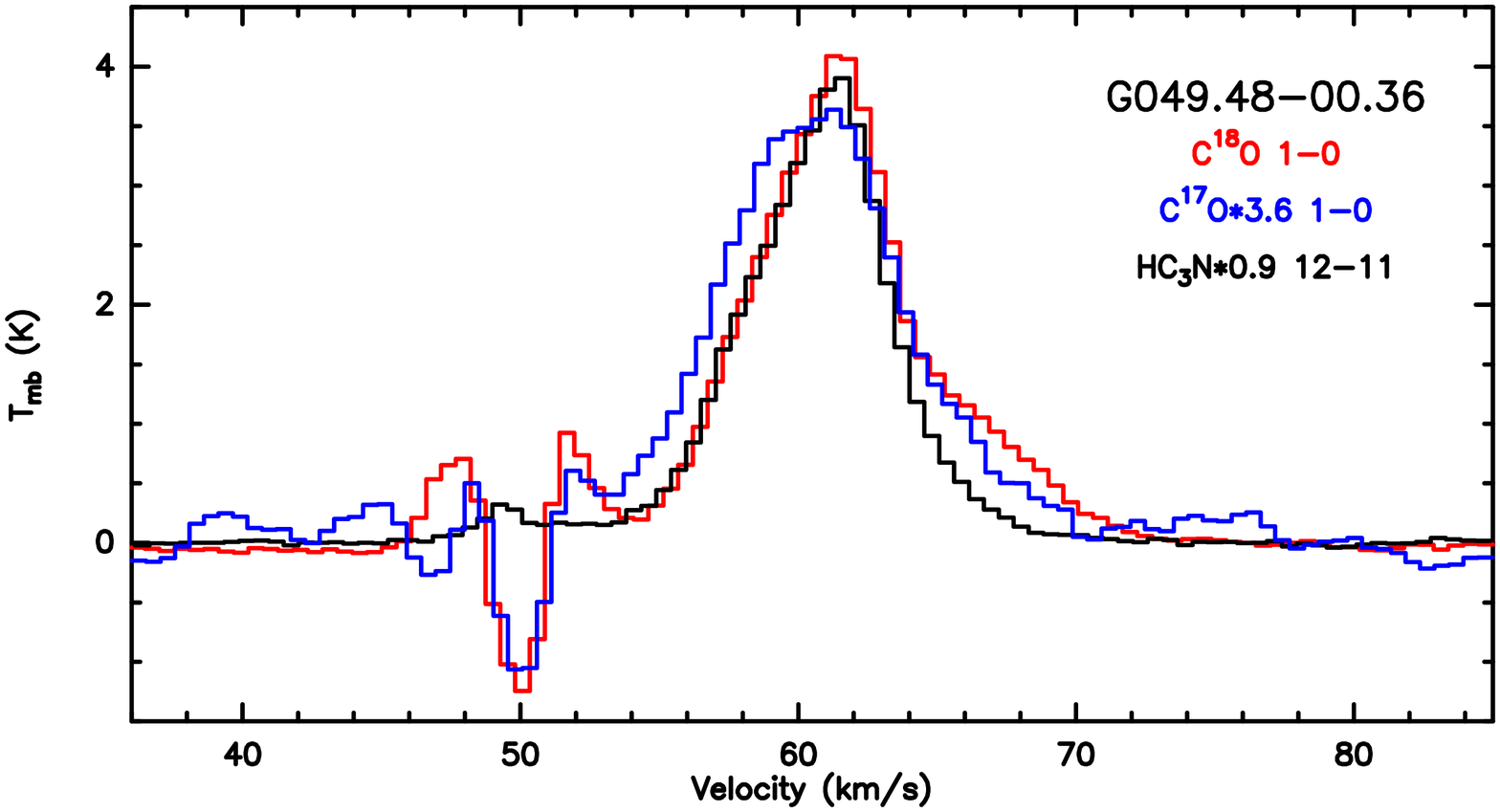}\\
\caption{Continued.}
\label{spectrum1}
\end{figure}
\begin{figure}
\centering
\addtocounter{figure}{-1}
\includegraphics[width=0.45\textwidth]{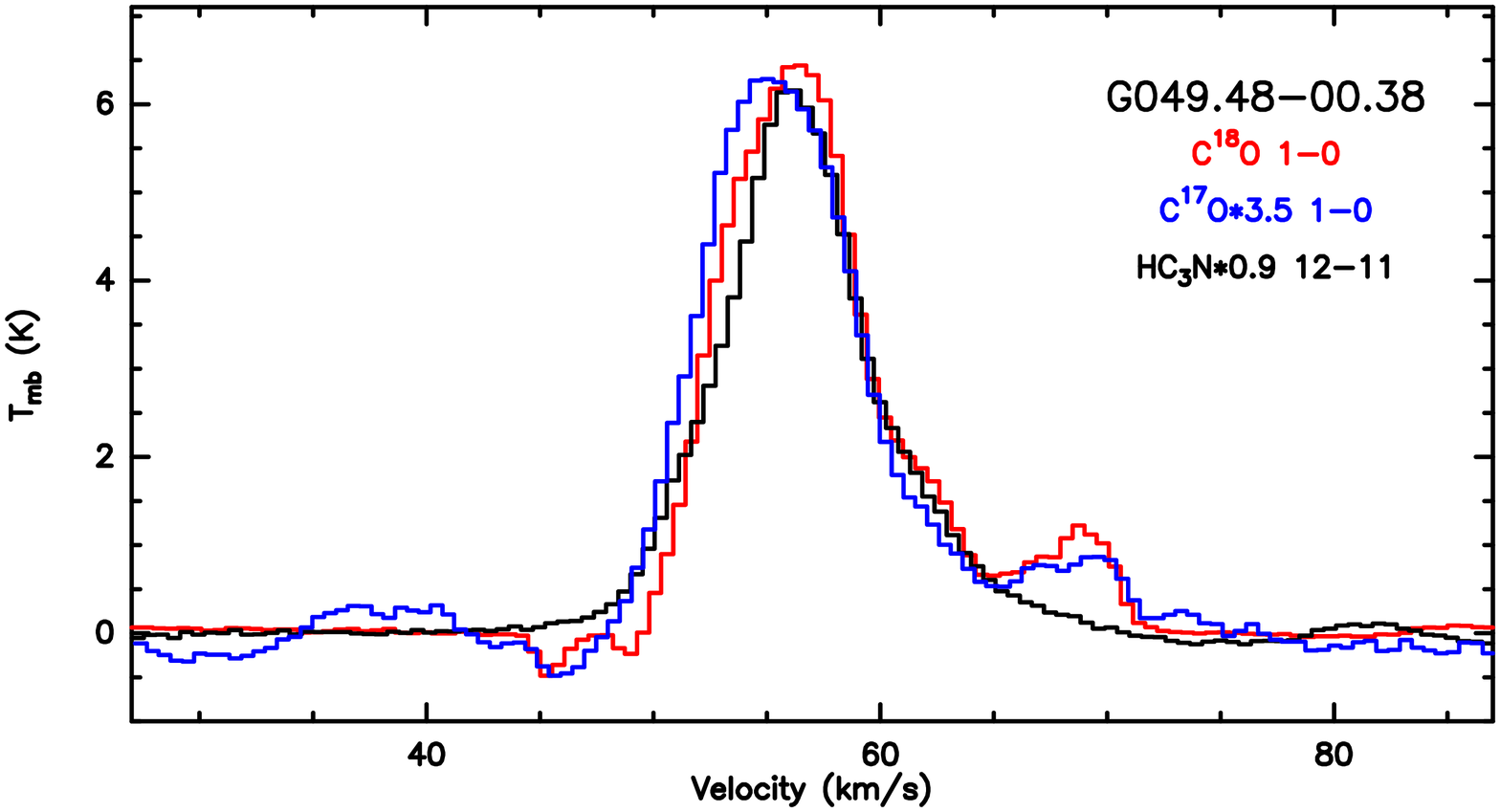}\includegraphics[width=0.45\textwidth]{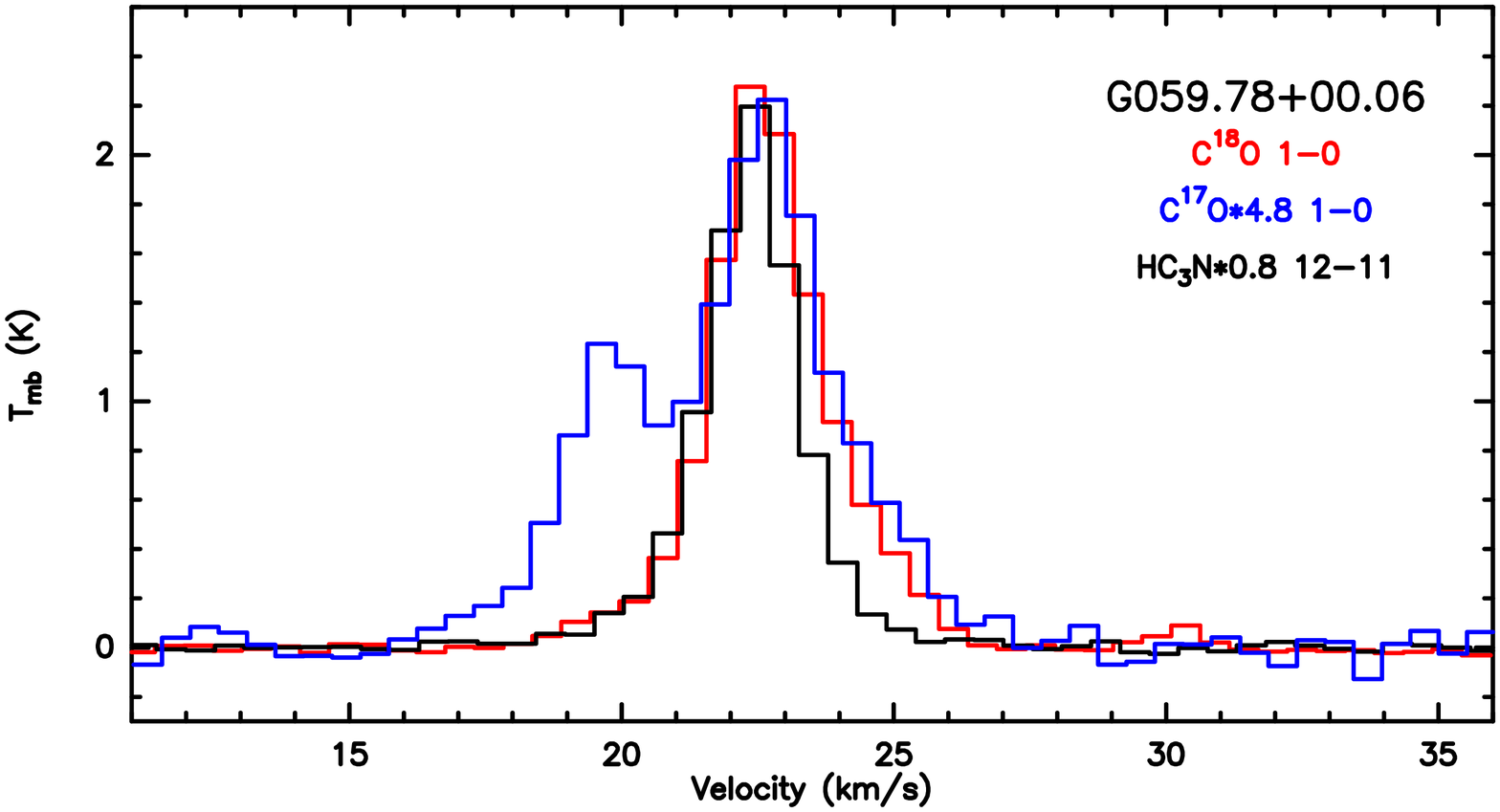}\\
\includegraphics[width=0.45\textwidth]{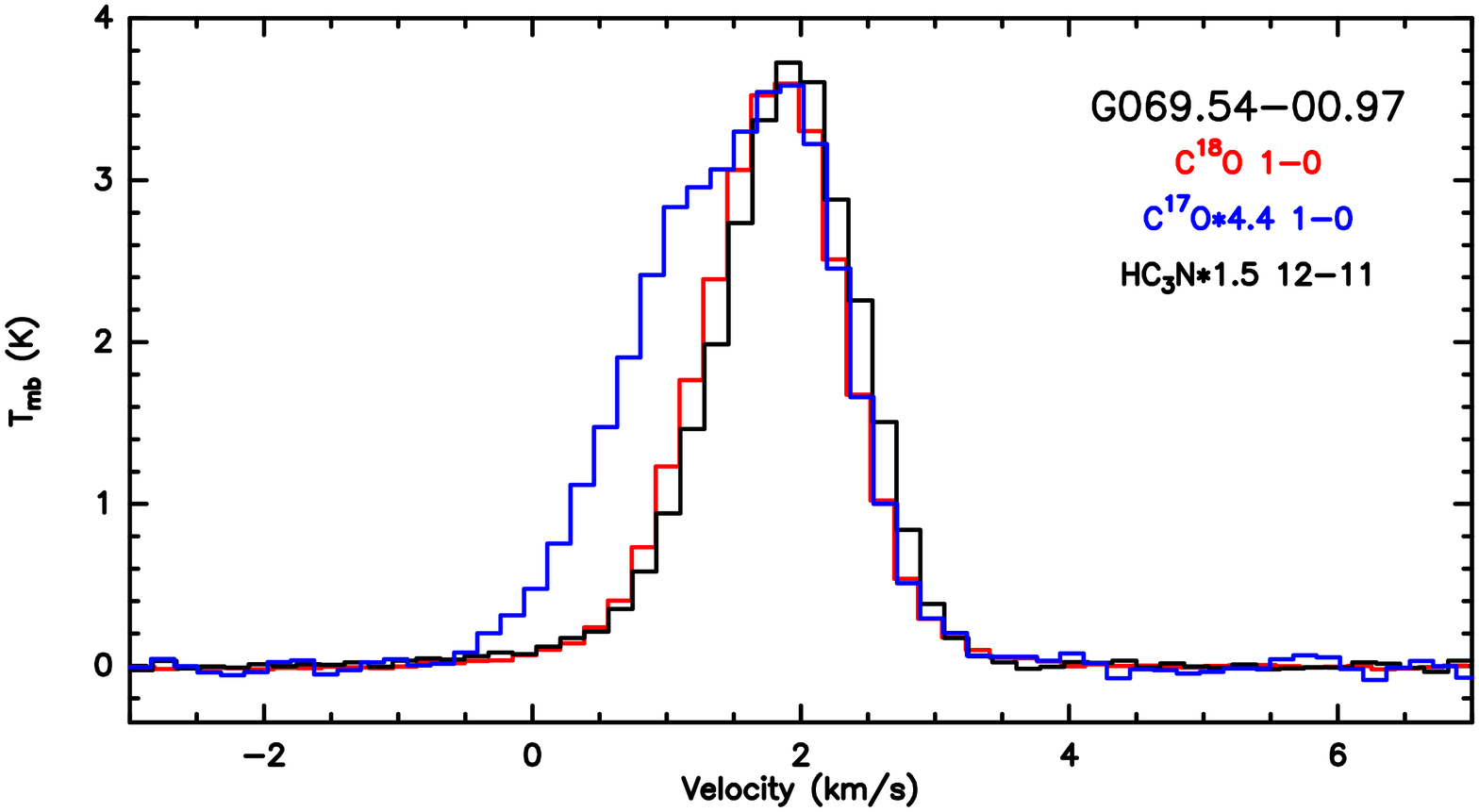}\includegraphics[width=0.45\textwidth]{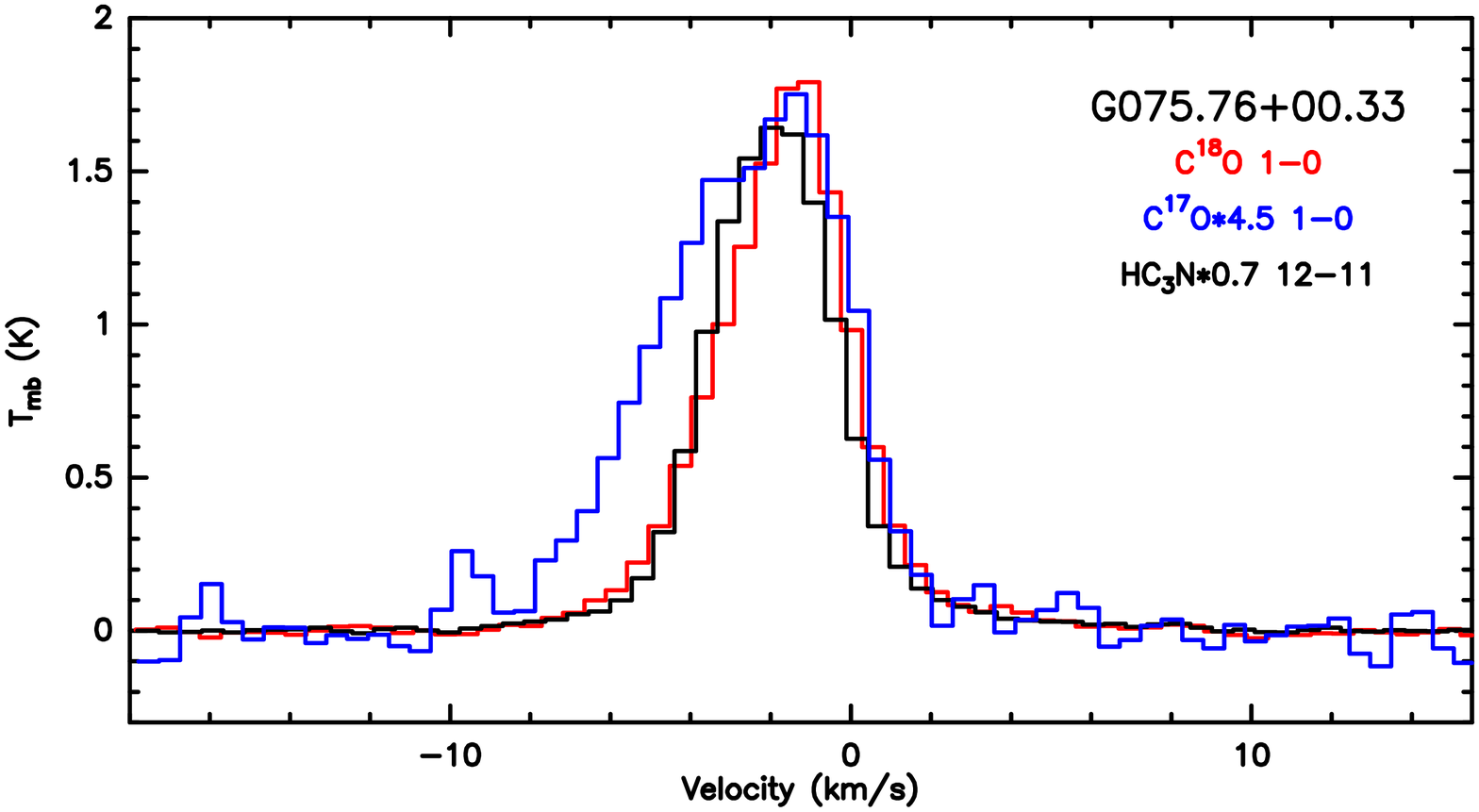}\\
\includegraphics[width=0.45\textwidth]{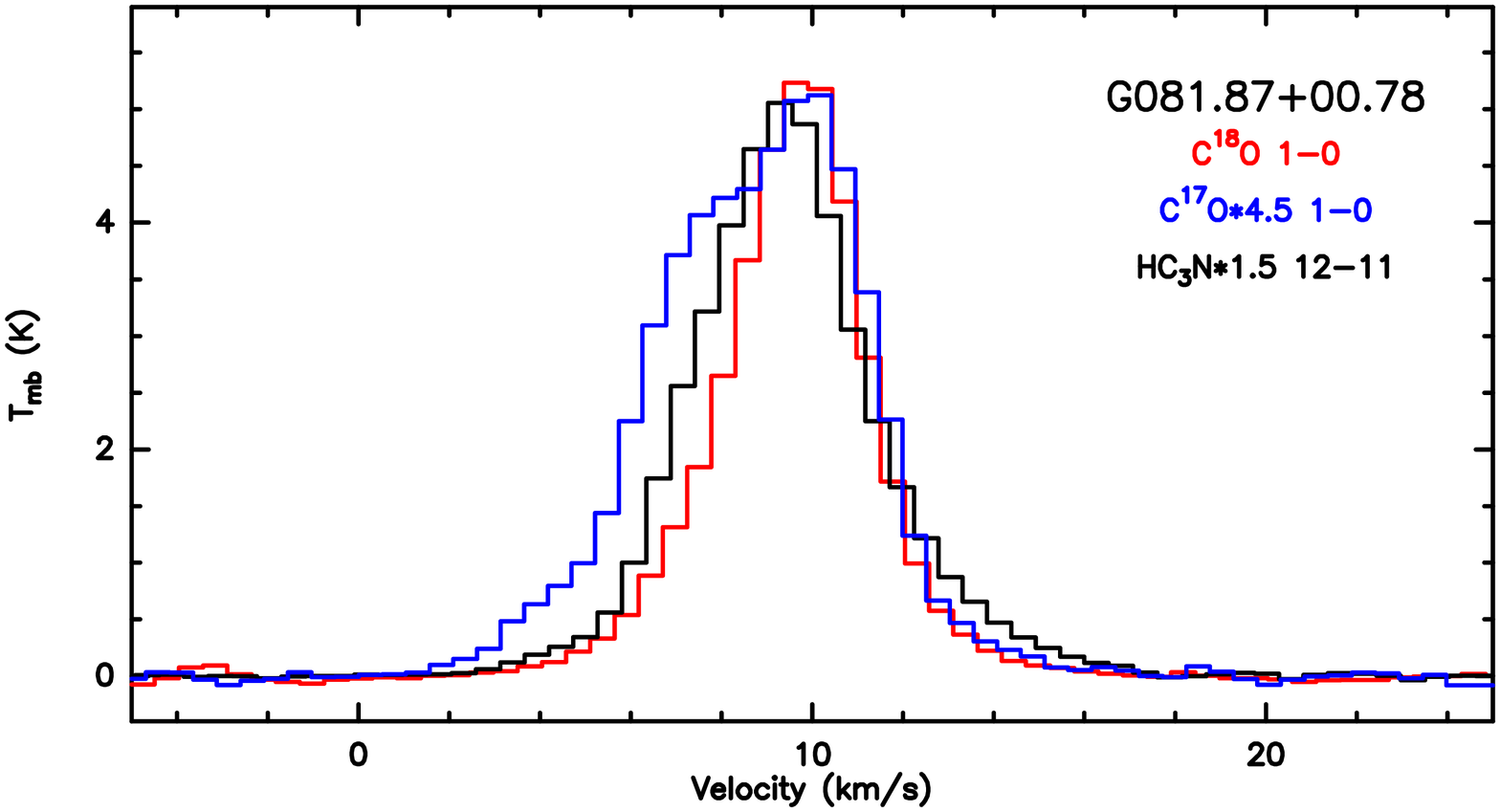}\includegraphics[width=0.45\textwidth]{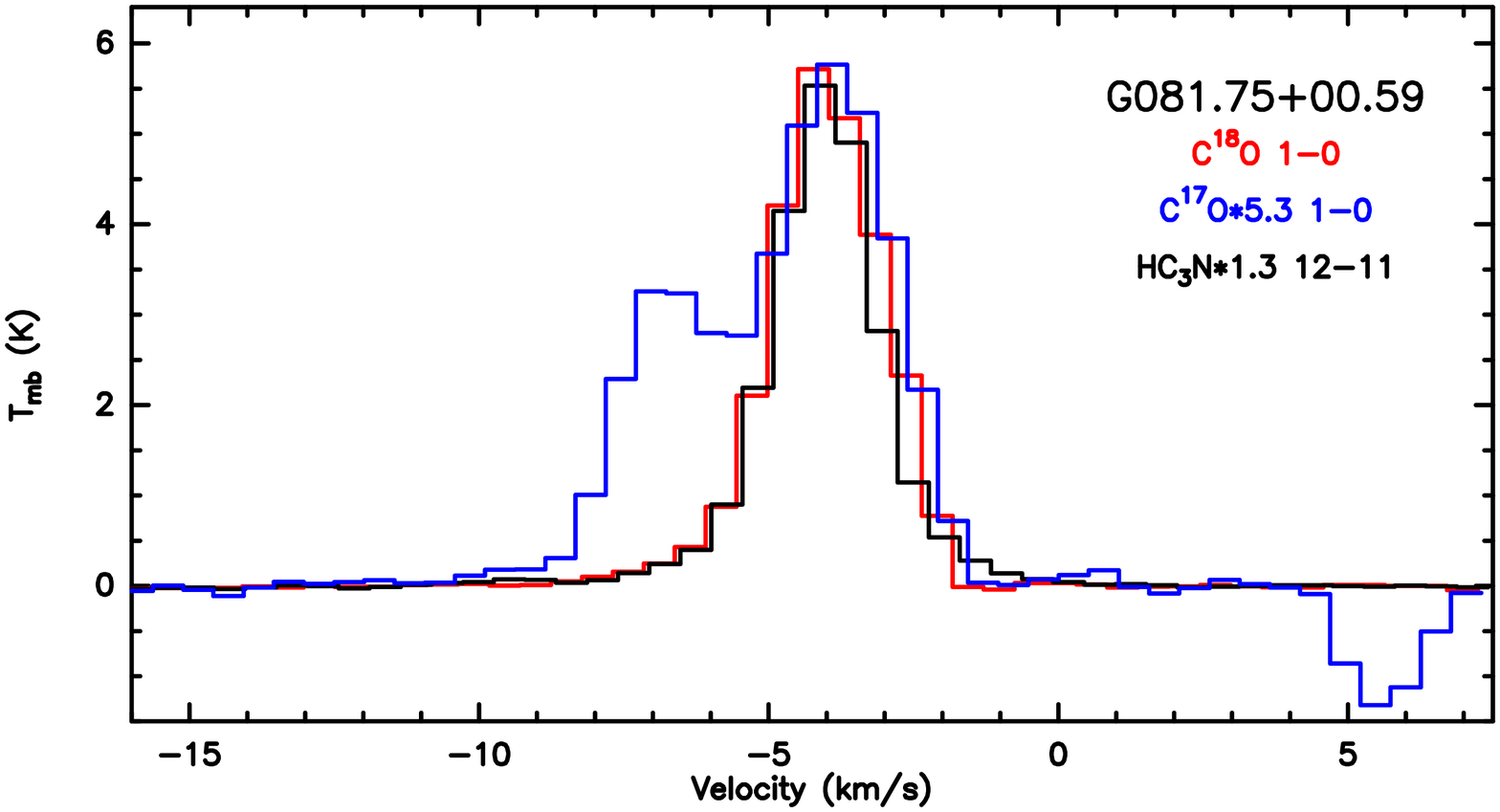}\\
\includegraphics[width=0.45\textwidth]{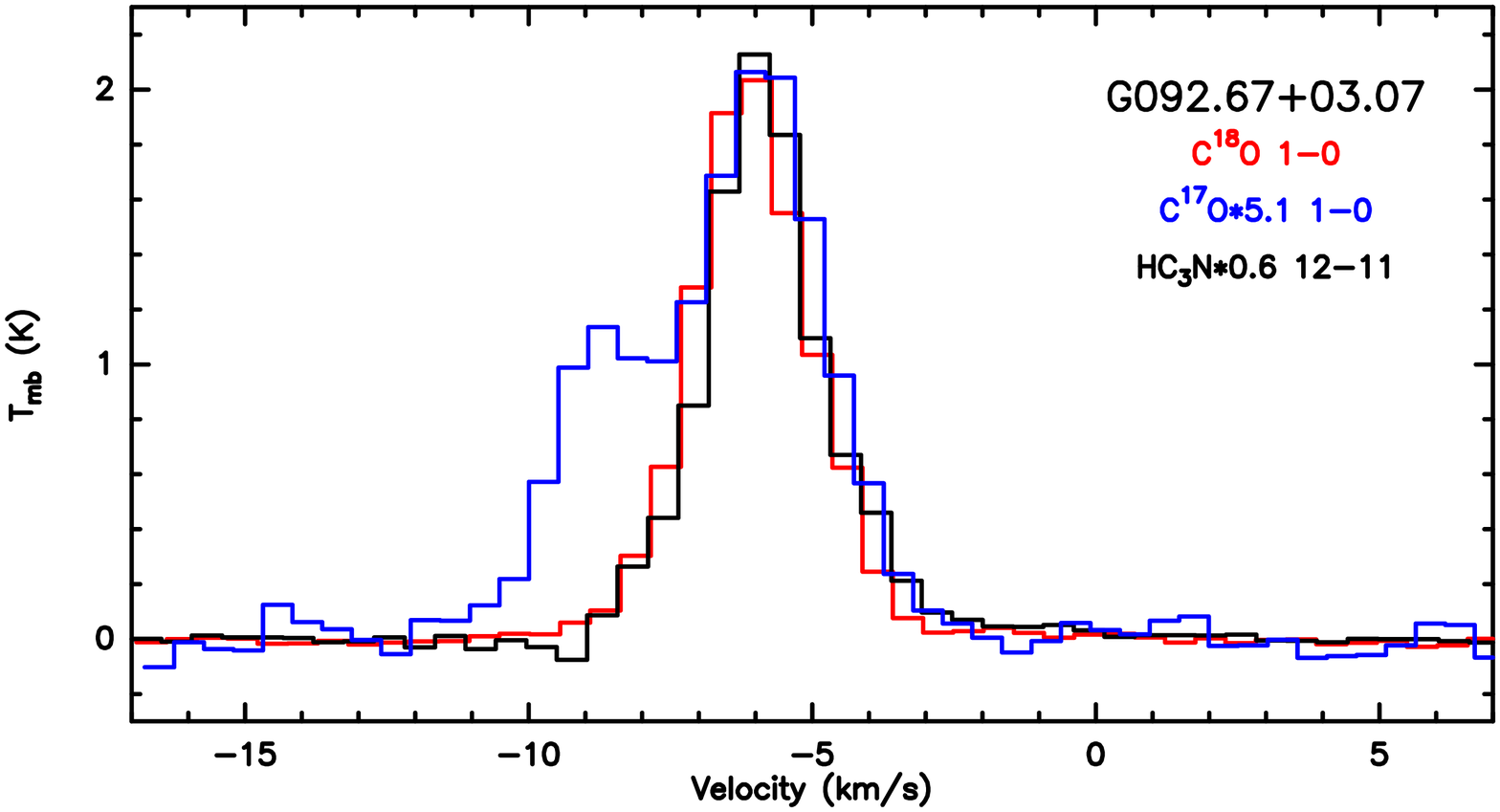}\includegraphics[width=0.45\textwidth]{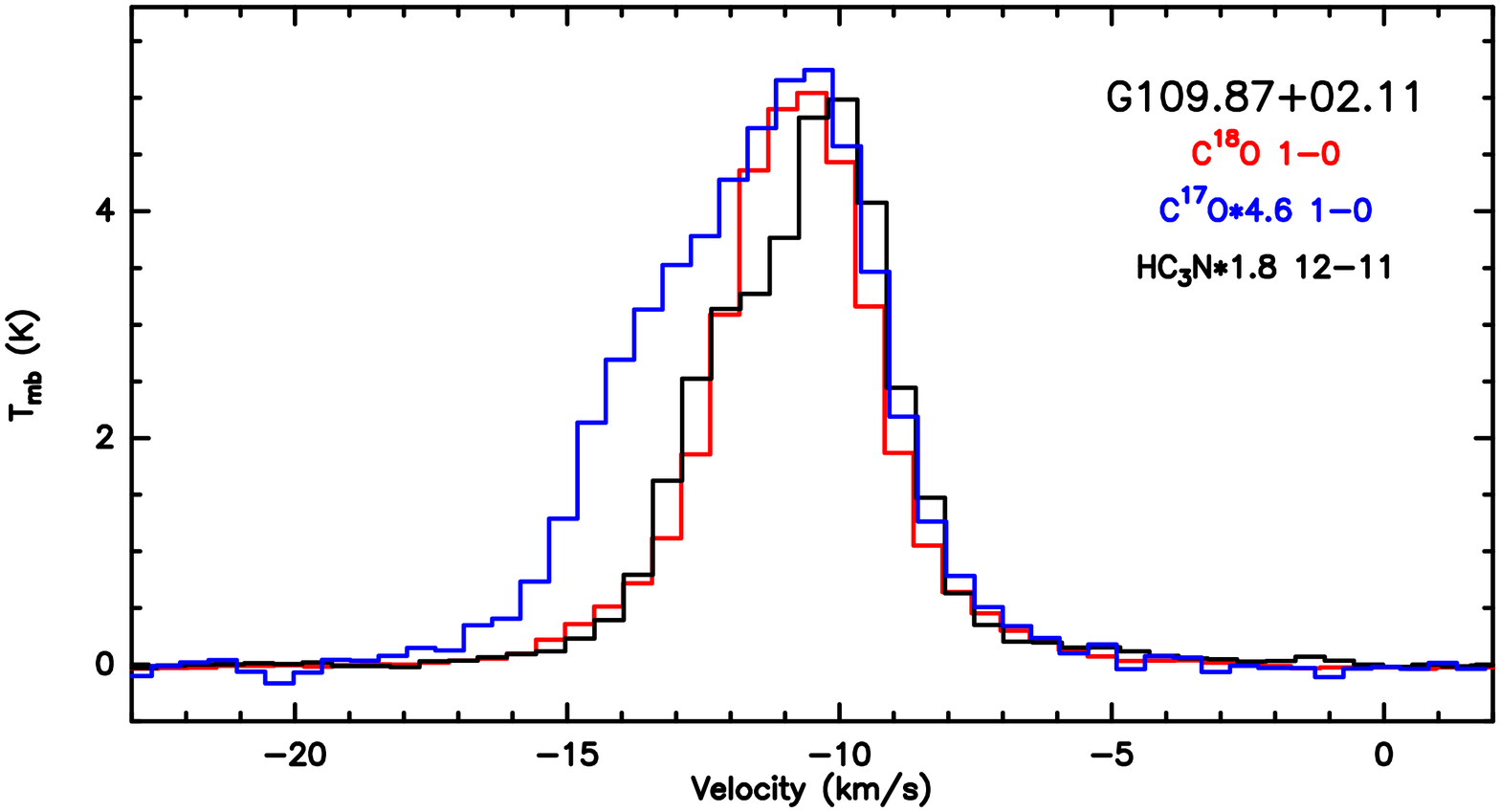}\\
\includegraphics[width=0.45\textwidth]{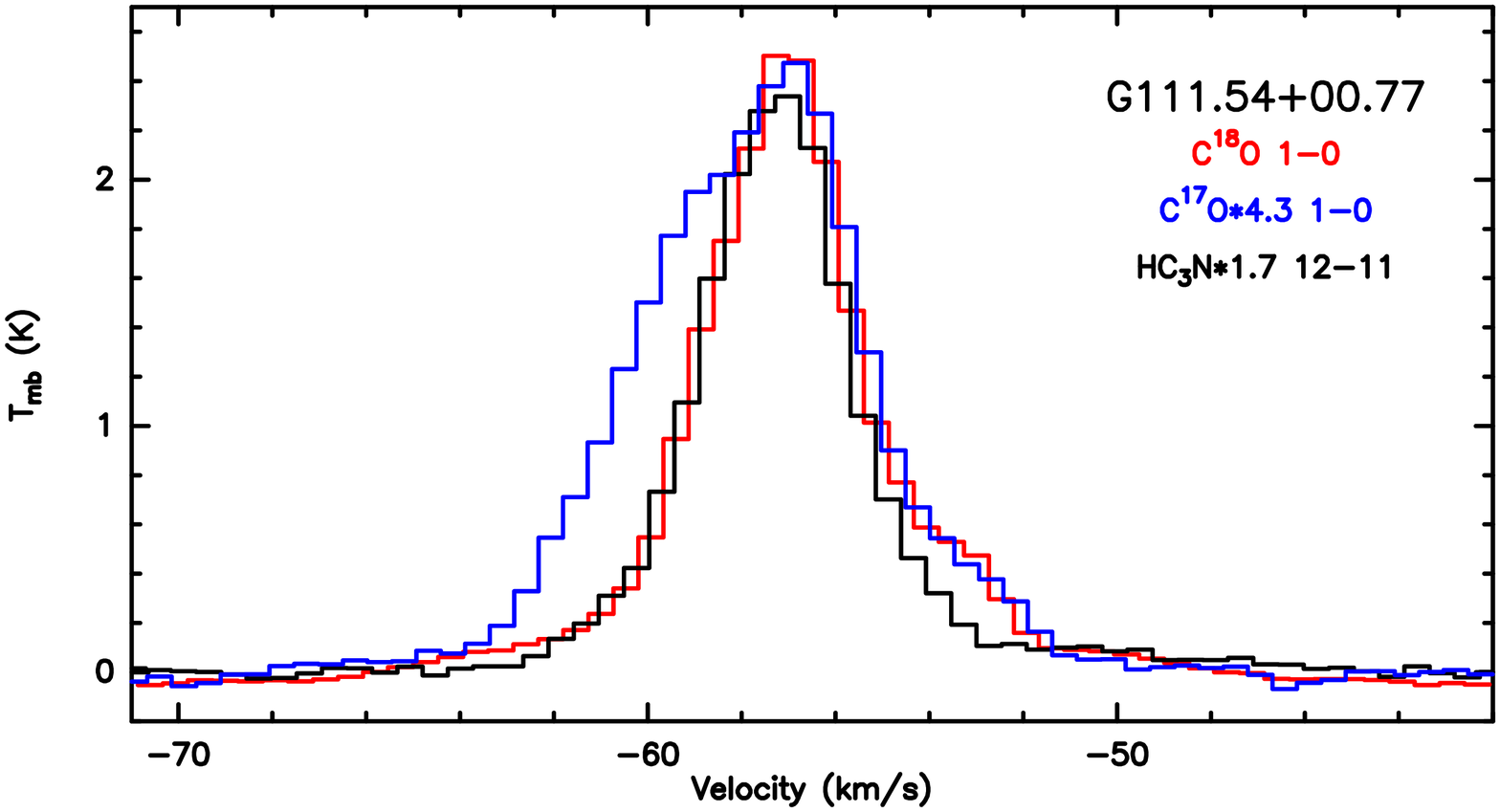}\\
\caption{Continued.}
\label{spectrum1}
\end{figure}

\begin{figure}
\centering
\includegraphics[width=0.45\textwidth]{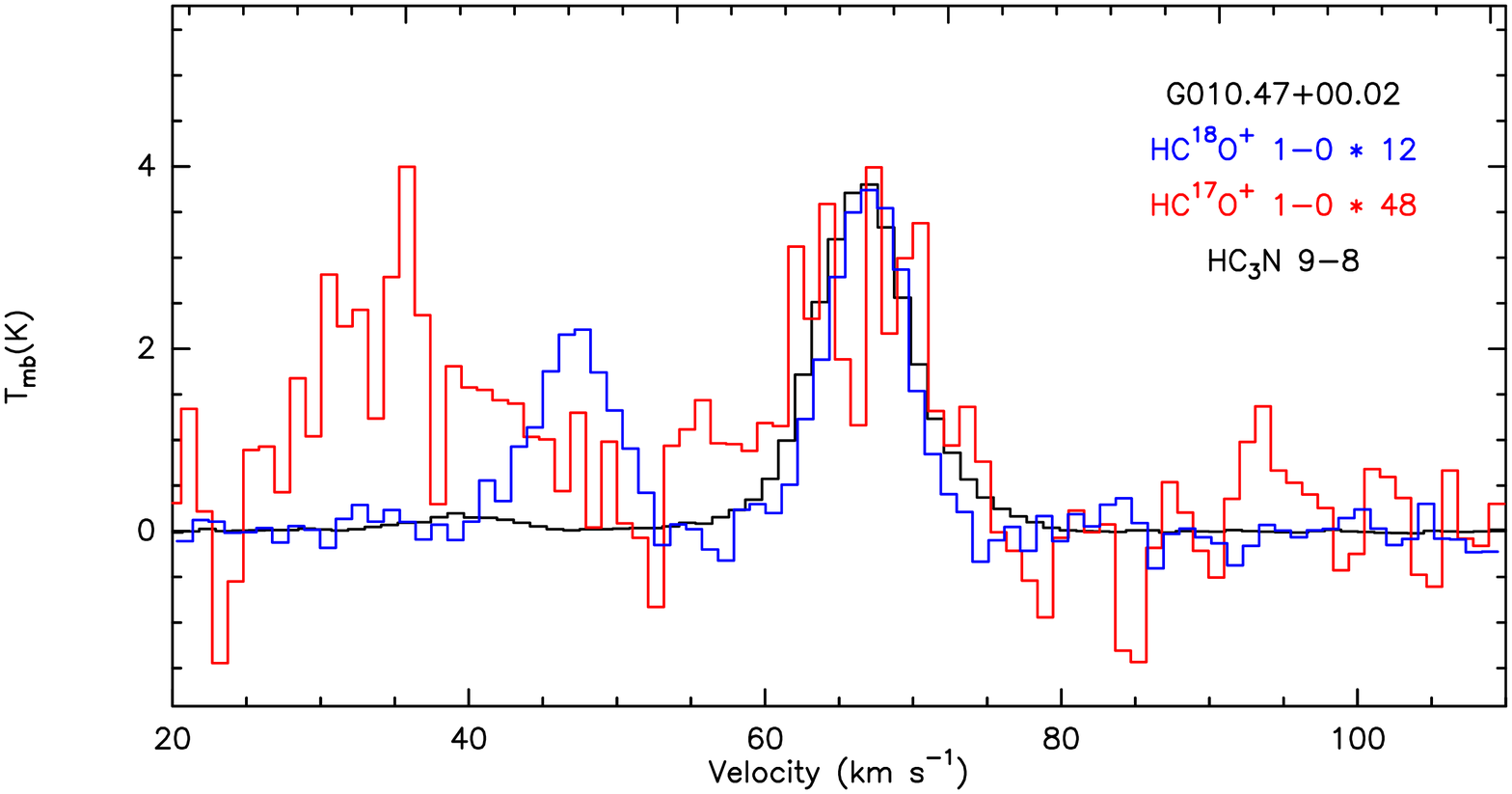}\includegraphics[width=0.45\textwidth]{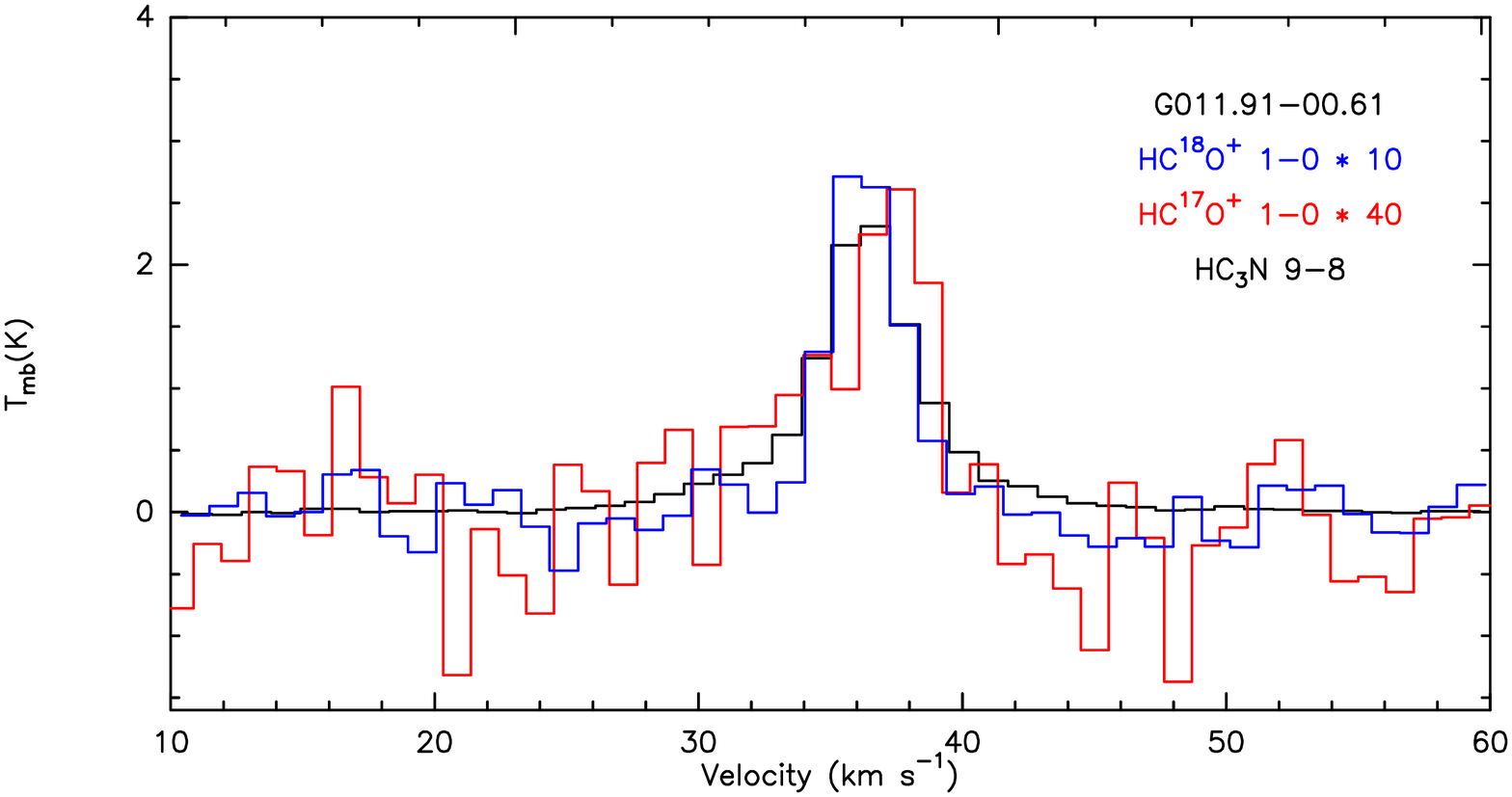}\\
\includegraphics[width=0.45\textwidth]{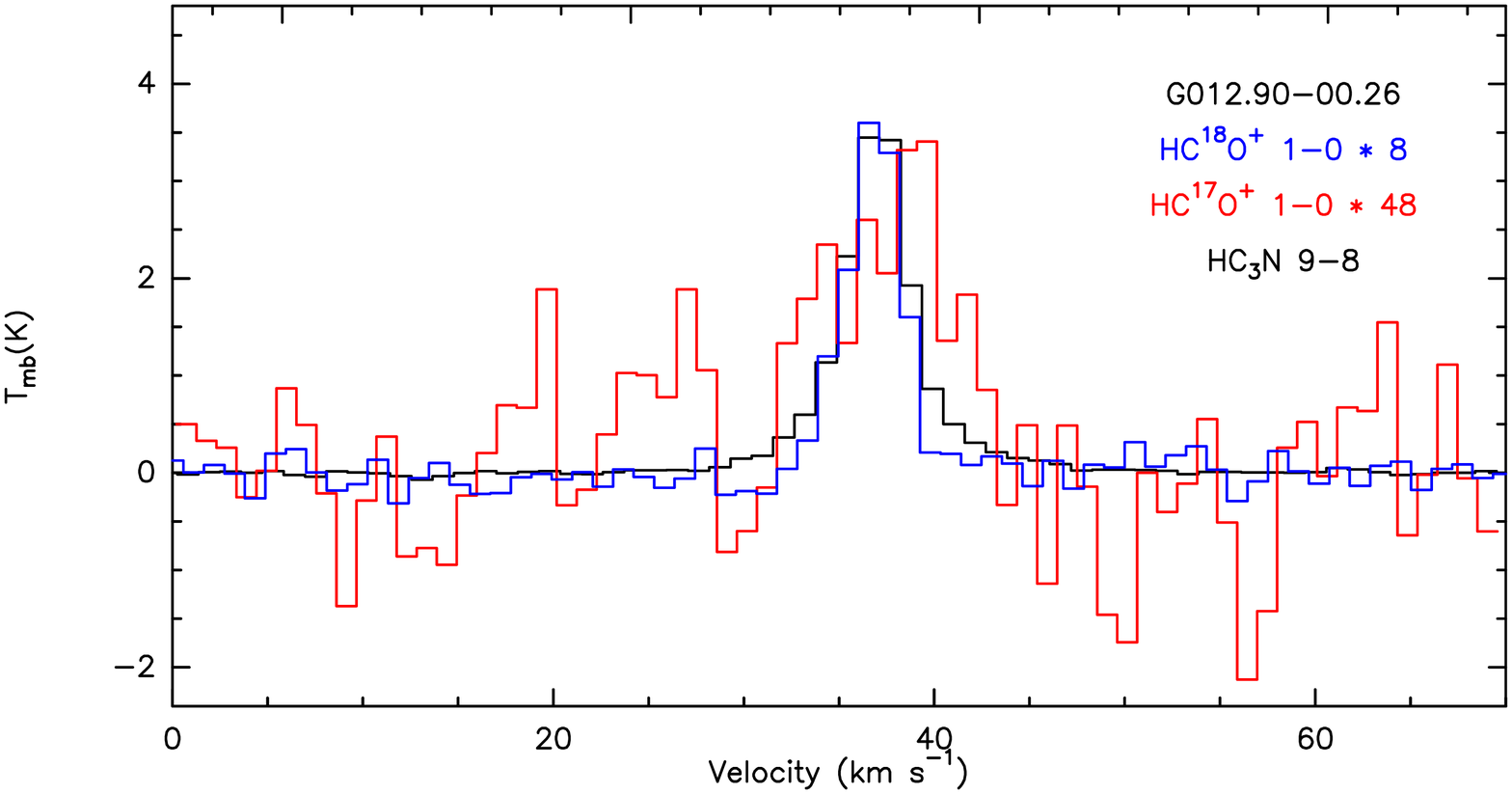}\includegraphics[width=0.45\textwidth]{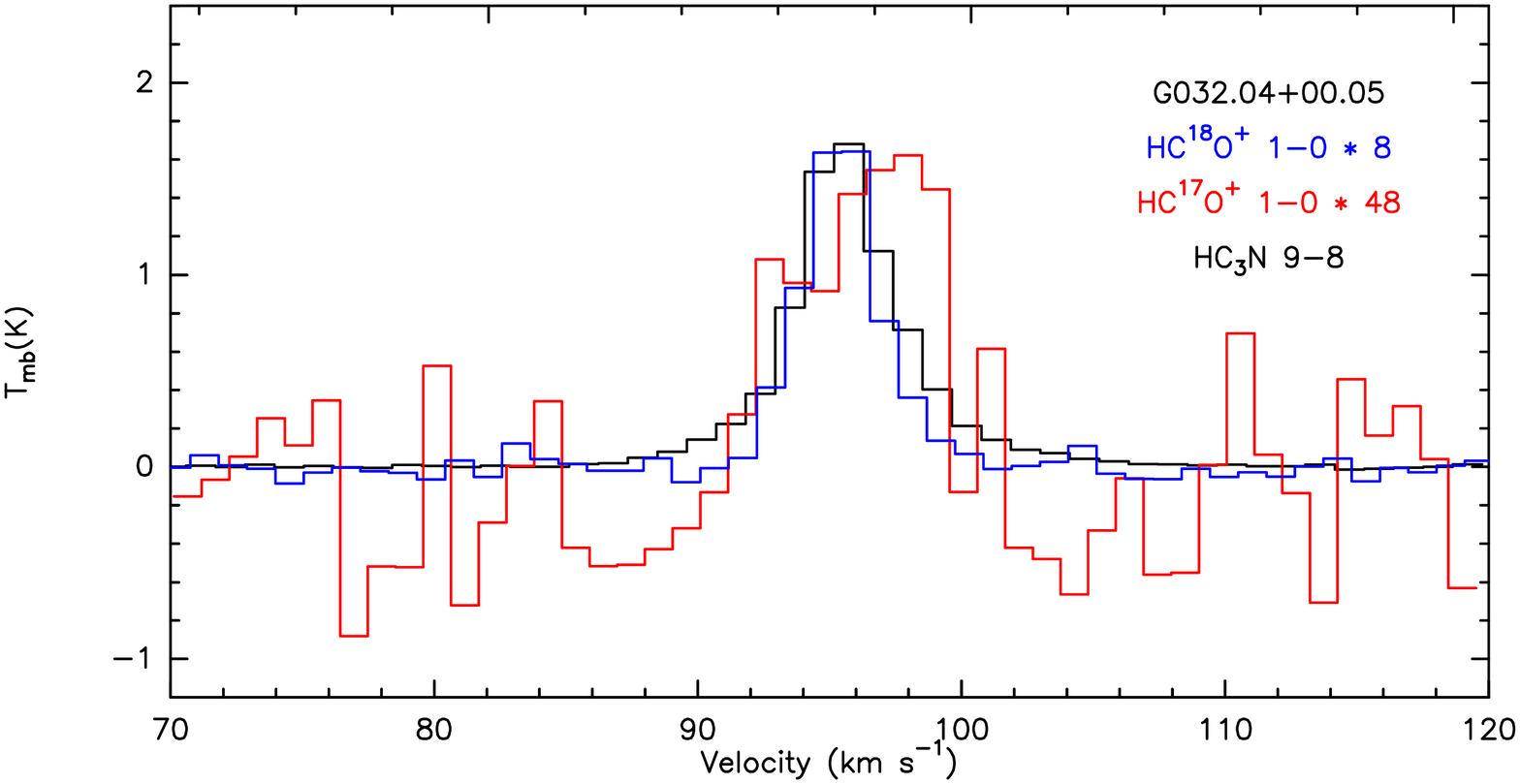}\\
\includegraphics[width=0.45\textwidth]{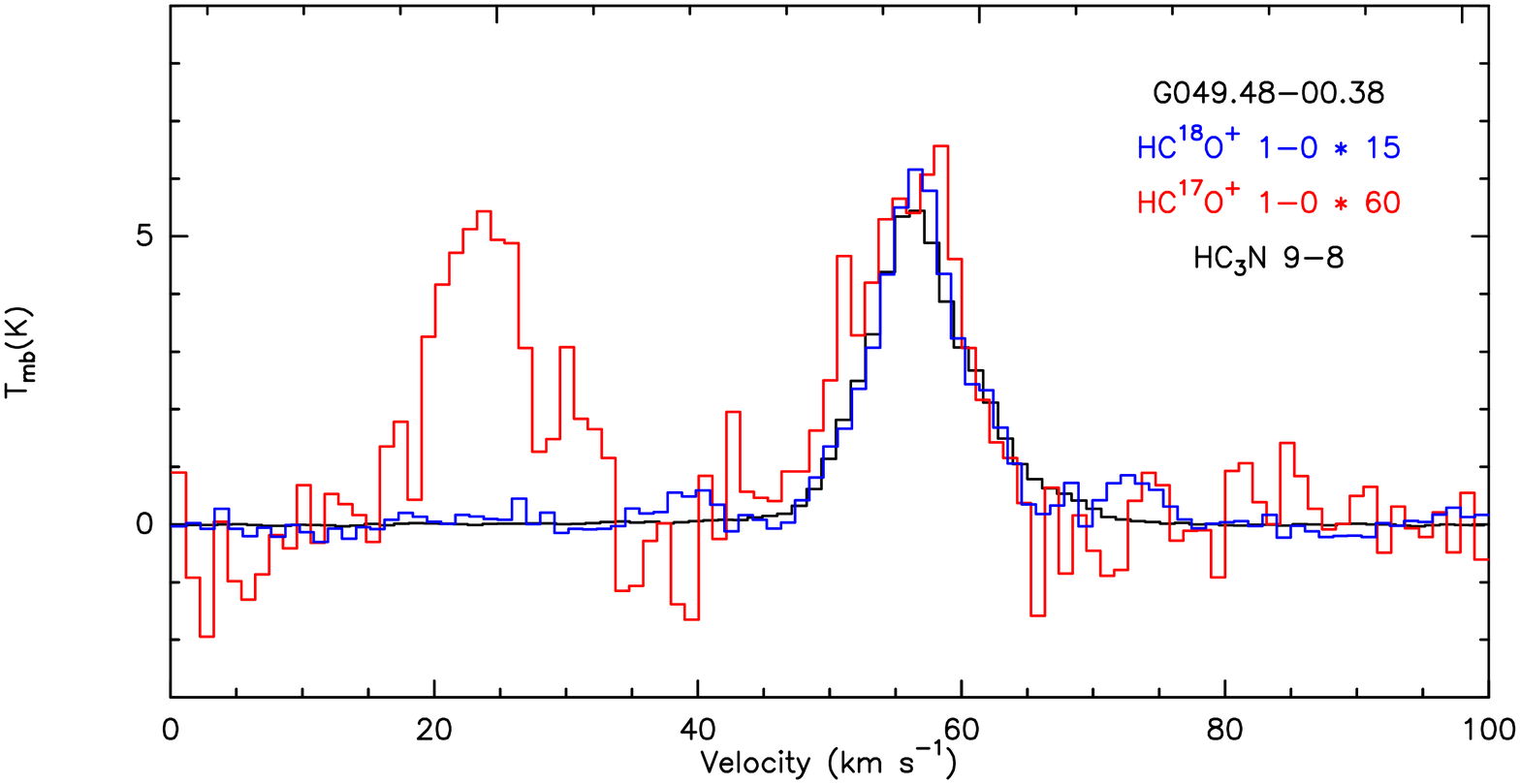}
\caption{The observational results of spectra for HC$^{18}$O$^{+}$ 1-0, HC$^{17}$O$^{+}$ 1-0 and HC$_{3}$N 9-8.}
\label{spectrum2}
\end{figure}


\bsp	
\label{lastpage}
\end{document}